\documentclass[nohyper,12pt,letterpaper]{JHEP3}
\usepackage{epsfig,youngtab}

\author{Robert de Mello Koch$^{1,2}$, Jelena Smolic$^{1}$ and Milena Smolic$^{1}$\\
\qquad \\
$^{1}$Department of Physics and Centre for Theoretical Physics,\\ 
University of the Witwatersrand,\\ 
Wits, 2050,\\ 
South Africa\\
\qquad\\
$^{2}$Stellenbosch Institute for Advanced Studies,\\
Stellenbosch,\\
South Africa\\
\qquad\\
E-mail: \email{robert@neo.phys.wits.ac.za, smolicj@science.pg.wits.ac.za, smolicm@science.pg.wits.ac.za}}

\abstract{
In this article, the free field theory limit of operators dual to giant gravitons with open strings attached, are studied. We 
introduce a graphical notation, which employs Young diagrams, for these operators. The computation of two point correlation 
functions is reduced to the application of three simple rules, written as graphical operations performed on the Young diagram 
labels of the operators. Using this technology, we have studied gravitational radiation by giant gravitons and bound states of 
giant gravitons, transitions between excited giant graviton states and joining of open strings attached to the giant. 
}

\preprint{WITS-CTP-031}

\title{Giant Gravitons - with Strings Attached (I)}


\def \Tr{\mbox{Tr\,}}
\def \threetwo{{}^3_2}
\def \twothree{{}^2_3}
\def \onetwo{{}^1_2}
\def \twoone{{}^2_1}
\def \oneone{{}^1_1}
\def \twotwo{{}^2_2}

\def \toonetwo{{}^\to_{{}^1_2}}
\def \totwoone{{}^\to_{{}^2_1}}

\def \onethree{{}^1_3}
\def \threeone{{}^3_1}
\def \totwo{{}^\to_2}
\def \toone{{}^\to_1}

\begin{document}

\section{Introduction}

The gauge theory/gravity correspondence\cite{Maldacena:1997re},\cite{Gubser:1998bc},\cite{Witten:1998qj} has provided 
deep and unexpected links between string theory on negatively curved spacetime, large $N$ quantum field 
theories and spin chains. For example, according to the correspondence, the masses of states in the string theory coincide
with anomalous dimensions of operators in the field theory. These anomalous dimensions are also given by the 
spectrum of a spin chain. In this article we will study the original example of the correspondence, which relates
${\cal N}=4$ super Yang-Mills theory to type IIB string theory on the AdS$_5\times$S$^5$ background. 

The correspondence is a strong/weak coupling duality in the 't Hooft coupling of the 
field theory. Therefore, computations that can be carried out on both sides of the 
correspondence necessarily compute quantities that are not corrected or receive 
small corrections, allowing weak coupling results to be extrapolated to strong 
coupling. As an example, an interesting set of observables to consider are dual to single graviton
perturbations of AdS$_5\times$S$^5$. These perturbations are members of a BPS multiplet so that their
energies are protected by supersymmetry. For these observables, one could also test that
AdS/CFT reproduces the correct field theory correlation functions.

Part of the difficulty in understanding the AdS/CFT correspondence is related to the fact that we do not as yet
have a good understanding of how to quantize the superstring in this background. There are however, interesting
geometric limits where the string can be quantized exactly\cite{pp}. On the field theory side, one takes a 
different large $N$ limit - the energy and ${\cal R}$-charges of the observables are also scaled in the new limit\cite{Berenstein:2002jq}.
The result is a systematic way to relax the supersymmetry enjoyed by a state. The operators which one obtains (BMN loops)
are nearly supersymmetric, and consequently although they are corrected, the perturbation theory is suppressed by
the large quantum numbers of the states in question. This allowed the use of perturbation theory to explore the
strong coupling regime of the gauge theory and consequently, truly stringy aspects of the AdS/CFT correspondence
could be probed. 

The single graviton observables could be studied, on the gravity side of the correspondence, using a supergravity approximation.
The states dual to BMN loops could be studied within the framework of perturbative string theory in the pp-wave limit. However,
we should be able to do even more. 
The AdS/CFT correspondence is a non-perturbative equivalence, and hence should also allow one to understand non perturbative objects
in string theory on the AdS geometry. Since we only have control over the weakly coupled limit of the field
theory, we still want to study objects that are protected. Giant gravitons are good examples of protected non perturbative objects.
Indeed, since they are half-BPS states, giant gravitons have proved to be the source 
of many valuable quantities that are accessible on both sides of the correspondence.

Giant graviton solutions describe branes extended in the sphere\cite{McGreevy:2000cw} or in the 
AdS space \cite{Grisaru:2000zn},\cite{Hashimoto:2000zp},\cite{Das:2000fu}
of the AdS$\times$S background. The giant gravitons are (classically) stable
due to the presence of the five form flux which produces a force that exactly
balances their tension. The dual description of giant gravitons in the Yang-Mills
theory has enjoyed some progress. Operators dual to giant gravitons (Schur polynomials in the Higgs fields) 
have been proposed in \cite{Corley:2001zk},\cite{Balasubramanian:2001nh}. However, 
despite compelling evidence (reviewed in section 2.1) for the identification of Schur polynomials as the
operators dual to giant gravitons, a detailed understanding of this correspondence has not yet been achieved.
For example, even a basic understanding of how the geometry of the giant graviton is coded in 
the dual operator, is not complete.

Excitations of giant gravitons can be obtained by attaching open strings to the giants.
The gauge theory operators dual to excited sphere giants are known and their anomalous
dimensions reproduce the expected open string spectrum\cite{Balasubramanian:2002sa}\footnote{See 
\cite{Berenstein:2006qk},\cite{Strings},\cite{Shanin}
for further studies of non-BPS excitations that have been interpreted as open strings attached to giant gravitons.}.
Recently, operators dual to an arbitrary system of excited giant gravitons have been 
proposed in the insightful paper\cite{Balasubramanian:2004nb}. The proposal has been 
tested by showing that the restrictions imposed on excitations of the system, by the Gauss 
law, are indeed satisfied. For the general case, it is an involved combinatoric task to 
compute the two point functions of these operators (even in the free field limit)
and further to compute their anomalous dimensions. Our goal in this article is to start 
developing the required technology.

The study of excitations of giant gravitons is likely to shed light on the precise details of
the Schur polynomial/giant graviton correspondence, which is one of the main motivations for our 
study. Concretely, we study two point functions in the free theory, of operators dual to excited 
giant graviton systems. With a specific choice for the open string excitations, we can factorize 
the problem of computing correlators into a factor involving only the open string words and a factor 
involving the Schur polynomial itself. The factor involving the open string words can be dealt with
using the usual ${1\over N}$ expansion. The factor involving the Schur polynomial contains $O(N)$
fields, so that huge combinatoric factors over power the usual ${1\over N^2}$ suppression of non-planar
diagrams and new techniques for evaluating these factors are needed. We develop the required technology.
Our result is a set of simple geometrical operations performed on the Young diagram labeling the operator,
allowing an efficient computation of the factor involving the Schur polynomials. In this article we
deal with the free field theory. However, it is possible to extend our rules, providing efficient
methods that allow the computation of the leading $g_{YM}^2$ correction coming from the $F$-terms. This
is reported in a separate paper\cite{jelena}.

The world volume theory of giant gravitons at low energy is a Yang-Mills theory. Since the world volume
of the giant gravitons does not coincide with the spacetime on which the original Yang-Mills theory is defined,
this new Yang-Mills theory (which ``emerges" from the original one)
will be local in a new space\cite{Balasubramanian:2004nb}, which arises from the 
matrix degrees of freedom of the original field theory. Our main application for the methods developed
in this article, is to explore this emergent Yang-Mills theory. In particular, we have studied gravitational radiation 
by giant gravitons and bound states of giant gravitons, transitions between excited giant graviton states and the joining 
and splitting of open strings attached to the giant. The results of our study seem to be consistent with an emergent gauge 
theory.

Our article is organized as follows: In section 2 we start by reviewing the Schur polynomial/giant graviton correspondence.
We then introduce the operators which are conjectured to be dual to excited giant gravitons. Section 2 concludes by 
summarizing three graphical rules which allow the computation of two point correlators. The technical details of the
derivation of the rules are summarized in the appendices. These technical details are not needed to master the rules,
which are presented (together with examples) in sections 2.5 and 2.6. A reader wishing only to use our results, need not read
more than sections 2.2, 2.3, 2.4 and 2.5. Since the computations involved are rather technical, we have taken the
quality assurance of our methods seriously. Towards this end, we have written a MATLAB code which numerically evaluates 
the correlation functions we study. In appendix J some of these numerical results are summarized. We have verified that
our methods correctly reproduce all correlators corresponding to Young diagrams with 5 boxes or less and with two or fewer
strings attached. 

Section 3 summarizes applications of our methods.
In section 3.1 we study gravitational radiation in a number of different settings. This allows us to probe locality in the bulk
and also the non-Abelian symmetry of the emergent gauge theory. In section 3.2 we consider transitions between excited giant
graviton states. In section 3.3 we study the splitting and joining of open strings attached to the giant graviton. The results
of this section seem to be consistent with a $3+1$ dimensional emergent gauge theory. Section 4 is used to discuss our results. 

\section{The Dual of Excited Giant Gravitons}

In this section the operators we study are defined. In the free field limit there is a simple factorization 
of the spacetime dependence and the factor arising from the color combinatorics\cite{Minahan:2002ve}. The spacetime dependence is
trivial and consequently, uninteresting. For this reason, we drop all spacetime dependence and focus on the color
combinatorics. This amounts to working in zero dimensions. The reader interested in a summary of our main results 
should consult section 2.2 for the definition of the operators we study and section 2.5 for a summary of
our technology.

We study the Lorentzian ${\cal N}=4$ super Yang-Mills theory on $R\times S^3$. 
The advantage of taking this approach is simply that $1/2$-BPS states (and systematically small deformations of
these states) of the theory on $R\times S^3$ can be described in the
s-wave reduction of the Yang-Mills theory, i.e. in a matrix quantum mechanics \cite{Berenstein:2004kk}.
According to the state-operator correspondence 
of conformal field theory, the generator associated to dilations on $R^4$ becomes the Hamiltonian for the theory on $R\times S^3$.
The action of ${\cal N}=4$ super Yang-Mills theory on $R\times S^3$ is
$$ S={N\over 4\pi \lambda}\int dt \int_{S^3} {d\Omega_3\over 2\pi^2}
\left( {1\over 2}(D\phi^i)(D\phi^i)+{1\over 4}\big(\big[\phi^i,\phi^j\big]\big)^2
- {1\over 2}\phi^i\phi^i \right),$$
where $\lambda =g_{YM}^2N$ is the 't Hooft coupling and $i,j=1,...,6$. We have not displayed the fermion or the gauge kinetic
terms in the action; in this article we only consider the scalar fields.  The mass term arises from 
conformal coupling to the metric of $S^3$. We group the six real scalars into three complex fields as follows
$$ Z=\phi^1+i\phi^2,\qquad Y=\phi^3+i\phi^4,\qquad X=\phi^5+i\phi^6 .$$
In what follows we will use these complex combinations. The free field theory propagators we use are
$$ \langle Z^\dagger_{ij}(t)Z_{kl}(t)\rangle = \delta_{il}\delta_{jk} = 
\langle Y^\dagger_{ij}(t)Y_{kl}(t)\rangle .$$
We should have a factor of ${4\pi\lambda\over N}$ multiplying the propagator. We have dropped this factor, which clutters formulas
and can easily be reinserted at the end of the computation; it contributes an overall factor of $\left({4\pi\lambda\over N}\right)^m$,
where $m$ is the total number of contractions performed in computing the correlator.

\subsection{Giants}

There are a number of important clues that can be used to identify the operators dual to giant
gravitons. We will study a class of half BPS giant gravitons. The half BPS chiral primary operators we focus on 
can be built from a single complex combination (we use $Z$ in what follows) of any two of the 
six Higgs fields appearing in the ${\cal N}=4$ super Yang-Mills theory. Using a total of $n$ $Z$s, 
there is a distinct operator for each partition of $n$. There is a one to one correspondence between 
these operators and half BPS representations of {\cal R} charge $n$\cite{Corley:2001zk}. For $n=O(1)$ these 
half BPS operators are dual to point like gravitons\cite{Witten:1998qj}, for $n=O(\sqrt{N})$ they are dual 
to strings\cite{Berenstein:2002jq} and for $n=O(N)$ they are dual to giant gravitons\cite{McGreevy:2000cw}. 
For the case that $n=O(1)$, operators composed of a product of a different number traces are orthogonal. This
naturally allows one to identify the number of traces with particle number and consequently, the supergravity Fock
space is clearly visible in the dual gauge theory. For $n=O(N)$, the usual suppression 
of non-planar diagrams is compensated by combinatoric factors, so that operators composed of a product of a 
different number of traces, are no longer orthogonal. Clearly then, giant gravitons are not simply dual to operators
with a fixed number of traces. One needs a new basis in which the two point functions are again diagonal.

Corley, Jevicki and Ramgoolam have developed a powerful 
machinery for the exact computation of correlators of Schur polynomials in the zero coupling limit of ${\cal N}=4$ 
super Yang-Mills theory with gauge group $U(N)$\cite{Corley:2001zk}. Their results show that the two point function is 
diagonalized by the Schur polynomials. It is thus tempting to identify Schur polynomials as the operators dual 
to giant gravitons\cite{Corley:2001zk}.

The Schur polynomial is defined by
\begin{equation}
\chi_R (Z)={1\over n!}\sum_{\sigma\in S_n}\chi_R (\sigma )\Tr (\sigma Z^{\otimes n}),
\label{Schur}
\end{equation}
$$\Tr (\sigma Z^{\otimes n}) =Z^{i_1}_{i_{\sigma (1)}}Z^{i_2}_{i_{\sigma (2)}}\cdots
Z^{i_{n-1}}_{i_{\sigma (n-1)}}Z^{i_n}_{i_{\sigma (n)}}.$$
The Schur polynomial label $R$ can be thought of as a Young diagram which has $n$ boxes. $\chi_R (\sigma )$ is the 
character of $\sigma\in S_n$ in representation $R$. These operators have conformal dimension $\Delta =n$ and ${\cal R}$
charge $J=n$ and transform in the $(0,n,0)$ of the $SU(4)$ ${\cal R}$ symmetry group. Of course, $\Delta =J$ as expected
for a half BPS operator. To state the known results for correlators of Schur polynomials, we need to recall the
definition of the weight of a box in a Young diagram. For a box in the $i^{th}$ row and the $j^{th}$ column the 
weight is $N-i+j$. Let $f_R$ denote the product of weights in the Young diagram. For example, if
$$ R_1=\yng(2,2,1)\qquad R_2=\yng(4),$$
we have
$$ f_{R_1}=N(N+1)(N-1)N(N-2),\qquad f_{R_2}=N(N+1)(N+2)(N+3).$$
The two and three point functions of Schur polynomials, which we use extensively in what follows, are\cite{Corley:2001zk}
$$\langle \chi_R (Z)\chi_S (Z^\dagger )\rangle =\delta_{RS} f_R ,$$
$$\langle \chi_R (Z)\chi_S (Z)\chi_T (Z^\dagger )\rangle = g(R,S;T) f_T ,$$
where $g(R,S;T)$ is the Littlewood-Richardson coefficient.
For an impressive development of the technology for Schur Polynomials of the $U(N)$ theory, 
see\cite{Corley:2002mj}. For an extension to the $SU(N)$ theory see \cite{Corley:2002mj},\cite{deMelloKoch:2004ws}.

Besides the fact that these operators are half BPS and diagonalize the two point function, they capture further features of
giant graviton dynamics. To see this, recall that a giant graviton expands to a radius proportional to the square root of its 
angular momentum\cite{McGreevy:2000cw}. If the giant is expanding in the S$^5$ of the
AdS$\times$S background, there is a limit on how large it can be - its radius must be less than the 
radius of the S$^5$\cite{McGreevy:2000cw}. This in turn implies a cut off on the angular momentum of the giant. Since angular momentum 
of the giant maps into ${\cal R}$ charge, there should be a cut off on the ${\cal R}$ charge 
of the dual operators. The Schur polynomials corresponding to totally antisymmetric representations do have a cut off on their
${\cal R}$ charge; this cut off exactly matches the cut off on the giants angular momentum\cite{Balasubramanian:2001nh}. 
Thus, it is natural to identify Schur polynomials for the completely antisymmetric representations as the operator
dual to sphere giant gravitons. Another class of Schur polynomials which are naturally singled out, are those corresponding to
totally symmetric representations. Since these representations are not cut off, they are naturally identified as operators
dual to AdS giant gravitons\cite{Corley:2001zk}, which, because they can expand in the AdS space, can expand to an arbitrarily 
large size and hence have no bound on their angular momentum. The Young diagram has at most 
$N$ rows, implying a cut off on the number of AdS giant
gravitons; the need for this cut off is clearly visible in the dual gravitational theory: it ensures
that the five form flux at the center of the AdS space does not become zero; a non-zero flux is needed to support an AdS giant. 
To obtain further support for these identifications, one can define a limit in 
which the dynamics of this half BPS sector decouples from the rest of the theory\cite{Berenstein:2004kk}, leading to a
description of the half-BPS states in terms of the eigenvalues of large $N$ 
gauged quantum mechanics for a single Hermitian matrix with quadratic potential. This quantum mechanics is the quantum 
mechanics of $N$ free non-relativistic fermions
in an external potential\cite{Brezin:1977sv}. The Schur polynomials map into energy eigenfunctions of the $N$
fermion system\cite{Corley:2001zk},\cite{Berenstein:2004kk}.
Recently, for arbitrary configurations of D-branes which preserve half of 
the supersymmetries, the full back reaction of the geometry in the supergravity limit 
was obtained\cite{Lin:2004nb}. The phase space structure of non-interacting
fermions plays a visible role in these solutions, providing further convincing support for the
proposal of \cite{Corley:2001zk},\cite{Berenstein:2004kk}. 

A number of interesting studies exploiting these ideas have appeared recently. For an attempt to recover gravitational thermodynamics 
see \cite{Balasubramanian:2005mg}; for an attempt to understand spacetime as an emergent phenomenon see\cite{Berenstein:2006yy}.
For a tantalizing suggestion of how to construct the metric of the ${1\over 2}$ BPS geometries directly in the ${\cal N}=4$ super Yang-Mills
theory, see \cite{SamMetric}.
The recent work \cite{Minwalla} suggests that many of the results that have been obtained for ${1\over 2}$-BPS giants may have an interesting
extension to the ${1\over 4}$ and ${1\over 8}$-BPS cases. For giants in a less supersymmetric background, that may also have a simple
field theory dual see\cite{jeff}.

\subsection{Excited Giants}

As reviewed above, there is significant evidence for the proposal of Corley, Jevicki and Ramgoolam that the dual of a giant
graviton is a Schur polynomial. The precise relation between Schur polynomials and giant gravitons however remains obscure.
The Schur polynomial corresponding to a Young diagram with one column, with of order $N$ boxes, is naturally interpreted as a
sphere giant graviton. Presumably, a Schur polynomial corresponding to a Young diagram with $O(1)$ columns, each with $O(N)$
boxes, corresponds to a bound state of sphere giant gravitons. Similarly, a Schur polynomial corresponding to a Young diagram with
$O(1)$ rows, each containing $O(N)$ boxes, is presumably dual to a bound state of AdS giant gravitons. The results of
\cite{Balasubramanian:2005mg} support this interpretation. However, to firmly establish (and define) the precise relation between 
Schur polynomials and giant gravitons, more detailed computations are required. It is with this goal in mind, that we study excited 
giant gravitons. Excitations of giant gravitons can be described by attaching
open strings to the giant graviton. The open string ends on the giant graviton and so it is natural to suspect that the open string
knows something about the giant graviton's geometry. In this subsection we will describe an 
attractive proposal for the operators dual to excited giant gravitons\cite{Balasubramanian:2004nb}.

The proposal of \cite{Balasubramanian:2004nb} amounts, roughly, to inserting words $(W^{(a)})^j_i$ describing the open strings
(one word for each open string) into the operator describing the system of giant gravitons
\begin{equation}
\chi_{R,R_1}^{(k)}(Z,W^{(1)},...,W^{(k)})={1\over (n-k)!}
\sum_{\sigma\in S_n}\Tr_{R_1}(\Gamma_R(\sigma))\Tr(\sigma Z^{\otimes n-k}W^{(1)}\cdots W^{(k)}),
\label{restrictedschur}
\end{equation}
$$\Tr (\sigma Z^{\otimes n-k}W^{(1)}\cdots W^{(k)})= Z^{i_1}_{i_{\sigma (1)}}Z^{i_2}_{i_{\sigma (2)}}\cdots
Z^{i_{n-k}}_{i_{\sigma (n-k)}}(W^{(1)})^{i_{n-k+1}}_{i_{\sigma (n-k+1)}}\cdots
(W^{(k)})^{i_{n}}_{i_{\sigma (n)}}.$$
The representation $R$ of the giant graviton system is a Young diagram with $n$ boxes, i.e. it is a representation of $S_n$.
$\Gamma_R(\sigma )$ is the matrix representing $\sigma$ in irreducible representation $R$ of the symmetric group $S_n$.
The representation $R_1$ is a Young Diagram with $n-k$ boxes, i.e. it is a representation of $S_{n-k}$. 
Imagine that the $k$ words above are all distinct, corresponding to the case that the open strings are distinguishable. 
Consider an $S_{n-k}\otimes (S_1)^k$ subgroup of $S_n$. The representation $R$ of $S_n$ will subduce into a (generically)
reducible representation of the $S_{n-k}\otimes (S_1)^k$ subgroup. One of the irreducible representations appearing in this
subduced representation is $R_1$. $\Tr_{R_1}$ is an instruction to trace only over the indices belonging to this irreducible
component. If the representation $R_1$ appears more than once, things are a little more subtle. The example discussed in
\cite{Balasubramanian:2004nb} illustrates this point nicely. Suppose $R\to R_1\oplus R_2\oplus R_2$ under restricting $S_n$ to 
$S_{n-2}\times S_1\times S_1$. Choose a basis so that
$$\Gamma_R (\sigma )=\left[
\matrix{\Gamma_{R_1}(\sigma)_{i_1 j_1} &0 &0\cr 0 &\Gamma_{R_2}(\sigma)_{i_2 j_2} &0\cr 0 &0 &\Gamma_{R_2}(\sigma)_{i_3 j_3}}
\right],\qquad \forall \sigma\in S_{n-2}\times S_1\times S_1 .$$
A generic element of $S_n$ need not belong to the $S_{n-2}\times S_1\times S_1$ subgroup and hence need not be block
diagonal
$$\Gamma_R (\sigma )=\left[
\matrix{A^{(1,1)}_{i_1 j_1} &A^{(1,2)}_{i_1 j_2} &A^{(1,3)}_{i_1 j_3}\cr 
        A^{(2,1)}_{i_2 j_1} &A^{(2,2)}_{i_2 j_2} &A^{(2,3)}_{i_2 j_3}\cr 
        A^{(3,1)}_{i_3 j_1} &A^{(3,2)}_{i_3 j_2} &A^{(3,3)}_{i_3 j_3}}
\right],\qquad \sigma\notin S_{n-2}\times S_1\times S_1 .$$
There are four suitable definitions for $\Tr_{R_2}(\Gamma_R(\sigma ))$: $\Tr (A^{(2,2)})$, $\Tr (A^{(2,3)})$, $\Tr (A^{(3,2)})$ or $\Tr (A^{(3,3)})$.
It is natural to interpret the operator obtained using $\Tr (A^{(2,3)})$ or $\Tr (A^{(3,2)})$ as dual to the system with the open
strings stretching between the giants and the operator obtained using $\Tr (A^{(2,2)})$ or $\Tr (A^{(3,3)})$ as dual to the system 
with one open string on each giant. In general, it is natural to identify the ``on the diagonal" blocks with states in which the
two open strings are on a specific giant and the ``off the diagonal" blocks as states in which the open strings stretch between two
giants. As a consequence of the fact that the representation $R_2$ appears with a multiplicity two, there is no unique way to extract
two $R_2$ representations out of $R$. The specific representations obtained will depend on the details of the subgroups used in
performing the restriction. There is an obvious generalization to the case that a representation $R_1$ appears $n$ times after restricting
to the subgroup. See the end of the section for an example in which a subgroup appears three times.

This prescription builds operators that are invariant under the $S_{n-k}\otimes (S_1)^k$ subgroup. This is natural from
the point of view of the quantum field theory: when Wick contracting we sum over all permutations of the $Z$ fields. From
a group theory point of view, this splitting seems a bit artificial. Indeed, notice that the sum in (\ref{restrictedschur})
runs over $S_n$ and not $S_{n-k}$. For elements $\sigma\in S_{n-k}$, $\Tr_{R_1}(\Gamma_R (\sigma ) )$ produces the
character of the group element. For elements $\sigma\notin S_{n-k}$, it seems that $\Tr_{R_1} (\Gamma_R (\sigma ))$ has no obvious 
group theoretic interpretation. In appendices B and C we give a more natural group theoretic realization of these traces,
by employing projection operators.

If any of the strings are identical, one needs to decompose with respect to a larger subgroup and to pick a representation
for the strings which are indistinguishable. For example, if two strings are identical, one would consider a
$S_{n-k}\otimes S_2\otimes (S_1)^{k-2}$ subgroup. The identical strings could be in the $\yng(2)$ or $\yng(1,1)$ representation.
Concretely, this means that when tracing over the indices associated with the identical open strings, we would restrict the
trace to the $\yng(2)$ or $\yng(1,1)$ subspaces.

As already discussed, the giant graviton system is dual to an operator containing a
product of order $N$ Higgs fields; the open strings are dual to an operator containing a product of order $\sqrt{N}$ fields.
We have in mind the case that $k$ is $O(1)$, $n$ is $O(N)$ and the words $(W^{(a)})^j_i$ are a product of $O(\sqrt{N})$ Higgs
fields.

There is already convincing evidence for this proposal\cite{Balasubramanian:2004nb}. The world volume of a giant 
graviton is a compact space. The endpoints of open strings ending on the giant graviton act as point charges in the
giant's world volume theory. Now, the Gauss law on a compact space implies that the total charge must sum to zero. This
is a severe restriction on the possible excitations of the giant graviton. In particular, it implies that the number of
open strings ending on the giant must equal the number of open strings leaving the giant. In a beautiful argument, 
\cite{Balasubramanian:2004nb} have convincingly demonstrated that (\ref{restrictedschur}) respects these constraints, by counting
the number of states consistent with the Gauss law constraint and showing that this matches with the number of possible
operators of the form (\ref{restrictedschur}).

We call the operator (\ref{restrictedschur}) a {\it restricted} Schur polynomial of representation $R$ with representation $R_1$ 
for the restriction. If the dimension of $R$ is equal to the dimension of $R_1$, we call this a {\it single restricted Schur
polynomial}. If the dimension of $R$ is not equal to the dimension of $R_1$, we call this a {\it multiple restricted Schur
polynomial}. The single restricted Schur polynomials are particularly simple, because tracing over the representation of the 
restricted Schur polynomial is the same as tracing over the representation of the restriction. For this reason, some of our
technology is most easily developed for the single restricted Schur polynomials, which is why we make this distinction.

We have developed a diagrammatic notation for the restricted Schur polynomials. The notation summarizes the strings that are
attached to the giant graviton system, the subgroups involved in the restriction and specifies whether we are tracing over an
``on the diagonal block" or over an ``off the diagonal block". As an example, consider
$$\chi_{\young({\,}{\,}1,{\,}2)}.$$
To define this operator, we first restrict from $S_5$ to $S_4$, and then to $S_3$. The $S_4$ subgroup is defined as the elements
of $S_5$ that leave the index of $W^{(1)}$ inert. The $S_3$ subgroup is defined as the elements of $S_5$ that leave the indices of 
$W^{(1)}$ and $W^{(2)}$ inert. The representations involved when we restrict from $S_5$ to $S_4$ and then to $S_3$ are obtained
by dropping the numbered boxes in order
$$\yng(3,2)\to\yng(2,2)\to\yng(2,1).$$
In this example, we trace over an ``on the diagonal block". When we trace over an ``off the diagonal block" the operator is denoted
with two numbers in the two boxes which are involved. For example, the operator with label
$$\chi_{\young({\,}{\,}\onetwo,{\,}\twoone)},$$
is constructed by tracing over the off diagonal block, with row index obtained from the representation produced with the following
restriction
$$\yng(3,2)\to\yng(2,2)\to\yng(2,1),$$
and with column index obtained from the representation produced with the following restriction
$$\yng(3,2)\to\yng(3,1)\to\yng(2,1).$$
This labelling reflects the fact that the two strings have opposite orientations, as required by the Gauss law.
The operator with label
$$\chi_{\young({\,}{\,}\twoone,{\,}\onetwo)},$$
is obtained by tracing over the block with the opposite row and column indices. As a last example, the operator
$$\chi_{\young({\,}{\,}{\onetwo},{\,}{\twothree},{\threeone})},$$
is obtained by tracing over the block with row label obtained from the representation produced with the restriction
$$\yng(3,2,1)\to\yng(2,2,1)\to\yng(2,1,1)\to\yng(2,1),$$
and column label obtained from the representation produced with the restriction
$$\yng(3,2,1)\to\yng(3,2)\to\yng(2,2)\to\yng(2,1).$$

\noindent
{\bf An Example:} We will consider the restricted Schur polynomials
$$\chi_{R,R_1}^{(2)}(Z,W^{(1)},W^{(2)})={1\over 3!}\sum_{\sigma\in S_5} \Tr_{R_1}(\Gamma_R (\sigma ))
Z^{i_1}_{i_{\sigma(1)}}Z^{i_2}_{i_{\sigma(2)}}Z^{i_3}_{i_{\sigma(3)}}
(W^{(2)})^{i_4}_{i_{\sigma(4)}}(W^{(1)})^{i_5}_{i_{\sigma(5)}},$$
where
$$R=\yng(3,2).$$
After subducing with respect to the $S_4$ subgroup ${\cal G}$ that leaves the index of $W^{(1)}$ inert
$${\cal G}=\{\sigma \in S_5 |\sigma (5)=5\},$$
the representation $R$ decomposes as
$$\yng(2,2)\oplus\yng(3,1).$$
Subducing next with respect to the $S_3$ subgroup ${\cal S}$ that leaves the index of $W^{(2)}$ inert
$${\cal S}=\{ \sigma\in{\cal G} |\sigma (4)=4\},$$
representation $R$ further decomposes as
$$\yng(2,1)\oplus\yng(2,1)\oplus\yng(3).$$
Labelling the indices of representation $R$ by these $S_3$ and $S_4$ irreducible representations, in a suitable
basis, we have
$$\Gamma_{\yng(3,2)}(\sigma)=\left[
\matrix{ A_{11} &A_{12} &A_{13}\cr & &\cr A_{21} &A_{22} &A_{23}\cr & & \cr A_{31} &A_{32} &A_{33} }\right]
\matrix{\}\yng(2,1)\cr \cr \}\yng(3)\cr \cr \}\yng(2,1)}
\matrix{\}\yng(2,2)\cr\cr\Big\}\yng(3,1)\cr\cr\cr }.$$
The restricted Schur polynomial obtained by tracing over the $A_{11}$ block is denoted by
$$\chi_{\young({\,}{\,}1,{\,}2)}.$$
The restricted Schur polynomial obtained by tracing over the $A_{33}$ block is denoted by
$$\chi_{\young({\,}{\,}2,{\,}1)}.$$
The restricted Schur polynomial obtained by tracing over the $A_{13}$ block is denoted by
$$\chi_{\young({\,}{\,}\onetwo,{\,}\twoone)}.$$
The restricted Schur polynomial obtained by tracing over the $A_{31}$ block is denoted by
$$\chi_{\young({\,}{\,}\twoone,{\,}\onetwo)}.$$
Finally, the restricted Schur polynomial obtained by tracing over the $A_{22}$ block is denoted by
$$\chi_{\young({\,}{\,}{\,},21)}.$$
Clearly, the labels in the boxes tell you how to construct the restrictions needed to define the row and column
indices of the block which was traced to produce the operator.
For further technical details, which are not needed to understand our results, the reader is referred to appendices C and D.

\subsection{Open Strings}

As we have already mentioned, the open string is described by a word with $O(\sqrt{N})$ letters. These letters can
in principle be fermions, Higgs fields or covariant derivatives of these fields. We will consider open strings moving 
with a large angular momentum on the $S^5$, in the direction corresponding to $Y$. The number of $Y$ fields tells us 
the spacetime angular momentum of the string state. To describe strings moving with a large angular momentum on the $S^5$,
take words with $O(\sqrt{N})$ $Y$ letters in the word. We can insert different letters into this word. In this article we will
consider only excitations corresponding to insertion of $X$ Higgs fields. In this second case,
the total number of $X$ fields appearing is $O(1)$. Since we do not allow $Z$'s to appear in the open string word, we are 
restricting ourselves to open strings that have no angular momentum in the direction of the giant. The  correlation functions 
in this case simplify dramatically, since there can be no contractions between Higgs fields in the open strings and the Higgs 
fields making up the giant. In fact, when computing two point functions in free field theory, as long as the number
of boxes in the representation $R$ is less than $O(N^2)$ and the numbers of $Z$'s in the open string is $O(1)$, the 
contractions between any $Z$s in the open string and the rest of the operator are suppressed in the large $N$ limit\cite{recent}.

Our labeling for the open string words is the following $(l_1\le l_2\le l_3\cdots\le l_k)$
$$ (W_{l_1,l_2,\cdots l_k})^i_j= (Y^{l_1}XY^{l_2-l_1}X Y^{l_3-l_1-l_2}\cdots Y^{l_k-\sum_{b=1}^{k-1}l_b}X)^i_j.$$
Geometrically we think of the $Y$'s as forming a lattice that is populated with $X$'s. The numbers $l_b$ give the locations
of the $X$'s in this lattice. The BMN loops are given by moving to momentum space on this lattice.

The most general form that the two point function of open string words can take is
$$ \langle (W)^{i}_j (W^\dagger)^{k}_{l}\rangle =F_0\delta^i_l\delta^k_j+F_1\delta^i_j\delta^k_l .$$
In appendix G we explicitly compute the two point function of the open string ground state of angular momentum $J$, that is,
$(W)^{i}_j=(Y^J)^{i}_j$. Expanding the exact result, we have 
$$F_0=N^{J-1}+{(J-1)(J-2)(J^2+5J+12)\over 24}N^{J-3}+O(J^8 N^{J-5}),$$
$$F_1=(J-1)N^{J-2}+{(J-1)(J-2)(J-3)(J^2+3J+4)\over 24}N^{J-4}+O(J^9 N^{J-6}).$$
A few comments are in order: In the large $N$ limit the term with coefficient $F_0$ dominates the term with coefficient 
$F_1$. We are interested in the case that we have $J$ fields in each word, with $J\sim O(\sqrt{N})$ such that $g_2\equiv {J^2\over N} <<1$.
With $J$ fields in each word we'd have $F_0=N^{J-1}(1+ O(g_2))$ and $F_1= (J-1)N^{J-2}(1 +O(g_2))$. Thus, we see that
$$ {F_1\over F_0}={(J-1)N^{J-2}\over N^{J-1}}(1+O(g_2))=O({\sqrt{g_2}\over\sqrt{N}}).$$
The index structure of the $F_0$ term mixes indices of the two words; the index structure of the $F_1$ term does not mix
indices from different words.

\subsection{A First Look at Two Point Functions}

Our strategy is to exploit the fact that since there are no contractions mixing Higgs fields coming from the open strings ($X$s and $Y$s) 
and Higgs fields coming from the giant graviton ($Z$s), the two point function factorizes as
\begin{eqnarray}
\langle\chi_{R,R_1}^{(k)}{\chi}_{R',R'_1}^{(k')\dagger}\rangle &=&{1\over (n-k)!(n'-k')!}
\sum_{\sigma\in S_n}\sum_{\sigma'\in S_{n'}}\Tr_{R_1}(\Gamma_R(\sigma ))
\Tr_{R'_1}(\Gamma_{R'}(\sigma' ))\nonumber\\
&\times &\langle\Tr(\sigma Z^{\otimes n-k}W^{(1)}\cdots W^{(k)})
\Tr(\sigma' (Z^\dagger )^{\otimes n'-k'}(W^\dagger)^{(1)}\cdots (W^\dagger)^{(k')})\rangle \nonumber\\
&=&{1\over (n-k)!(n'-k')!}
\sum_{\sigma\in S_n}\sum_{\sigma'\in S_{n'}}\Tr_{R_1}(\Gamma_R(\sigma ))
\Tr_{R'_1}(\Gamma_{R'}(\sigma' ))\nonumber\\
&\times &\langle
Z^{i_1}_{i_{\sigma (1)}}        \cdots       Z^{i_{n-k}}_{i_{\sigma (n-k)}}
(Z^\dagger)^{i'_1}_{i'_{\sigma' (1)}}\cdots (Z^\dagger)^{i'_{n'-k'}}_{i'_{\sigma' (n'-k')}}\rangle\nonumber\\
&\times & \langle (W^{(1)})^{i_{n-k+1}}_{i_{\sigma (n-k+1)}}\cdots
(W^{(k)})^{i_{n}}_{i_{\sigma (n)}}
((W^\dagger)^{(1)})^{i'_{n'-k'+1}}_{i'_{\sigma' (n'-k'+1)}}\cdots
((W^\dagger)^{(k')})^{i'_{n'}}_{i'_{\sigma' (n')}} \rangle \nonumber,
\end{eqnarray}
For simplicity, specialize to attaching a single word to each giant. In this case, using the results of the previous section, we have
(we now set $n=n'$; for $n\ne n'$ the correlator vanishes)
\begin{eqnarray}
\langle\chi_{R,R_1}^{(1)}(\chi^{(1)}_{R',R'_1})^\dagger\rangle &=&{1\over ((n-1)!)^2}
\sum_{\sigma ,\sigma'\in S_{n}}\Tr_{R_1}(\Gamma_R(\sigma ))\Tr_{R'_1}(\Gamma_{R'}(\sigma' ))\nonumber\\
&\times &\langle\Tr(\sigma Z^{\otimes n-1}W^{(1)})\Tr(\sigma' (Z^\dagger )^{\otimes n-1}(W^{(1)})^\dagger)\rangle \nonumber\\
&=&{1\over ((n-1)!)^2}
\sum_{\sigma ,\sigma'\in S_{n}}\Tr_{R_1}(\Gamma_R(\sigma ))\Tr_{R'_1}(\Gamma_{R'}(\sigma' ))\label{Comp}\\
&&\langle
Z^{i_1}_{i_{\sigma (1)}}        \cdots       Z^{i_{n-1}}_{i_{\sigma (n-1)}}
(Z^\dagger)^{i'_1}_{i'_{\sigma' (1)}}\cdots (Z^\dagger)^{i'_{n-1}}_{i'_{\sigma' (n-1)}}\rangle
(F_0\delta^{i'_n}_{i_{\sigma (n)}}\delta^{i_n}_{i'_{\sigma'(n)}}+
F_1\delta^{i_n}_{i_{\sigma (n)}}\delta^{i'_n}_{i'_{\sigma'(n)}}) \nonumber ,
\end{eqnarray}
Split the computation into two terms, one with coefficient $F_0$ and one with coefficient $F_1$.
The term with coefficient $F_0$ can be written as
$$
{1\over ((n-1)!)^2}
\sum_{\sigma ,\sigma'\in S_{n}}\Tr_{R_1}(\Gamma_R( \sigma))\Tr_{R'_1}(\Gamma_{R'}(\sigma' ))\langle
Z^{i_1}_{i_{\sigma (1)}}\cdots Z^{i_{n-1}}_{i_{\sigma (n-1)}}Z^{i_n}_{i_{\sigma (n)}}\times$$
$$
\times (Z^\dagger)^{i'_1}_{i'_{\sigma' (1)}}\cdots (Z^\dagger)^{i'_{n-1}}_{i'_{\sigma' (n-1)}}
(Z^\dagger)^{i'_n}_{i'_{\sigma'(n)}}\rangle \Big|_n F_0
$$
$$
\equiv F_0 \left({n!\over (n-1)!}\right)^2\langle\chi_{R,R_1} (Z)(\chi_{R',R'_1}(Z))^\dagger \rangle|_n ,$$
where the notation $\Big|_n$ is an instruction to only sum over the Wick contractions that contract $Z^{i_n}_{i_{\sigma (n)}}$
with $(Z^\dagger)^{i'_n}_{i'_{\sigma'(n)}}$. The factor of $\left({n!\over (n-1)!}\right)^2$ is needed because the 
Schur polynomial is defined with a coefficient of ${1\over n!}$ while the restricted Schur polynomial 
with $k$ strings attached has coefficient
${1\over (n-k)!}$. These restricted correlators will be denoted graphically with numbered arrows in the boxes corresponding to
fields which are to be contracted. For example,
$$\langle\chi_{\young({\,}{\,}\toone,{\,}{\,})}\chi_{\young({\,}{\,}\toone,{\,}{\,})}^\dagger\rangle
=\left({5!\over (5-1)!}\right)^{2}\langle\chi_{R,R_1}(Z)(\chi_{R,R_1}(Z))^\dagger \rangle|_5,$$
$$\langle\chi_{\young({\,}{\,}\toone,{\,}{\totwo})}\chi_{\young({\,}{\,}\toone,{\,}{\totwo})}^\dagger\rangle
=\left({5!\over (5-2)!}\right)^{2}\langle\chi_{R,R_2}(Z)(\chi_{R,R_2}(Z))^\dagger\rangle|_{5,4},$$
where
$$R=\yng(3,2)\qquad R_1=\yng(2,2),\qquad R_2=\yng(2,1).$$
The term with coefficient $F_1$ can be manipulated to give
$$ F_1 {1\over ((n-1)!)^2}
\sum_{\sigma ,\sigma'\in S_{n}}\Tr_{R_1}(\Gamma_R(\sigma ))\Tr_{R'_1}(\Gamma_{R'}(\sigma' ))\times$$
$$\langle Z^{i_1}_{i_{\sigma (1)}}        \cdots       Z^{i_{n-1}}_{i_{\sigma (n-1)}}
(Z^\dagger)^{i'_1}_{i'_{\sigma' (1)}}\cdots (Z^\dagger)^{i'_{n-1}}_{i'_{\sigma' (n-1)}}\rangle
\delta^{i_n}_{i_{\sigma (n)}}\delta^{i'_n}_{i'_{\sigma'(n)}} $$
$$=F_1 \langle D_{W^{(1)}}\chi_{R,R_1}^{(1)}(D_{W^{(1)}}\chi^{(1)}_{R',R'_1})^\dagger\rangle ,$$
where
$$ D_{W^{(1)}}\equiv {d\over d(W^{(1)})^i_i},$$
is the reduction operator introduced in \cite{deMelloKoch:2004ws}. One of the results of this article is
to develop efficient methods to compute the reduction of restricted Schur polynomials. Using our graphical notation, an explicit
example of (\ref{Comp}) is
$$\langle\chi_{\young({\,}{\,}1,{\,}{\,})}\chi_{\young({\,}{\,}1,{\,}{\,})}^\dagger\rangle
=F_0\langle\chi_{\young({\,}{\,}\toone,{\,}{\,})}\chi_{\young({\,}{\,}\toone,{\,}{\,})}^\dagger\rangle
+F_1\langle D_{W^{(1)}}\chi_{\young({\,}{\,}1,{\,}{\,})}(D_{W^{(1)}}\chi_{\young({\,}{\,}1,{\,}{\,})})^\dagger\rangle.
$$

For multi string excitations of giants we will consider mainly pair wise contractions of strings
$$ \langle (W^{(1)})^{i_{n-k+1}}_{i_{\sigma (n-k+1)}}\cdots
(W^{(k)})^{i_{n}}_{i_{\sigma (n)}}
(W^{\dagger(1)})^{i'_{n'-k'+1}}_{i'_{\sigma' (n'-k'+1)}}\cdots
(W^{\dagger(k')})^{i'_{n'}}_{i'_{\sigma' (n')}} \rangle =$$
$$\langle (W^{(1)})^{i_{n-k+1}}_{i_{\sigma (n-k+1)}}
(W^{\dagger (1)})^{i'_{n'-k'+1}}_{i'_{\sigma' (n'-k'+1)}}\rangle
\cdots
\langle (W^{(k)})^{i_{n}}_{i_{\sigma (n)}}
(W^{\dagger (k')})^{i'_{n'}}_{i'_{\sigma' (n')}}\rangle +{\rm all \,\, possible\,\, pairings} .$$
The terms which have been dropped correspond to contractions which mix four (or more) words and hence correspond
to open string interactions. We therefore expect these terms to be sub leading in $N$. We will assume that
$$\langle (W^{(k)})^i_j(W^{\dagger (l)})^{i'}_{j'}\rangle \propto\delta^{kl},$$
so that only a single pairing contributes.

We will end this section by considering  a concrete example for the case $k=k'=2$ in detail. The
extension to higher $k$ is straight forward. The example is

\begin{eqnarray}
\nonumber
\langle\chi_{\young({\,}{\,}1,{\,}2)}\chi_{\young({\,}{\,}1,{\,}2)}^\dagger\rangle
&=&F_0^1\langle\chi_{\young({\,}{\,}\toone,{\,}2)}\chi_{\young({\,}{\,}\toone,{\,}2)}^\dagger\rangle
+F_1^1\langle D_{W^{(1)}}\chi_{\young({\,}{\,}1,{\,}2)}(D_{W^{(1)}}\chi_{\young({\,}{\,}1,{\,}2)})^\dagger\rangle\\
\nonumber
&=&F_0^1F_0^2\langle\chi_{\young({\,}{\,}\toone,{\,}\totwo)}\chi_{\young({\,}{\,}\toone,{\,}\totwo)}^\dagger\rangle
+F_0^1F_1^2\langle D_{W^{(2)}}\chi_{\young({\,}{\,}\toone,{\,}2)}(D_{W^{(2)}}\chi_{\young({\,}{\,}\toone,{\,}2)})^\dagger\rangle\\
\nonumber
&+&F_1^1F_0^2\langle D_{W^{(1)}}\chi_{\young({\,}{\,}1,{\,}\totwo)}(D_{W^{(1)}}\chi_{\young({\,}{\,}1,{\,}\totwo)})^\dagger\rangle\\
&+&F_1^1F_1^2\langle D_{W^{(2)}}D_{W^{(1)}}\chi_{\young({\,}{\,}1,{\,}2)}
(D_{W^{(2)}}D_{W^{(1)}}\chi_{\young({\,}{\,}1,{\,}2)})^\dagger\rangle
\nonumber
\end{eqnarray}
The coefficient $F_0^i$ is the $F_0$ contraction of the $i$th word;
the coefficient $F_1^i$ is the $F_1$ contraction of the $i$th word.

The operators used in our examples would of course, not represent physically realistic operators for the description of 
excited giant gravitons. To be physically realistic, we would need to consider representations with $O(N)$ blocks and $O(1)$ 
strings attached to the giant.

In this subsection, we have managed to reduce the computation of the two point function, to the computation of the two point
function of restricted Schur polynomials where only a subset of contractions are summed and to the 
computation of reductions of the restricted Schur polynomial. Our reorganization of the computation is useful because we have 
been able to find efficient methods to compute these quantities. Our methods are summarized in the next subsection.

\subsection{Summary of Results}

In this section we will state a set of rules that allows a simple computation of reductions of restricted Schur polynomials
as well as the computation of the two point function of restricted Schur polynomials where a subset of contractions are summed.
These rules are proved in the appendices.

{\bf Rule 1: The reduction of a restricted Schur polynomial} is simplest if one reduces with respect to the open string with the smallest
label. The reduction is simply given by dropping the box associated with the open string and multiplying by the weight of the
removed box. As an example we have 
$$ D_{W^{(2)}} D_{W^{(1)}}\chi_{\young({\,}{\,}1,{\,}2)}=(N+2) D_{W^{(2)}}\chi_{\young({\,}{\,},{\,}2)}=
N(N+2)\chi_{\young({\,}{\,},{\,})}.$$
Recall that the specific subgroups used in the restriction play an important role in determining the operator. This is why
the reduction with respect to the smallest string label is simplest. If we wanted to first compute the reduction with respect 
to $W^{(2)}$, we need to use a subgroup swap rule, which will tell us how polynomials constructed using different restrictions
are related. The reduction with respect to a word with indices belonging to an ``off the diagonal block" vanishes
$$ D_{W^{(1)}}\chi_{\young({\,}{\,}\onetwo,{\,}\twoone)}=0.$$
It is incorrect to conclude that
$$ D_{W^{(2)}}\chi_{\young({\,}{\,}\onetwo,{\,}\twoone)}=0.$$
After employing the subgroup swap rule, this reduction is non-zero.

{\bf Rule 2: The subgroup swap rule} can be used if we need to reduce with respect to an open string which does not have the smallest
label. Let us first discuss the case that only two strings or less are attached to the giant.
In this case, to compute a two point function, one reduces the result of the subgroup swap implying that stretched string state
contributions can be dropped.
The restricted Schur polynomials defined above were reduced with respect to the subgroup that leaves the index of
$W^{(1)}$ inert, and then further with respect to the subgroup that leaves the index of $W^{(2)}$ inert. For the sake of discussing
rule 2, we will denote this as
$$\chi_{\young({\,}{\,}1,{\,}2)}|_1|_2 .$$
The restricted Schur polynomials defined by reducing with respect to the subgroup that leaves the index of
$W^{(2)}$ inert, and then with respect to the subgroup that leaves the index of $W^{(1)}$ inert will be denoted by
$$\chi_{\young({\,}{\,}1,{\,}2)}|_2|_1 .$$
Consider the restricted Schur polynomial $\chi_{R,R_1}^{(2)}$. Let $c_i$ denote the weight associated with the box
occupied by $W^{(i)}$. The subgroup swap rule says that, up to stretched string state contributions,
$$\chi_{R,R_1}^{(2)}|_2|_1 = (1-a)\chi_{R,R_1}^{(2)}|_1|_2+a\chi_{R,R_1}^{(2)}|_1|_2( 1\leftrightarrow 2)$$
with $\chi_{R,R_1}^{(2)}|_1|_2(1\leftrightarrow 2)$ the restricted Schur polynomial obtained by swapping
the labels $1$ and $2$, and
$$ a={1\over (c_1-c_2)^2}.$$
Using the subgroup swap rule, we easily find
$$ D_{W^{(2)}}\chi_{\young({\,}{\,}1,{\,}2)}=
   {3N\over 4}\chi_{\young({\,}{\,}1,{\,})}+{N+2\over 4}\chi_{\young({\,}{\,},{\,}1)}.$$
The subgroup swap rule can be used to swap any two strings $n$ and $n+1$ for any $n$. To swap $n$ and $n+2$ say,
we would first swap $n$ and $n+1$, and then $n$ and $n+2$. See appendix C for explicit examples.

If more than two strings are attached to the giant, extra contributions 
coming from twisted string states need to be included.
To state the subgroup swap rule in its full generality, we will modify our notation slightly.
Stretched strings are denoted by placing a pair of indices in the box of the stretched string.
Up to now, if a string was attached to a single brane, a single number would appear in the box;
we will repeat that number so that every box has both an upper and a lower label. Thus, for
example, 
$$\chi_{\young({\,}{\,}1,{\,}2)}\to \chi_{\young({\,}{\,}{\oneone},{\,}{\twotwo})}.$$
Denote the weight of the box that has the upper label equal to the index that is fixed first, before the swap, by $c^U_1$.
Denote the weight of the box that has the lower label equal to the index that is fixed first, before the swap, by $c^L_1$.
Denote the weight of the box that has the upper index equal to the index that is fixed next, before the swap, by $c^U_2$. 
Denote the weight of the box that has the lower index equal to the index that is fixed next, before the swap, by $c^L_2$. 
The upper and lower swap factors are given by
$$ S^U={c^U_1-c^U_2 \over (c^U_1-c^U_2)^2},\qquad S^L={c^L_1-c^L_2 \over (c^L_1-c^L_2)^2}.$$
The upper and lower no-swap factors are given by
$$ N^U=\sqrt{1-{1\over (c^U_1-c^U_2)^2}},\qquad N^L=\sqrt{1-{1\over (c^L_1-c^L_2)^2}}.$$
In full generality, the subgroup swap rule says that when two subgroups are swapped, all possible swaps of
the upper indices of the two boxes and the lower indices of the two boxes are allowed. For each swap of
indices (upper or lower) we include a factor $S^U$ or $S^L$; if the indices are not swapped, we include a 
factor $N^U$ or $N^L$.
Thus, for example,
$$ \chi_{\young({\,}{\,}{1},{\,}{\twothree},{\threetwo})}|_1|_3|_2 =
 N^U N^L\chi_{\young({\,}{\,}{1},{\,}{\twothree},{\threetwo})}|_3|_1|_2
+N^U S^L\chi_{\young({\,}{\,}{\onethree},{\,}{\twoone},{\threetwo})}|_3|_1|_2
+S^U N^L\chi_{\young({\,}{\,}{\threeone},{\,}{\twothree},{\onetwo})}|_3|_1|_2
+S^U S^L\chi_{\young({\,}{\,}{3},{\,}{\twoone},{\onetwo})}|_3|_1|_2 $$
$$ = {3\sqrt{5}\over 8}\chi_{\young({\,}{\,}{1},{\,}{\twothree},{\threetwo})}|_3|_1|_2
+    {\sqrt{15}\over 8}\chi_{\young({\,}{\,}{\onethree},{\,}{\twoone},{\threetwo})}|_3|_1|_2
+    {\sqrt{3}\over 8}\chi_{\young({\,}{\,}{\threeone},{\,}{\twothree},{\onetwo})}|_3|_1|_2
+    {1\over 8}\chi_{\young({\,}{\,}{3},{\,}{\twoone},{\onetwo})}|_3|_1|_2 .$$

We can only swap ``adjacent indices" - in the above example, for the operator on the left hand side
we could swap $1\leftrightarrow 3$ or $2\leftrightarrow 3$, but not $1\leftrightarrow 2$\footnote{i.e.
using a single application of the the subgroup swap rule we can 
relate $\chi_R|_1|_3|_2$ to $\chi_R|_3|_1|_2$ or $\chi_R|_1|_2|_3$ but not to
$\chi_R|_2|_3|_1$.}; for the operators on the right hand side
we could swap $1\leftrightarrow 3$ or $2\leftrightarrow 1$, but not $3\leftrightarrow 2$.
These pairs of labels in each box are naturally identified as Chan-Paton factors of the open strings.
The subgroup swap rule shows that the specific labelling of the states can be changed by choosing a
different chain of subgroups for the restriction. Perhaps this freedom in labelling is responsible
for the expected emergent gauge symmetry.
This general rule is derived in appendix D.2.

{\bf Rule 3: The restricted correlator rule} states that
$$\langle\chi_{R,R'}(Z)(\chi_{S,S'}(Z))^\dagger \rangle\Big|_{n,n-1,...,n-k}=
{(n-k-1)!d_{R'}f_R\over n! d_R}\delta_{R\to R',S\to S'}.$$
for a trace running over an ``on the diagonal block" and
$$\langle\chi_{R,R'T'}(Z)(\chi_{S,U'S'}(Z))^\dagger\rangle\Big|_{n,n-1,...,n-k}=
{(n-k-1)!d_{R'} f_R\over n! d_R}\delta_{R\to (R'T'),S\to (U'S')}\delta_{T'U'}\delta_{R'S'}.$$
for a trace running over an ``off the diagonal block". The indices of the ``off the diagonal block" are explicitely displayed
in this last formula. The delta function $\delta_{R\to R',S\to S'}$ indicates that the all representations appearing in
intermediate steps of restricting $R$ to $R'$ must match all representations appearing in
intermediate steps of restricting $S$ to $S'$. Similarly, the delta function $\delta_{R\to (R'T'),S\to (U'S')}$
indicates that the all representations appearing in
intermediate steps of restricting $R$ to $(R'T')$ must match all representations appearing in
intermediate steps of restricting $S$ to $(U'S')$.
An example of an application of this rule, using our graphical notation, is
\begin{eqnarray}
\nonumber
\langle\chi_{\young({\,}{\,}\toone,{\,}\totwo)}\chi_{\young({\,}{\,}\toone,{\,}\totwo)}^\dagger\rangle
&=&{n!\over (n-k)!}{d_R\over d_{R'}}f_R\delta_{R\to R',S\to S'}\cr
\nonumber
&=& {5!\over 3!}{2 \over 5} N^2 (N-1)(N+1)(N+2).
\end{eqnarray}
We call this operation ``gluing". For the example just discussed, we say that both boxes 1 and 2 were glued.

The restricted correlator rule determines the leading order contribution to the two point function, if we are in the 
correct regime to interpret this operator as a string attached to a giant graviton. 
It is thus clear that, in this regime,
the operators introduced in \cite{Balasubramanian:2004nb} are orthogonal at large $N$.

{\bf Computing correlators using the rules:} Imagine we have an operator with $n$ strings attached. Starting from box 1,
sum the term obtained by gluing and the term obtained by reducing. Then for each of these two terms generate two new terms by
gluing box 2 or reducing it. Continue in this way till box $n$ itself is glued or reduced. This process generates a total of
$2^n$ terms. Evaluate each term using rules 1 to 3.

$$ $$

Using these rules, we can now complete the computation of the two point function we considered in the previous section
\begin{eqnarray}
\langle\chi_{\young({\,}{\,}1,{\,}2)}\chi_{\young({\,}{\,}1,{\,}2)}^\dagger\rangle
&=&F_0^1F_0^2\langle\chi_{\young({\,}{\,}\toone,{\,}\totwo)}\chi_{\young({\,}{\,}\toone,{\,}\totwo)}^\dagger\rangle
+F_0^1F_1^2\langle D_{W^{(2)}}\chi_{\young({\,}{\,}\toone,{\,}2)}(D_{W^{(2)}}\chi_{\young({\,}{\,}\toone,{\,}2)})^\dagger\rangle\cr
&+&F_1^1F_0^2\langle D_{W^{(1)}}\chi_{\young({\,}{\,}1,{\,}\totwo)}(D_{W^{(1)}}\chi_{\young({\,}{\,}1,{\,}\totwo)})^\dagger\rangle\cr
&+&F_1^1F_1^2\langle D_{W^{(2)}}D_{W^{(1)}}\chi_{\young({\,}{\,}1,{\,}2)}
(D_{W^{(2)}}D_{W^{(1)}}\chi_{\young({\,}{\,}1,{\,}2)})^\dagger\rangle\cr
&=& 8N^2 (N^2-1)(N+2)F_0^1F_0^2 +4N^2 (N^2-1)(N+2)^2F_1^1F_0^2\cr
&&+({7\over 4}N^6 +4N^5 -{3\over 4}N^4-4N^3 -N^2)F_0^1 F_1^2\cr
&&+N^3 (N+2)^2(N^2-1)F_1^1F_1^2 .
\nonumber
\end{eqnarray}

\subsection{Two Point Function of a Giant with 3 Strings Attached}

We end this section with a final example, which illustrates the subgroup swap rule in its full complexity. Consider computing
the correlator
$$\langle\chi_{\young({\,}{\,}{1},{\,}{2},{3})}\chi_{\young({\,}{\,}{1},{\,}{2},{3})}^\dagger\rangle .$$
We will only discuss how to evaluate
the coefficient of the term $F_0^1 F_0^2 F_1^3$, which is
$$C_1=\langle D_{W^{(3)}}\chi_{\young({\,}{\,}{\toone},{\,}{\totwo},{3})}
         (D_{W^{(3)}}\chi_{\young({\,}{\,}{\toone},{\,}{\totwo},{3})})^\dagger\rangle .$$
To evaluate this reduction, we first need to swap $3\leftrightarrow 2$ and then $3\leftrightarrow 1$ in each operator. 
Under the $3\leftrightarrow 2$ swap we have
$$ C_2=D_{W^{(3)}}\chi_{\young({\,}{\,}{1},{\,}{2},{3})}\to {3\over 4}D_{W^{(3)}}\chi_{\young({\,}{\,}{1},{\,}{2},{3})}
+{1\over 4}D_{W^{(3)}}\chi_{\young({\,}{\,}{1},{\,}{3},{2})}+{\sqrt{3}\over 4}
D_{W^{(3)}}\chi_{\young({\,}{\,}{1},{\,}{\twothree},{\threetwo})}
+{\sqrt{3}\over 4} D_{W^{(3)}}\chi_{\young({\,}{\,}{1},{\,}{\threetwo},{\twothree})}.$$
Finally, after the $3\leftrightarrow 1$ swap we have
$$C_2\to {45\over 64}D_{W^{(3)}}\chi_{\young({\,}{\,}{1},{\,}{2},{3})}
+{3\over 16}D_{W^{(3)}}\chi_{\young({\,}{\,}{1},{\,}{3},{2})}+
{3\over 64}D_{W^{(3)}}\chi_{\young({\,}{\,}{3},{\,}{2},{1})}$$
$$+{1\over 16}D_{W^{(3)}}\chi_{\young({\,}{\,}{3},{\,}{1},{2})}
+{\sqrt{3}\over 32}D_{W^{(3)}}\chi_{\young({\,}{\,}{3},{\,}{\twoone},{\onetwo})}
+{\sqrt{3}\over 32}D_{W^{(3)}}\chi_{\young({\,}{\,}{3},{\,}{\onetwo},{\twoone})}.$$
After these swaps, we can reduce to obtain
$$C_2 = {45(N-2)\over 64} \chi_{\young({\,}{\,}{1},{\,}{2})}
+{3N\over 16}\chi_{\young({\,}{\,}{1},{\,},{2})}+
{3(N+2)\over 64}\chi_{\young({\,}{\,},{\,}{2},{1})}$$
$$+{N+2\over 16}\chi_{\young({\,}{\,},{\,}{1},{2})}
+{(N+2)\sqrt{3}\over 32}\chi_{\young({\,}{\,},{\,}{\twoone},{\onetwo})}
+{(N+2)\sqrt{3}\over 32}\chi_{\young({\,}{\,},{\,}{\onetwo},{\twoone})}.$$
After the gluing
$$ C_1 = \left({45(N-2)\over 64}\right)^2
\langle\chi_{\young({\,}{\,}{\toone},{\,}{\totwo})}^\dagger \chi_{\young({\,}{\,}{\toone},{\,}{\totwo})}\rangle
+\left({3N\over 16}\right)^2
\langle\chi_{\young({\,}{\,}{\toone},{\,},{\totwo})}^\dagger \chi_{\young({\,}{\,}{\toone},{\,},{\totwo})}\rangle +
\left({3(N+2)\over 64}\right)^2
\langle\chi_{\young({\,}{\,},{\,}{\totwo},{\toone})}^\dagger \chi_{\young({\,}{\,},{\,}{\totwo},{\toone})}\rangle $$
$$+\left({N+2\over 16}\right)^2
\langle\chi_{\young({\,}{\,},{\,}{\toone},{\totwo})}^\dagger \chi_{\young({\,}{\,},{\,}{\toone},{\totwo})}\rangle
+\left({(N+2)\sqrt{3}\over 32}\right)^2
\langle\chi_{\young({\,}{\,},{\,}{\totwoone},{\toonetwo})}^\dagger \chi_{\young({\,}{\,},{\,}{\totwoone},{\toonetwo})}\rangle
+\left({(N+2)\sqrt{3}\over 32}\right)^2
\rangle\chi_{\young({\,}{\,},{\,}{\toonetwo},{\totwoone})}^\dagger \chi_{\young({\,}{\,},{\,}{\toonetwo},{\totwoone})}\rangle ,$$
which is easily evaluated to give
$$C_1 ={1097\over 256}N^7-{247\over 32}N^6-{5485\over 256}N^5+{1235\over 32}N^4+{1097\over 64}N^3-{247\over 8}N^2 .$$

\section{Applications}

The operators we are studying in this article, are conjectured to be dual to giant gravitons with open strings attached.
Since the giant gravitons have finite volume, these operators need to satisfy non-trivial constraints implied by the Gauss law.
They do indeed satisfy these constraints\cite{Balasubramanian:2004nb}, providing 
convincing evidence for the proposed duality. The low energy dynamics of the 
open strings attached to the giant gravitons will give rise to a Yang-Mills theory. This new {\it emergent} $3+1$
dimensional Yang-Mills theory is not described as a local field theory on the $S^3$ on which the original Yang-Mills theory
is defined - it is local on the world volume of the giant gravitons\cite{Balasubramanian:2004nb}\footnote{See also \cite{Balasubramanian:2001dx}
for an extremely interesting, related set of ideas. Motivated by results from black hole dynamics, this article is the first to suggest
that in the presence of very heavy D-brane states the low energy dynamics of a Yang-Mills theory may enjoy a duality with an
emergent gauge theory.}. This world volume emerges from the matrix degrees of freedom
participating in the Yang-Mills theory. Understanding how to reconstruct this emergent gauge theory may be simpler and
provide important clues into the problem of reconstructing the full AdS$_5\times$S$^5$ quantum gravity. Using the technology developed 
in the previous section, we can explore a number of interesting physical questions which allow us to explore these issues. 

In particular, by studying the amplitudes for closed string emission from an excited giant graviton we find convincing 
evidence that physics in the emergent spacetime is local. We also find indications that the emergent world volume theory 
of a bound state of giant gravitons is a non-Abelian theory with the expected gauge group and we provide some evidence 
that the quark-antiquark potential in the emergent gauge theory comes out correctly.

For a general framework which obtains the connection between CFT correlation functions and probability amplitudes in the dual quantum
gravity see\cite{withTom}. In the language of \cite{withTom} we compute amplitudes using the overlap normalization.

Since we are working in the free Yang-Mills theory, we are strictly speaking, probing the limit of tensionless strings.

\subsection{Gravitational Radiation}

In this subsection we will study the amplitudes for closed string emission from an excited D-brane and from bound states
of D-branes. This allows us to test locality in the bulk spacetime as well as to provide some evidence that the emergent
gauge theory has the correct (non-Abelian) gauge group.

\subsubsection{Closed String Radiation from Sphere Giants}

Sphere giant gravitons (and bound states of them) are conjectured to be dual to operators that correspond to representations
with $O(N)$ rows and $O(1)$ columns. 
The operator dual to a sphere giant of momentum $p$, with an open string attached is (we use the notation $(1^p)$ for the antisymmetric
representation with $p$ boxes)
$$\chi^{(1)}_{(1^{p+1}),(1^p)}(Z,W),$$
where the open string word is given by
$$ (W)^i_j=(Y^J)^i_j $$
We take $p$ to be $O(N)$.
The operator dual to the (unexcited) D-brane together with the emitted closed string is 
$$\Tr (Y^J)\chi_{(1^p)}(Z).$$
Thus, the normalized interaction amplitude is given by
$${\cal A}_{(1^{p+1}),(1^p)}=
{\langle \Tr (Y^{\dagger J})\chi_{(1^{p})}^\dagger \chi^{(1)}_{(1^{p+1}),(1^p)}\rangle\over 
\sqrt{\langle \Tr (Y^{\dagger J})\chi_{(1^{p})}^\dagger \Tr (Y^J)\chi_{(1^{p})}\rangle
\langle (\chi^{(1)}_{(1^{p+1}),(1^p)})^\dagger \chi^{(1)}_{(1^{p+1}),(1^p)}\rangle}}.$$
It is straight forward to obtain
$$\langle (\chi^{(1)}_{(1^{p+1}),(1^p)})^\dagger \chi^{(1)}_{(1^{p+1}),(1^p)}\rangle
=N^{J-1}p\prod_{i=1}^{p+1}(N-i+1)\left( 1+O\left({J^4\over N^2}\right)\right) .$$
In estimating the size of the correction, we have assumed that $N-p$ is $O(N)$; this correction is of the same
size if $N-p$ is $O(1)$. Further,
$$\langle \Tr (Y^{\dagger J})\chi_{(1^{p})}^\dagger \Tr (Y^J)\chi_{(1^{p})}\rangle =
JN^J\prod_{i=1}^{p}(N-i+1)\left(1+O\left({J^4\over N^2}\right)\right) ,$$
$$\langle \Tr (Y^{\dagger J})\chi_{(1^{p})}^\dagger \chi^{(1)}_{(1^{p+1}),(1^p)}\rangle
=JN^{J-1}\prod_{i=1}^{p+1}(N-i+1)\left(1+O\left({J^4\over N^2}\right)\right) .$$
In all expressions we quote, the corrections are due to the fact that although we treat the contractions of
the $Z$'s exactly, when contracting the open string words (i.e. the $Y$ fields), only the planar graphs
are summed.  Putting these results together, it is now straight forward to obtain
\begin{equation}
{\cal A}_{(1^{p+1}),(1^p)}=\sqrt{J(N-p)\over pN}\left(1+O\left({J^4\over N^2}\right)\right) .
\label{giantdecay}
\end{equation}
This is in agreement with the result obtained in \cite{Balasubramanian:2004nb}. To interpret this amplitude,
we will review the D-brane instability discovered in \cite{Berenstein:2006qk}. Our sphere giant is moving
in a non-zero Ramond-Ramond background flux. The giant, which will feel a Lorentz force, is accelerated 
with respect to geodesic free fall. The string which does not couple to the Ramond-Ramond background and hence
if free would undergo geodesic free fall, is pulled by the giant along the accelerated path. Thus, a non-zero 
force is exerted on the end points of the string. This force will do two things; it will tend to stretch the string
and it will tend to bring the end points of the string closer together. By bringing the end points of the open
string closer together, the force will drive the gravitational radiation.
What is the magnitude of this force exerted on the end points of the string?
We can estimate this force using classical reasoning.
Each element of the giant is moving along a circular path of radius
$$ r=R\sqrt{N-p\over N},$$
(where $R$ is the radius of the S$^5$) and with an angular velocity $\dot{\phi}=\omega ={1\over R}$. The mass of the string is 
set by $J$, $M= {J\over R}$. To drag a mass $M$ along a circular path of radius $r$, in flat space, we need to exert a force 
$\vec{F}=-F\hat{r}$ on it, with
$$ F=M\omega^2 r = {J\over R^2} \sqrt{N-p\over N}.$$
Notice that when $N-p$ is $O(N)$, $F\sim {J\over R^2}$; when $N-p$ is $O(1)$, the force is supressed by a factor of
$\sqrt{N}$ $F\sim {1\over\sqrt{N}}{J\over R^2}$. This is exactly how the amplitude ${\cal A}_{(1^{p+1}),(1^p)}$ behaves:
if $N-p$ is $O(N)$\, ${\cal A}_{(1^{p+1}),(1^p)}\sim\sqrt{J\over p}$; when $N-p$ is $O(1)$, the force is supressed by a factor
of $\sqrt{N}$, ${\cal A}_{(1^{p+1}),(1^p)}\sim {1\over\sqrt{N}}\sqrt{J\over p}$.

We can easily generalize the above result in two ways. First, to consider a boundstate of sphere giant 
gravitons, the relevant amplitude is given by
$${\cal A}_{R,R'}=
{\langle \Tr (Y^{\dagger J})\chi_{R'}^\dagger \chi^{(1)}_{R,R'}\rangle\over 
\sqrt{\langle \Tr (Y^{\dagger J})\chi_{R'}^\dagger \Tr (Y^J)\chi_{R'}\rangle
\langle (\chi^{(1)}_{R,R'})^\dagger \chi^{(1)}_{R,R'}\rangle}}.$$
where $R$ and $R'$ are defined in figure 1. The computation proceeds exactly as for the single sphere giant; we
will not show all of the details. The amplitude for gravitational radiation from a boundstate of $n$ sphere giants,
each of angular momentum $p$ is
$${\cal A}_{R,R'}={1\over\sqrt{n}}\sqrt{J(N-p)\over pN}.$$ 
\begin{figure}[t]\label{fig:cgraph1}
\begin{center}
\includegraphics[height=6cm,width=12cm]{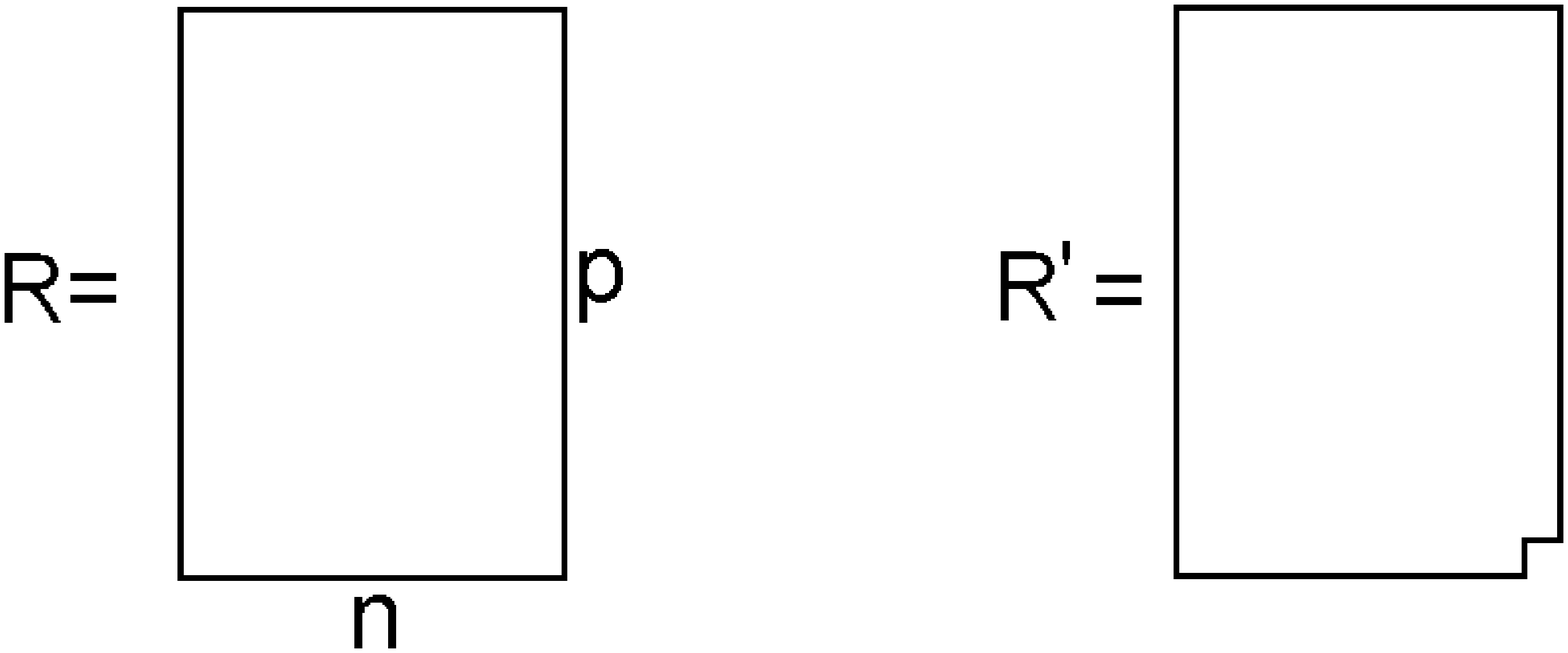}
\caption{The representations used to compute the amplitude for gravitational radiation from a bound state of $n$ sphere giants.
$R$ has $p$ rows and $n$ columns where $p$ is $O(N)$ and $n$ is $O(1)$. $R'$ is obtained by removing a single box.} 
\end{center}
\end{figure}
For a possible interpretation of
this amplitude, note that there is good evidence (see for example \cite{Balasubramanian:2005mg}) that the giant
gravitons behave as bosons.  In this case, perhaps we should extract a factor of $\sqrt{n}$ from the amplitude; this factor is the 
usual enhancement expected as a consequence of the fact that we deal with $n$ bosons. Concretely, the interaction
(the string plays no role in this argument and hence is supressed)
$$ H_{int}= g a^\dagger_{p-1}a_p $$ 
that allows a state with $n$ identical giants of momentum $p$ to decay into a state with $n-1$ giants of momentum $p$ and 
one giant of momentum $p-1$, gives a matrix element proportional to $\sqrt{n}g$ 
$$ \langle 0|{(a_p)^{n-1}a_{p-1}\over \sqrt{(n-1)!}}H_{int}{(a_p^\dagger)^n\over \sqrt{n!}}|0\rangle =\sqrt{n}g ,$$
and not $g$ as one might naively have expected. This enhancement can be understood as 
a consequence of the fact that the second quantized boson creation operators automatically produces
a correctly symmetrized state. Our excited giant operator is symmetric under swapping boxes in the same row and
hence it too builds a correctly symmetrized state. The remaining piece of the amplitude is 
$O\left({1\over n}\right)$. It is naturally interpreted as a non-planar correction in the expected emergent $U(n)$ gauge theory, 
which should arise as the low energy world volume description of $n$ co-incident D-branes. Of course, the $U(n)$ theory does not
implement this symmetrization (it is a ``first quantized description" with fixed $n$) 
which is why the $\sqrt{n}$ factor must be included before comparing.

The amplitudes we have considered here, correspond to the situation where the endpoints of the open string join and the
whole open string is emitted as a single closed string. A second generalization of the above amplitude that we can consider,
is to allow a piece of the open string to pinch off, leaving a smaller open string attached to the giant. The relevant amplitude is
$${\cal A}_{(1^{p+1}),(1^p)}(J_1,J_2)=
{\langle \Tr (Y^{\dagger J_1})(\chi_{(1^{p+1}),(1^{p})}^{(1)}(Z,W^{(2)}))^\dagger \chi^{(1)}_{(1^{p+1}),(1^p)}(Z,W^{(1)})\rangle\over 
\sqrt{{\cal N}_1 {\cal N}_2}},$$
where 
$${\cal N}_1=\langle \Tr (Y^{\dagger J_1})(\chi_{(1^{p+1}),(1^{p})}^{(1)}(Z,W^{(2)}))^\dagger \Tr (Y^{J_1})
\chi_{(1^{p+1}),(1^{p})}^{(1)}(Z,W^{(2)})\rangle,$$
$${\cal N}_2=\langle (\chi^{(1)}_{(1^{p+1}),(1^p)}(Z,W^{(1)}))^\dagger \chi^{(1)}_{(1^{p+1}),(1^p)}(Z,W^{(1)})\rangle ,$$
and the open strings $W^{(1)}$ and $W^{(2)}$ are
$$ (W^{(1)})^i_j=(Y^{J_1+J_2})^i_j\qquad (W^{(2)})^i_j=(Y^{J_2})^i_j .$$
A straight forward computation gives
$${\cal A}_{(1^{p+1}),(1^p)}(J_1,J_2)= {\sqrt{J_1}(J_2+1)\over N}\left( 1 +O\left({J^4\over N^2}\right)\right) .$$
This amplitude is maximized when $J_2$ is a maximum. Evidently it is easier for small bits of the string to break off.
At this maximum value, $J_1=O(1)$, $J_2=O(J)$ with $J=J_1+J_2$ and the amplitude is
of order $O\left({J\over N}\right)$. When $N-p$ is $O(N)$, the amplitude (\ref{giantdecay}) is $O\left(\sqrt{J\over N}\right)$.
Since ${J\over N}\ll 1$, the dominant decay process is the one in which the complete open string is emitted as a single
closed string. When $N-p$ is $O(1)$, the amplitude (\ref{giantdecay}) is $O\left({\sqrt{J}\over N}\right)$. In this case,
the dominant decay process is the one in which small pieces of the open string pinch off.

\subsubsection{Closed String Radiation from AdS Giants}

The computations for AdS giants are a straight forward generalization of the computations of the previous subsection. For this 
reason, we do not provide all of the details. 
AdS giant gravitons (and bound states of them) are conjectured to be dual to operators that correspond to representations
with $O(N)$ columns and $O(1)$ rows. The normalized interaction amplitude for the emission of a closed string by an excited AdS giant
is given by (we use the notation $(p)$ for the symmetric representation with $p$ boxes)
$${\cal A}_{(p+1),(p)}=
{\langle \Tr (Y^{\dagger J})\chi_{(p)}^\dagger \chi^{(1)}_{(p+1),(p)}\rangle\over 
\sqrt{\langle \Tr (Y^{\dagger J})\chi_{(p)}^\dagger \Tr (Y^J)\chi_{(p)}\rangle
\langle (\chi^{(1)}_{(p+1),(p)})^\dagger \chi^{(1)}_{(p+1),(p)}\rangle}}.$$
It is now a simple matter to compute
$${\cal A}_{(p+1),(p)}=\sqrt{J(N+p)\over pN}\left( 1 +O\left({J^4\over N^2}\right)\right) .$$
For small $p$ we find that this agrees with the amplitude computed for the sphere giant. This is exactly as expected: for
small values of $p$ we essentially have a point like graviton and the AdS and sphere giants are identical. However,
as $p$ is increased the above amplitude decreases more slowly than the amplitude for the sphere giant. This is what we should
expect. As $p$ increases it couples more strongly to the Ramond-Ramond field and the giant blows up, pushing further from the 
origin of AdS space. The open string attached to the giant will thus feel a greater force\footnote{Recall that the geodesic of a particle moving in
AdS space is driven towards the origin.}. This slower fall off of the AdS amplitude is again a manifestation of the
D-brane instability discovered in \cite{Berenstein:2006qk}.

We can again generalize this result to a boundstate of $n$ giants
$${\cal A}={1\over\sqrt{n}}\sqrt{J(N+p)\over pN}\left( 1 +O\left({J^4\over N^2}\right)\right) .$$ 
Again, after factoring out the bosonic enhancement factor\footnote{The AdS giants do not behave as bosons. In particular, we have an
upper limit of $N$ on the number of AdS giants that we can create. However, since $n<<N$, treating the AdS giants as 
bosons is an excellent approximation.} of $\sqrt{n}$, we obtain an amplitude of $O\left({1\over n}\right)$
which is of the correct size to be identified with the first non-planar corrections of a non-Abelian gauge theory with gauge
group $U(n)$. This is consistent with the expected emergent $U(n)$ gauge theory, which should arise as the low energy world volume 
description of $n$ co-incident AdS giant gravitons.

\subsubsection{Probing Locality Using Closed String Radiation}

Consider the representations $R$, $S$ and $T$ shown in figure 2. We are interested in the case that $b_1$ and $b_2$ are both $O(N)$
and $a_1$ and $a_2$ are both $O(1)$. The Schur polynomial $\chi_R(Z)$ should be dual to a bound state of $a_1$ giant gravitons of angular momentum 
$b_1+b_2$ and $a_2$ giant gravitons of angular momentum $b_1$. The radius of the giant graviton $R_{gg}$ is determined by its angular
momentum $p$, in terms of the radius $R$ of the $S^5$ in the AdS$_5\times$S$^5$ background as
$$ R_{gg}=R\sqrt{p\over N}.$$
The sphere giants wrap an S$^3$ within the S$^5$. Decompose the S$^5$ as S$^3\times$D$_2$ where S$^3$ is the sphere that the giant graviton
wraps and D$_2$ is a disk. The giant traces out a circular orbit on the disk, of radius $\sqrt{R^2-R_{gg}^2}$.
Thus, the $a_1$ sphere giants are separated from the $a_2$ sphere giants by a radial distance of more than\footnote{This is the distance
which separates them on the D$_2$; the fact that their worldvolumes are S$^3$s with different radii also contributes to the separation.}
$$  R\left[ \sqrt{N- b_1 \over N} -\sqrt{N-(b_1+b_2)\over N}\right] =O(R).$$
Thus, for large $R$ (we work in units of the string length) we have two well separated bound states of sphere giants. The large 
$R$ limit corresponds to the large 't Hooft coupling limit of the super Yang-Mills theory. Here we work in the opposite limit of
zero 't Hooft coupling. However, since we are working with operators that are nearly BPS, it is not unreasonable
to hope that our results can safely be extrapolated to the strong coupling limit. For this reason, even though we
work in the free field theory limit, we will still look for signals that the operator $\chi_R(Z)$ is dual to 
two well separated bound states of sphere giants. We interpret any such evidence as signals of locality in the
bulk spacetime that emerges from the matrix degrees of freedom of the original Yang-Mills theory.

To start, we will compute the amplitude
$${\cal A}_{R,T}=
{\langle \Tr (Y^{\dagger J})\chi_{T}^\dagger \chi^{(1)}_{R,T}\rangle\over 
\sqrt{\langle \Tr (Y^{\dagger J})\chi_{T}^\dagger \Tr (Y^J)\chi_{T}\rangle
\langle (\chi^{(1)}_{R,T})^\dagger \chi^{(1)}_{R,T}\rangle}}.$$
The natural interpretation of $\chi^{(1)}_{R,T}$ is as a bound state of $a_1$ sphere giants of angular momentum $b_1+b_2$
and an excited boundstate of $a_2$ sphere giants of angular momentum $b_1$. The above amplitude is easily evaluated to give
$${\cal A}_{R,T}={1\over\sqrt{a_2}}\sqrt{J(N-b_1)\over Nb_1 }\left( 1 +O\left({J^4\over N^2}\right)\right).$$
This is exactly what would have been obtained if the $a_1$ sphere giant bound state was not present. Thus, the $a_1$ sphere
giants of momentum $b_1+b_2$ do not interact with the $a_2$ sphere giants of momentum $b_1$, which is exactly the behavior we
expect from two well separated bound states.

\begin{figure}[t]\label{fig:cgraph2}
\begin{center}
\includegraphics[height=8cm,width=12cm]{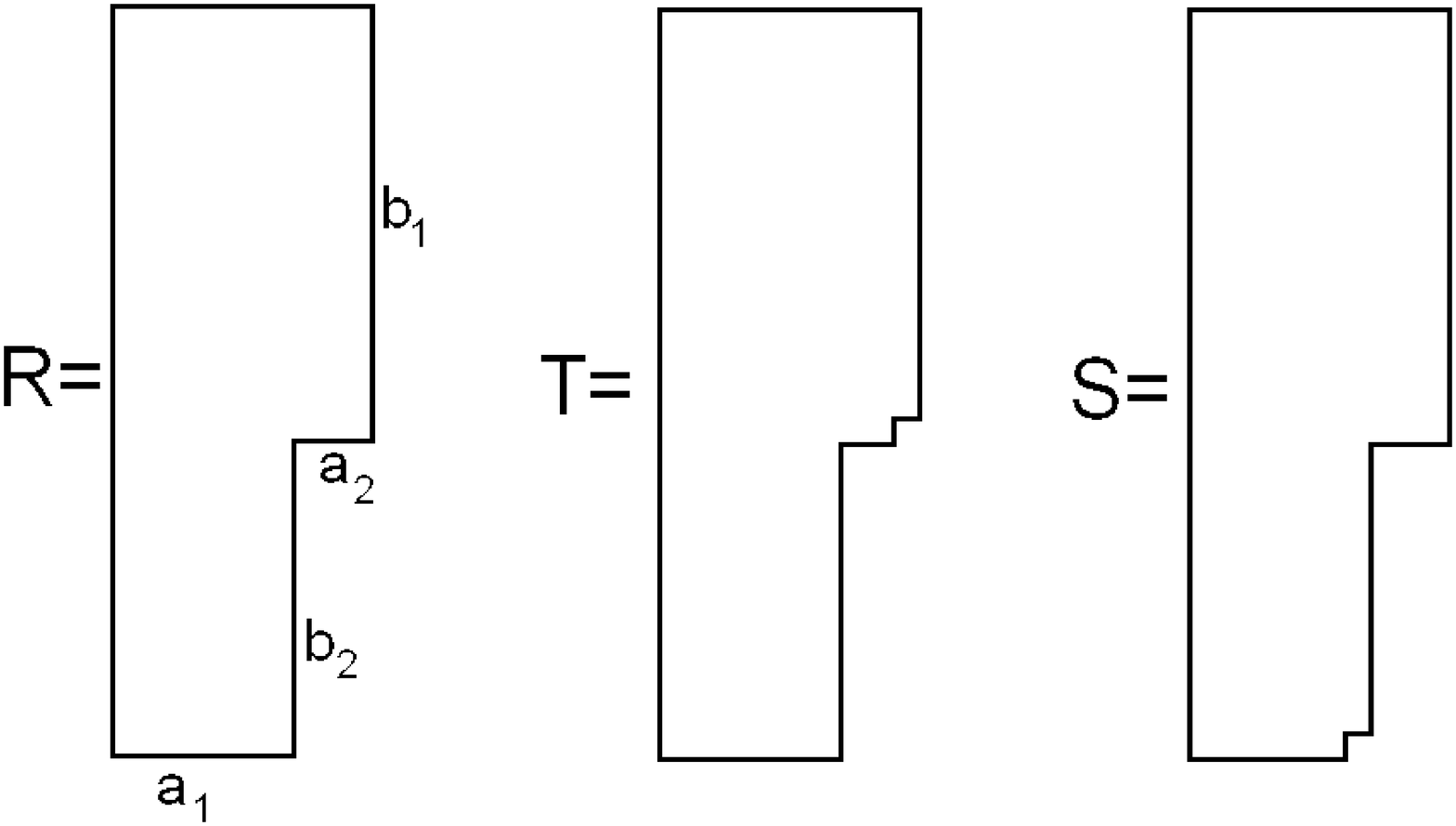}
\caption{The representations used to test locality of sphere giants.} 
\end{center}
\end{figure}

We can also consider the amplitude
$${\cal A}_{R,S}=
{\langle \Tr (Y^{\dagger J})\chi_{S}^\dagger \chi^{(1)}_{R,S}\rangle\over 
\sqrt{\langle \Tr (Y^{\dagger J})\chi_{S}^\dagger \Tr (Y^J)\chi_{S}\rangle
\langle (\chi^{(1)}_{R,S})^\dagger \chi^{(1)}_{R,S}\rangle}}.$$
In this case, it is the bound state of $a_1$ sphere giants of angular momentum $b_1+b_2$ that is excited. Evaluating
the above amplitude we obtain
$${\cal A}_{R,S}={1\over\sqrt{a_1}}\sqrt{J(N-b_1-b_2)\over (b_1+b_2)N}\left( 1 +O\left({J^4\over N^2}\right)\right) .$$
Again, this is what would have been obtained if the $a_2$ sphere giant bound state was not present, which is 
again the behavior we expect from two well separated bound states.

Using the sphere giants, we have been able to probe the question of locality in the S$^5$ of the AdS$_5\times$S$^5$ background.
We will now explore locality in the AdS$_5$ part of the background using the AdS giants. As in the case of the sphere giants,
the radius of the giant graviton $R_{gg}$ is determined by its angular
momentum $p$, in terms of the radius $R$ of the AdS$_5$ in the AdS$_5\times$S$^5$ background as
$$ R_{gg}=R\sqrt{p\over N}.$$
Thus, the $b_1$ AdS giants are separated from the $b_2$ AdS giants by a radial distance
$$  R\sqrt{a_1 + a_2 \over N} -R\sqrt{a_1\over N} =O(R).$$
The representations we will use are shown in figure 3. We assume that $a_1$ and $a_2$ are $O(N)$ and $b_1$ and $b_2$
are $O(1)$.
\begin{figure}[t]\label{fig:cgraph3}
\begin{center}
\includegraphics[height=8cm,width=12cm]{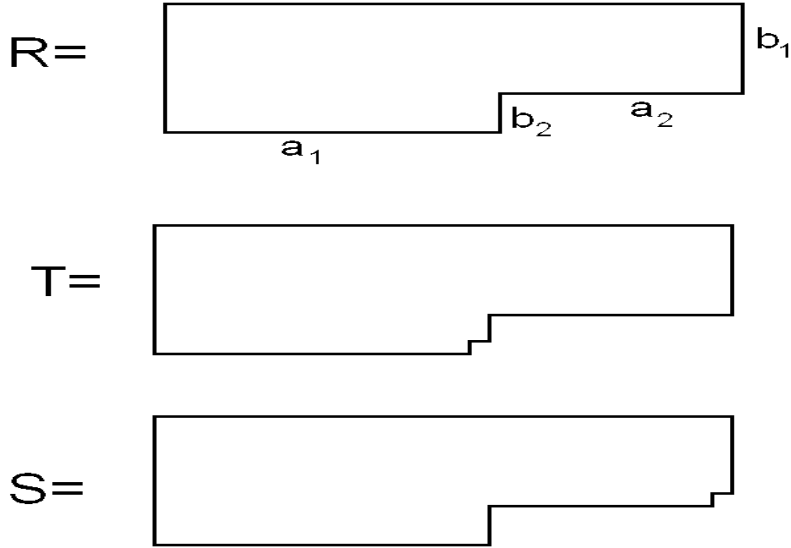}
\caption{The representations used to test locality of AdS giants.} 
\end{center}
\end{figure}
Evaluating the relevant amplitudes, we obtain
$${\cal A}_{R,T}={1\over\sqrt{b_2}}\sqrt{J(N+a_1)\over Na_1 } \left( 1 +O\left({J^4\over N^2}\right)\right),$$
$${\cal A}_{R,S}={1\over\sqrt{b_1}}\sqrt{J(N+a_1+a_2)\over (a_1+a_2)N} \left( 1 +O\left({J^4\over N^2}\right)\right) .$$
These amplitudes are manifestly consistent with locality in the AdS space. 

\subsubsection{Closed versus Open Strings}

There are two types of string excitations that can be created: we can attach an open string to the membrane bound state, or
we can excite a closed string. In the dual fermion language, the AdS space corresponds to a droplet. Single KK graviton excitations of
this droplet corresponds to ripples on the edge of the drop. Sphere giants correspond to holes deep in the droplet, while AdS
giants correspond to fermions excited well above the Fermi level. Using this dual fermion interpretation, in the Young
diagram language, Berentsein has given a beautiful translation\cite{Berenstein:2004kk} of the shape of the Young diagram into
membrane plus string excitations. This interpretation is summarized in figure 4.
\begin{figure}[t]\label{fig:cgraph3}
\begin{center}
\includegraphics[height=6cm,width=9cm]{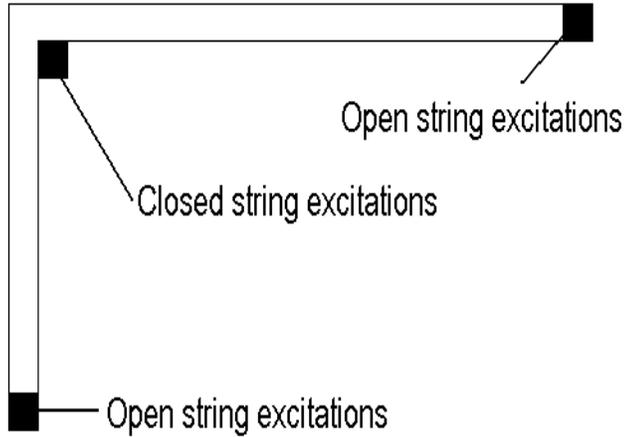}
\caption{Translation of a Young diagram into membrane plus string excitations. The string excitations correspond to the
darkened boxes.} 
\end{center}
\end{figure}
In this section we will explore and provide further evidence for this interpretation.

\begin{figure}[t]\label{fig:cgraph3}
\begin{center}
\includegraphics[height=8cm,width=12cm]{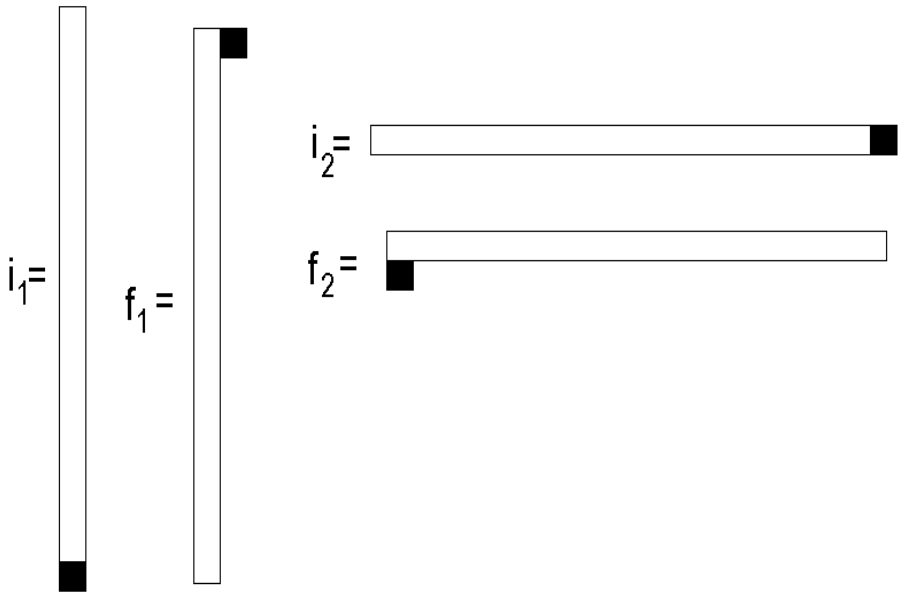}
\caption{The representations used to test the interpretation of the $F_0$ and $F_1$ contributions to the two point functions.
The darkened boxes represent the position of the string. There are $p$ white boxes in each diagram shown.} 
\end{center}
\end{figure}
Consider the amplitude to make a transition from the operator corresponding to the initial state $i_1$ to the operator described
by the final state $f_1$ (see figure 5)
$$ {\cal A}=\sqrt{J(N-p)\over Np}\left( 1 +O\left({J^4\over N^2}\right)\right).$$
This amplitude only recieves a contribution from the $F_1$ term.
This is identical to the amplitude for the initial excited sphere giant state to decay into an unexcited sphere giant plus a closed 
string. Thus, the final state is behaving as if it is a closed string plus sphere membrane state. This is in perfect agreement
with Berenstein's picture. We can also
consider the amplitude to make a transition from the operator corresponding to the initial state $i_2$ to the operator described
by the final state $f_2$
$$ {\cal A}=\sqrt{J(N+p)\over Np}\left( 1 +O\left({J^4\over N^2}\right)\right).$$
Again, this amplitude only recieves a contribution from the $F_1$ term.
This is identical to the amplitude for the initial excited AdS giant state to decay into an unexcited AdS giant plus a closed 
string. Thus, the final state is behaving as if it is a closed string plus AdS membrane state. Again this is in perfect agreement
with Berenstein's picture.

\subsubsection{Spacetime Foam}

In the subsection 3.1.3 we have managed to find some circumstantial evidence for locality in the bulk spacetime that emerges from the matrix 
degrees of freedom of the original Yang-Mills theory. It is interesting (and comforting) to see this locality emerge in a situation
where it is expected. It is equally interesting to ask how locality breaks down, a question that should provide some insight into what
happens at the Planck scale in a theory of quantum gravity. In the regime in which locality does break down, we expect that quantum gravity
corrections become important. As a first step towards probing these issues, in this section we will study the correlation functions of 
operators which are described in the dual quantum gravity by a ``spacetime foam". The low energy effective description of a 
foam\cite{Balasubramanian:2005mg} with given global charges is in terms of the superstar geometry\cite{Myers:2001aq}, which is singular.
For very interesting and insightful related work see\cite{Silva}.

We have already mentioned that as the number of boxes in the Young diagram changes, the interpretation of the operator in the dual
gravitational theory changes: for $\sim 1$ boxes the operator is dual to a supergravity state, for $\sim\sqrt{N}$ boxes the operator
is dual to a string state and for $\sim N$ boxes the operator is dual to a giant graviton. In this section we consider operators with
$\sim N^2$ boxes. In this case, the operator is dual to a new geometry.

The two point function of an excited giant with a single string attached receives two contributions, depending on how the associated open
string words contract. We have denoted these two open string contributions as $F_0$ and $F_1$. In the preceding subsections, the term
proportional to $F_0$ has been the dominant contribution to the normalization factor of the amplitudes we computed. Given
the way the indices of the string contract, it is natural to interpret this term as an open string {\it overlap}. $F_0$ enters when
we compute the {\it normalization} of an operator.
By the same logic, the term proportional to $F_1$ is naturally interpreted as a 
closed string {\it interaction}. $F_1$ naturally enters when we compute the {\it transition amplitude} between two states: the open string
``peels" off the boundstate to form a closed string and then ``reattaches" as an open string in a new position.
In this section, we will ask if there are operators $\chi_{R,R'}^{(1)}$, such that the contributions from the $F_1$ and $F_0$ 
terms to the two point
function are of the same order of magnitude. Presumably, the geometries dual to these operators have Planck scale features implying high
curvatures, so that the corrections to the leading result are important.

Consider attaching a single string to our boundstate of giants.
Using the rules from section 2.5, we know that the two point function takes the form
$$\langle (\chi_{R,R'}^{(1)})^\dagger \chi_{R,R'}^{(1)}\rangle = nF_0 f_R {d_{R'}\over d_R} + F_1 c_{R,R'}^2 f_{R'},$$
where $n$ is the number of boxes in the Young diagram, $c_{R,R'}$ is the weight of the box that must be removed from $R$
to obtain $R'$ and $d_R,d_{R'}$ are the dimensions of the $R,R'$ representations of the symmetric groups $S_n,S_{n-1}$.
Now, we know that $f_R=c_{R,R'}f_{R'}.$ Thus, we are looking for representations $R$ such that
$$ nF_0 {d_{R'}\over d_R} \sim  F_1 c_{R,R'}.$$
In the limit that we consider ($J^2 N\ll 1$) $F_0$ dominates $F_1$. Thus, we need to find representations $R$ for which
$${d_{R'}\over d_R}\ll 1.$$
This will be the case if representation $R$ can subduce many different representations $R'$. The set of all possible
representations that can be subduced from a given Young diagram $R$ is given by the set of all Young diagrams that can be 
obtained by removing a single box from $R$. Since we can remove any box that is a corner, this naturally suggests that we
should consider representations $R$ which correspond to Young diagrams that have a very large number of corners. An example of
such a Young diagram is given in figure 6.
\begin{figure}[t]\label{fig:cgraph5}
\begin{center}
\includegraphics[height=6cm,width=12cm]{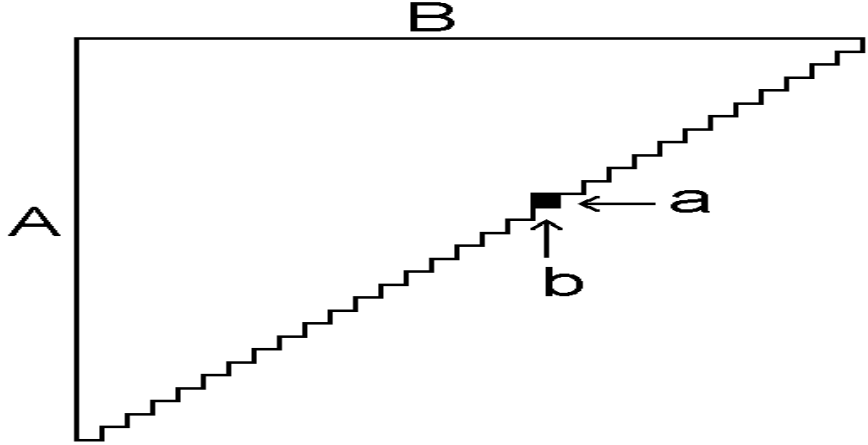}
\caption{The representation $R$ shown has the maximum number of ``corners" possible. As a result, it subduces a large number of representations
when a single box is removed. $R'$ is obtained by removing the filled box. The filled box is in row $a$ and column $b$.
We consider the case that $A$ and $B$ are both $O(N)$.} 
\end{center}
\end{figure}
The quantities appearing in our two point function are
\begin{eqnarray}
n{d_{R'}\over d_R}&=&{(A-a+b)!!\over (A-a+b-1)!!}{(B+a-b)!!\over (B+a-b-1)!!}\cr
&=& {\left[(A-a+b)!!\right]^2\over (A-a+b)!}{\big[(B+a-b)!!\big]^2\over (B+a-b)!},\cr
& &\cr
c_{R,R'}&=&N-a+b.
\nonumber
\end{eqnarray}
We will consider the case that $b=O(N)$, $a=O(1)$ so that $A-a+b=O(N)$ and $B+a-b=O(1)$.
Using Stirling's approximation to evaluate the above factorials we find
$$ n{d_{R'}\over d_R} ={e^2\sqrt{2} \over \sqrt{\pi}}\sqrt{A-a+b} .$$
Clearly, $c_{R,R'}$ is $O(N)$. Taking the open string word
$$ (W)^i_j=(Y^J)^i_j $$
with $J^2 N$ fixed, we easily find
$$ nF_0 {d_{R'}\over d_R}\sim \sqrt{A-a+b} N^{J-1}, \qquad  F_1 c_{R,R'}\sim (J-1)N^{J-1}.$$
Thus, the corrections to the leading term are the same size as the leading term. It is interesting to note that triangular
Young diagrams were studied in \cite{Balasubramanian:2005mg},\cite{Silva} where they were argued to be relevant for a description of the geometry
of the superstar\cite{Myers:2001aq} and further were used to show that they are described by an effective geometry that is singular. This singular
geometry was interpreted as an effective description of microstates that differ from each other by Planck scale structures.
See \cite{GubserH} for a connection to ${\cal R}$ charged black holes in AdS$_5\times$S$^5$.

Based on the results of this section, it is clear that the number of corners in the Young diagram provides important information about
the dual effective geometry. The importance of the corners in Young tableaux is clear from the LLM solution: the
number of corners is the number of edges of LLM geometries with circular symmetry.
If the diagram has $O(1)$ corners and $O(N)$ boxes, it should be described by a smooth effective geometry.  
If the diagram has $O(N)$ corners, the corresponding microstates exhibit Planck scale structure and should be described by a singular 
effective geometry. The number of corners translates roughly into the number of distiguishable bunches of D-branes. More
corners implies more distinguishable D-branes, and hence more possible open strings excitations. Thus, brane systems described by
Young diagrams with many corners have many nearby states that can be explored, implying a large entropy, signaling that one is getting 
nearer to a black hole state\footnote{We thank David Berenstein for suggesting this explanation to us.}. 

\subsection{Interacting Giants}

In this section we will consider the process in which an excited giant graviton makes a transition from one excited state
to another excited state. The amplitude for this process is given by
$${\cal A}=
{{\langle (\chi_{R,R'}^{(1)}})^\dagger \chi^{(1)}_{T,T'}\rangle\over 
\sqrt{\langle (\chi_{R,R'}^{(1)})^\dagger \chi_{R,R'}^{(1)}\rangle
\langle (\chi^{(1)}_{T,T'})^\dagger \chi^{(1)}_{T,T'}\rangle}}.$$
The representations $R,R'$ and $T,T'$ are defined in figure 7. When we consider the case that $b_1$ and $b_2$ are $O(N)$ and
$a_1$ and $a_2$ are $O(1)$, this process looks like the open string attached to the bound state of $a_1$ sphere giants (of angular
momentum $b_1+b_2$) is emitted and then absorbed by the bound state of $a_2$ sphere giants (of angular momentum $b_1$). The leading
contribution to the amplitude for this process is given by
$$ {\cal A}={1\over\sqrt{a_1}}\sqrt{J(N-b_1-b_2)\over N(b_1+b_2)}\times {1\over\sqrt{a_2}}\sqrt{J(N-b_1)\over Nb_1 }
\left( 1 +O\left({J^4\over N^2}\right)\right).$$
Notice that this amplitude is a product of the amplitude for the first bound state to emit a string times the amplitude for the
second bound state to absorb the string. This is what one would guess for the amplitude if one assumes a local cubic interaction,
and if in addition the amplitude for a closed string to propagate from the first bound state to the second bound state is 1. This
second assumption is very natural, since in the free field theory limit that we consider, the background is small in units of the 
string length.

We could also consider the case that $a_1$ and $a_2$ are $O(N)$ and $b_1$ and $b_2$ are $O(1)$. In this case, the process we are studying
looks like an open string attached to the bound state of $b_1$ AdS giants (of angular momentum $a_1+a_2$) is emitted and then absorbed by 
the bound state of $b_2$ AdS giants (of angular momentum $a_1$). The leading
contribution to the amplitude for this process is given by
$$ {\cal A}={1\over\sqrt{b_1}}\sqrt{J(N+a_1+a_2)\over N(a_1+a_2)}\times {1\over\sqrt{b_2}}\sqrt{J(N+a_1)\over N a_1 }
\left( 1 +O\left({J^4\over N^2}\right)\right).$$
Again, this amplitude is a product of the amplitude for the first bound state to emit a string times the amplitude for the
second bound state to absorb the string. Both transition amplitudes received a contribution only from the $F_1$ term.
\begin{figure}[t]\label{fig:cgraph4}
\begin{center}
\includegraphics[height=6cm,width=12cm]{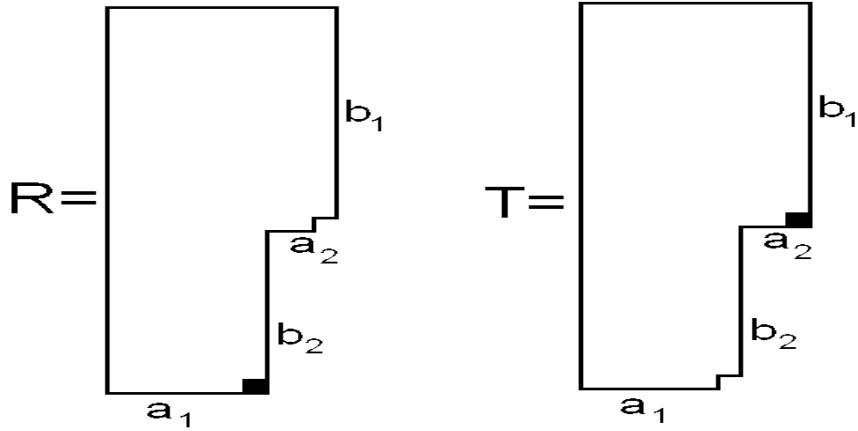}
\caption{The representations $R$ and $T$ used to compute the amplitude for a transition between two excited giant graviton
states. $R'$ and $T'$ are obtained by removing the filled box from $R$ and $T$ respectively.} 
\end{center}
\end{figure}

\subsection{Splitting and Joining}

The open strings that live on the world volume of the giant gravitons interact by splitting and joining. In this subsection we consider
the simplest possible process of two strings in their ground states joining into one. The operators with one and two string excitations 
are defined using the representations given in figure 8. The normalized amplitude for the string joining process is given by
$${\cal A}={\langle (\chi^{(1)}_{S,S'})^\dagger \chi^{(2)}_{R,R''}\rangle\over
\sqrt{\langle (\chi^{(1)}_{S,S'})^\dagger \chi^{(1)}_{S,S'}\rangle
\langle (\chi^{(2)}_{R,R''})^\dagger \chi^{(2)}_{R,R''}\rangle}}.$$
The correlator in the numerator will be related to the open string field theory vertex for the open string field theory describing
the strings attached to a giant graviton. A discussion of this point for the case of BMN operators can be found in\cite{huang}.
The open strings attached to $\chi^{(2)}_{R,R''}$ are given by
$$ W^{(1)}=Y^{J_1},\qquad W^{(2)}=Y^{J_2}.$$
The open string attached to $\chi^{(1)}_{S,S'}$ is given by
$$ W=Y^{J_1+J_2}.$$ 

The computation of the denominator is straight forward using the rules given in section 2.5. The computation of the 
correlator in the numerator
factorizes into a contribution from the open string words and a contribution from contracting the $Z$s. The open string correlator
is treated in appendix G. We keep only the leading contribution, that is, the terms labeled $e$ and $f$ in appendix G. The
contribution coming from contracting the $Z$s is treated in appendix H. Using these results, we find that the leading contribution 
to the amplitude is given by
\begin{figure}[t]\label{fig:cgraph6}
\begin{center}
\includegraphics[height=8cm,width=12cm]{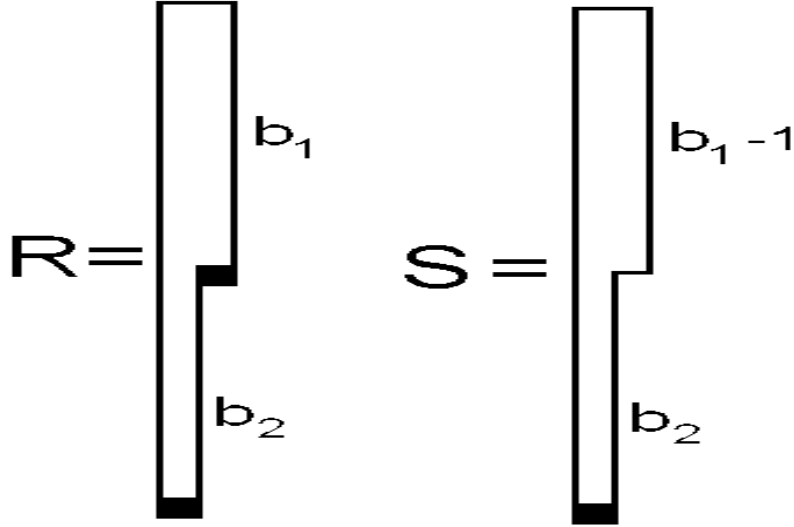}
\caption{The representations used in the computation of the string joining amplitude. The representation $R'$ is obtained by
dropping the filled box in the second column of $R$. The representation $R''$ is obtained by dropping both filled boxes from 
$R$. The representation $S'$ is obtained by dropping the filled box from $S$. We consider the situation that $b_1$
is $O(N)$ and $b_2$ is $O(1)$ and $b_2\gg 1$.} 
\end{center}
\end{figure}
$${\cal A}=2\sqrt{N-b_1\over Nb_1}{1\over b_2+1}\left( 1 +O\left({J^4\over N^2}\right)\right).$$
We take $b_1$ to be $O(N)$ and $b_2$ to be $O(1)$. Thus, the two strings which join are on branes that are nearby in spacetime.
For this reason, we expect that we are probing the dynamics of the brane world volume theory. Notice that this amplitude is
independent of the angular momentum of the open strings which join.
To interpret this amplitude, note that arguing exactly as in section 3.1.3, the two open strings that are
interacting are separated by a distance of ($R$ is the radius of the S$^5$)
$$ r=R\left[\sqrt{N-b_1\over N}-\sqrt{N-b_1-b_2\over N}\right]\approx {Rb_2\over 2\sqrt{N(N-b_1)}}.$$
The distance $r$ between the two interacting strings is determined by $b_2$. $b_2$ itself is obtained by counting how
many boxes on the Young diagram we pass through when moving on the right most edge of the Young diagram, between the two
strings that are interacting\footnote{We find that this is generally the case for string
splitting/joining amplitudes: for general choices of representations $R$ and $S$, the joining/splitting amplitude falls off
as the inverse of the number of boxes on the Young diagram we pass through when 
moving on the right most edge of the Young diagram, between the two
strings that are interacting.}. This is another indication that the geometry
is coded into the Young diagram labeling the operator.
There is a connection between the coordinate that we have identified here and the coordinates employed by LLM\cite{Lin:2004nb}.
Recall that the boundary conditions for the LLM solutions are specified by droplets on the $y=0$ plane. The radial coordinate
on the $y=0$ plane is the distance $r$ we have just introduced.
The fact that the string sigma model dynamics simplifies when expressed using these LLM
coordinates was emphasized in \cite{clifford}.
In terms of this distance, we can write the amplitude for string joining as (we approximate $b_2+1\approx b_2$)
$$ {\cal A}={R\over N\sqrt{b_1}}{1\over r}. $$
To reproduce this amplitude in Born approximation we would need a ${1\over r}$ (quark-antiquark) potential in the brane world volume theory.
Assume that the dominant contribution to this potential will arise from the exchange of massless particles. To reproduce the
potential implied by our amplitude we see that the emergent Yang-Mills theory is a 3+1 dimensional theory.

Although the fact that we have obtained a ${1\over r}$ potential is encouraging, our argument is not as clean as we would like. Indeed, 
if it is possible, we would have liked to separate the two open strings in the directions belonging to the emergent world volume,
and place them at an equal radial distance. The separation appearing in our calculation is, if our naive interpretation is
correct, seems to be in the radial direction. 

\section{Discussion}

In this paper we have initiated a systematic study of the operators dual to giant gravitons with open strings attached to
them. We have introduced a graphical notation, which employs Young diagrams, for these operators. The computation of two 
point correlation functions has been reduced to the application of three simple rules, which were summarized in section 2.5. 
The rules themselves are written as graphical operations performed on the Young diagram labels of the operators and the final 
result for the correlation function is read directly off these labels. As a test of our results, we have written code to
numerically compute the values of a number of correlators. Our graphical rules are in complete agreement with this 
``experimental data". Using this technology, we have
studied gravitational radiation by giant gravitons and bound states of giant gravitons, transitions between excited giant
graviton states and joining of open strings attached to the giant. The results of our study suggest a number of interesting
conclusions:

\begin{itemize}

\item The Young diagram labeling of the operators dual to giant gravitons originally introduced to study the $c=1$ matrix model
      by Jevicki\cite{Antal} and in the ${\cal N}=4$ super Yang-Mills context by Corley, Jevicki and Ramgoolam,
      continues to provide a useful labeling in the more general situation where the operators are dual to excited
      giant gravitons.

\item We have studied operators labeled by Young diagrams with $m=O(1)$ columns, each having $p=O(N)$ rows. These are 
      expected to be dual to a bound state of $m$ coincident sphere giants. The expected world volume theory of these $m$
      sphere giants is a 3+1 dimensional Yang-Mills theory with gauge group $U(m)$. This emergent theory should be local
      on a space built out of the matrix degrees of freedom of the original Yang-Mills theory. Thus this is a concrete
      toy model that can be used to study how extra dimensions arise from matrix models. We have found some evidence for this 
      conjectured $U(m)$ emergent gauge theory. After extracting the usual bosonic
      enhancement factor of $\sqrt{m}$ from the amplitude for gravitational radiation from a bound state of $m$ giant gravitons,
      we find an amplitude of $O\left({1\over m}\right)$, which is the correct size to be interpreted as a non-planar correction
      in a Yang-Mills theory with gauge group $U(m)$. Further, by studying string joining and splitting we have found evidence
      for a quark-antiquark potential that drops as ${1\over r}$, with $r$ the distance between the quark and the antiquark.
      This may be a signal that the emergent gauge theory has three spatial dimensions. The distance $r$ is defined by counting how
      many boxes on the Young diagram we pass through when moving on the right most edge of the Young diagram, between the two
      strings that are interacting. This is just one of many examples where we have an indication that the geometry
      is coded into the Young diagram labeling the operator.

\item Operators labeled by Young diagrams with $m=O(1)$ rows, each having $p=O(N)$ columns are dual to a bound state of
      AdS giants. We have again found evidence that the world volume theory of these $m$ AdS giants is described by
      a 3+1 dimensional Yang-Mills theory with gauge group $U(m)$.

\item We have found signals of locality in the bulk spacetime. To probe this issue, the basic process we have considered is the emission of
      a closed string from a bound state of excited giant gravitons that are separated from a second (unexcited) bound state.
      We have some evidence that widely separated bound states of giant gravitons do not interact with each other.

\item We have studied transitions between excited giant graviton states. The operators we 
      use for the study of transitions are naturally interpreted as dual to 
      states of widely separated bound states of giant gravitons. The transitions we study correspond to the process where an open string
      attached to one of the bound states is emitted and reabsorbed by the second bound state. The amplitude can be written neatly as the
      product of the amplitude for the first bound state to emit a closed string with the amplitude for the second bound state to absorb
      a closed string. This suggests that a description of the dynamics employing the giant graviton and string degrees of freedom
      will be a local theory with a cubic interaction vertex. It is also interesting to ask if, using amplitude calculations of the type
      we have explored, such a dynamical description can be constructed. For relevant related work directed at (unexcited) giant gravitons
      see for example\cite{Mandal} and for work directed at strings see for example \cite{friends}.

\item Our two point correlators for giants with a single string attached receive two contributions, distinguished by how the open string
      words are contracted. We have seen that these two contributions are naturally interpreted as a leading (open string overlap) term 
      and a (closed string interaction) correction term. The leading term determines the normalizations of our amplitudes. The correction
      determines transition amplitudes. We have been able to describe the circumstances in which the correction is the same size as the leading
      term. 
      In these situations we expect that quantum gravity corrections become important and classical notions, such as spacetime and
      locality, may not be applicable. Our results show
      that the number of corners in the Young diagram provides important information about the dual effective geometry. If the diagram has 
      $O(1)$ corners and $O(N)$ boxes, it should be described by a smooth effective geometry. If the diagram has $O(N)$ corners, the corresponding
      microstates exhibit Planck scale structure and should be described by a singular effective geometry. This is natural in view of the LLM
      solutions: a Young diagram with many corners corresponds to having many concentric, thin black rings on the $y=0$ plane. 
      Since thin rings will effectively be averaged in a low energy description, this leads to a gray disk on the $y=0$ plane
      which does lead to a singular geometry. 
      These conclusions are consistent with the notion of the quantum foam described in \cite{Balasubramanian:2005mg}.

\item We find a nice confirmation of the physical picture developed in \cite{Berenstein:2004kk}. This allows us to translate a given
      restricted Schur polynomial into giant graviton bound state plus 
      a specific set of string excitations. String words added at the end of a long row or a long
      column correspond to open string excitations of the giant graviton bound state. String words added in row $i$ and column $j$
      with both $i$ and $j$ $O(1)$ describe closed strings.

\end{itemize}

One deficiency of our work, is that our results apply only in the zero coupling limit of the theory.
In a second article \cite{jelena}, we will show that the action of the $F$-terms can also be summarized by a simple
graphical rule. This allows us to account for the first perturbative correction to the free field theory answer.

$$ $$

\noindent
{\it Acknowledgements:} 
We would like to thank Michael Abbott, David Bekker, Sera Cremonini, Aristomenis Donos, Antal Jevicki, 
Jeff Murugan, Sanjaye Ramgoolam, Joao Rodrigues, Michael Stephanou, Alex Welte and especially David Berenstein,  
for pleasant discussions and/or helpful correspondence. 
This work is supported by NRF grant number Gun 2047219.

\appendix

\section{Reduction Formula for Schur Polynomials}

In this appendix we prove the reduction formula discovered in \cite{deMelloKoch:2004ws}. Our strategy is to exploit
known recursion relations obeyed by Schur polynomials corresponding to the completely symmetric and antisymmetric representations
to prove the reduction formula for these representations. We then recall the rewriting of an arbitrary 
Schur polynomial in terms of the determinant of a matrix whose elements are Schur polynomials of only the symmetric or antisymmetric 
representations. Using this relation and the already established result, we extend the proof of the reduction rule to an arbitrary Schur 
polynomial.

\subsection{Statement of the Reduction Rule}

Consider a Schur polynomial $\chi_{(\lambda_1,\lambda_2,\cdots ,\lambda_r)}$ with $r$ rows
and $\lambda_i$ boxes in the $i^{th}$ row. The reduction rule we wish to prove says that the
reduction of the Schur polynomial is given by summing all Schur polynomials that can be obtained by removing a single box to leave a
valid Young diagram, and weighting each such term by the weight of the removed box\footnote{Recall that the box in the $i^{th}$
row (counting the top row as 1 and increasing by one for each row below it) and $j^{th}$ column (counting the leftmost column 
as 1 and increasing by one for each column to the right) has weight $N-i+j$.}
\begin{eqnarray}
D\chi_{(\lambda_1,\lambda_2,\cdots ,\lambda_r)}&=&(N+\lambda_1-1)\chi_{(\lambda_1 -1,\lambda_2,\cdots ,\lambda_r)}
  +(N+\lambda_2-2)\chi_{(\lambda_1,\lambda_2-1,\cdots ,\lambda_r)}+\cdots\cr
& &\quad  +(N+\lambda_r-r)\chi_{(\lambda_1,\lambda_2,\cdots ,\lambda_r-1)}.
\nonumber
\end{eqnarray}

\subsection{Symmetric Representations}

The Schur polynomials corresponding to the completely symmetric representations $\chi_{(k)}(Z)$ obey Brioschi's formulae\cite{Lederman}
$$\chi_{(k)}(Z)={1\over k}\sum_{p=1}^{k}\Tr (Z^p)\chi_{(k-p)}(Z).$$
In the above relation, $\chi_{(0)} (Z)=1$ and $\chi_{(1)} (Z)=\Tr Z$. Clearly,
$$ D\chi_{(0)} (Z) =0,\qquad D\chi_{(1)} (Z) =N=N\chi_{(0)}(Z).$$
Make the inductive hypothesis 
\begin{equation}
D\chi_{(l)}=(N+l-1)\chi_{(l-1)}(Z),\qquad k > l\ge 1 ,
\label{SymmRule}
\end{equation}
which we have proved for $l=1$. Reducing Brioschi's formulae we have
$$ D\chi_{(k)}(Z)={1\over k}\sum_{p=1}^{k}\left(D\Tr (Z^p)\chi_{(k-p)}(Z)+\Tr (Z^p)D\chi_{(k-p)}(Z)\right).$$
Use the inductive hypothesis and $D\Tr (Z^p)=p\Tr (Z^{p-1})$ to obtain ($\chi_{(-1)}(Z)\equiv 0$)
\begin{eqnarray}
D\chi_{(k)}(Z)&=&{1\over k}\sum_{p=1}^{k}\left(p\Tr (Z^{p-1})\chi_{(k-p)}(Z)+\Tr (Z^p)(N+k-p-1)\chi_{(k-p-1)}(Z)\right)\cr
&=&{N+k\over k}\sum_{p=1}^{k-1}\Tr (Z^p)\chi_{(k-p-1)}(Z)+{N\over k}\chi_{(k-1)}(Z)\cr
&=& {N+k\over k}(k-1)\chi_{(k-1)}+{N\over k}\chi_{(k-1)}=(N+k-1)\chi_{(k-1)}(Z) .
\nonumber
\end{eqnarray}
To get the last line above, we again used Brioschi's formulae. This furnishes a proof by induction of (\ref{SymmRule}).

\subsection{Antisymmetric Representations}

The Schur polynomials corresponding to the completely antisymmetric representations $\chi_{(1^k)}(Z)$ obey Newton's formulae\cite{Lederman}
$$\chi_{(1^k)}(Z)={1\over k}\sum_{p=1}^{k}(-1)^{p-1}\Tr (Z^p)\chi_{(1^{k-p})}(Z).$$
In the above relation, $\chi_{(1^0)} (Z)=1$ and $\chi_{(1^1)} (Z)=\Tr Z$. Clearly,
$$ D\chi_{(1^0)} (Z) =0,\qquad D\chi_{(1^1)} (Z) =N=N\chi_{(1^0)}(Z).$$
Make the inductive hypothesis 
\begin{equation}
D\chi_{(1^l)}=(N-l+1)\chi_{(1^{l-1})}(Z),\qquad k > l\ge 1 ,
\label{AntiSymmRule}
\end{equation}
which we have proved for $l=1$. Reducing Newton's formulae we have
$$ D\chi_{(1^k)}(Z)={1\over k}\sum_{p=1}^{k}(-1)^{p-1}
\left(D\Tr (Z^p)\chi_{(1^{k-p})}(Z)+\Tr (Z^p)D\chi_{(1^{k-p})}(Z)\right).$$
Use the inductive hypothesis and $D\Tr (Z^p)=p\Tr (Z^{p-1})$ to obtain ($\chi_{(1^{-1})}(Z)\equiv 0$)
\begin{eqnarray}
D\chi_{(1^k)}(Z)&=&{1\over k}\sum_{p=1}^{k}(-1)^{p-1}\left(p\Tr (Z^{p-1})\chi_{(1^{k-p})}(Z)\right.\cr
&&\qquad \left. +\Tr (Z^p)(N-k+p+1)\chi_{(1^{k-p-1})}(Z)\right)\cr
&=&{N-k\over k}\sum_{p=1}^{k-1}(-1)^{p-1}\Tr (Z^p)\chi_{(1^{k-p-1})}(Z)+{N\over k}\chi_{(1^{k-1})}(Z)\cr
&=& {N-k\over k}(k-1)\chi_{(1^{k-1})}+{N\over k}\chi_{(1^{k-1})}=(N-k+1)\chi_{(1^{k-1})}(Z) .
\nonumber
\end{eqnarray}
To get the last line above, we again used Newton's formulae. This furnishes a proof by induction of (\ref{AntiSymmRule}).

\subsection{General Representations}

An arbitrary Schur polynomial can be expressed as\cite{Lederman}
\begin{equation}
\chi_{(\lambda_1,\lambda_2,\cdots ,\lambda_r)}=\det (\chi_{(\lambda_i-i+j)}).
\label{MasterFormula}
\end{equation}
As an example of this formula, it is simple to verify that 
$$\chi_{\yng(3,2)}=\det\left[\matrix{\chi_{\yng(3)} &\chi_{\yng(4)}\cr \chi_{\yng(1)} &\chi_{\yng(2)}}\right].$$
Using the already established reduction formula for the Schur polynomials of the symmetric representations
we obtain
$$ D\chi_{(\lambda_1,\lambda_2,\cdots ,\lambda_r)}=\sum_{k=1}^r\det (\chi^{k,a}_{(\lambda_i-i+j)})
+\sum_{k=1}^r\det (\chi^{k,b}_{(\lambda_i-i+j)}),$$
where
\begin{eqnarray}
\chi^{k,a}_{(\lambda_i-i+j)}&=&\chi_{(\lambda_i-i+j)},\qquad i\ne k\cr
&=& (N+\lambda_i-i)\chi_{(\lambda_i-i+j-1)}\qquad i=k,
\nonumber
\end{eqnarray}
\begin{eqnarray}
\chi^{k,b}_{(\lambda_i-i+j)}&=&\chi_{(\lambda_i-i+j)},\qquad i\ne k\cr
&=& (j-1)\chi_{(\lambda_i-i+j-1)}\qquad i=k,
\nonumber
\end{eqnarray}
The term $\sum_{k=1}^r\det (\chi^{k,b}_{(\lambda_i-i+j)})$ can be organized to give a sum of terms 
$$\sum_{k=1}^r\det (\chi^{k,b}_{(\lambda_i-i+j)})=\sum_{i=2}^{r} \, T_i ,$$
with the following structure ($T_i$ is a sum of $r$ terms)
$$T_i=\det \left[
\matrix{
0       &0       &\cdots &0         &A_{1,i-1}  &A_{1,i} &\cdots &A_{1,r-1}\cr
A_{2,1} &A_{2,2} &\cdots &A_{2,i-1} &A_{2,i}  &A_{2,i+1} &\cdots &A_{2,r}  \cr
A_{3,1} &A_{3,2} &\cdots &A_{3,i-1} &A_{3,i}  &A_{3,i+1} &\cdots &A_{3,r}  \cr
:       &:       &\cdots &:         &:        &:         &\cdots &:        \cr
A_{r,1} &A_{r,2} &\cdots &A_{r,i-1}  &A_{r,i} &A_{r,i+1} &\cdots &A_{r,r} }\right]$$
$$\, +\det\left[
\matrix{
A_{1,1} &A_{1,2} &\cdots &A_{1,i-1} &A_{1,i}  &A_{1,i+1} &\cdots &A_{1,r}  \cr
0       &0       &\cdots &0         &A_{2,i-1}&A_{2,i}   &\cdots &A_{2,r-1}\cr
A_{3,1} &A_{3,2} &\cdots &A_{3,i-1} &A_{3,i}  &A_{3,i+1} &\cdots &A_{3,r}  \cr
:       &:       &\cdots &:         &:        &:         &\cdots &:        \cr
A_{r,1} &A_{r,2} &\cdots &A_{r,i-1}  &A_{r,i} &A_{r,i+1} &\cdots &A_{r,r} }\right]+\cdots $$
$$+\, \det\left[
\matrix{
A_{1,1} &A_{1,2} &\cdots &A_{1,i-1} &A_{1,i}  &A_{1,i+1} &\cdots &A_{1,r}  \cr
A_{2,1} &A_{2,2} &\cdots &A_{2,i-1} &A_{2,i}  &A_{2,i+1} &\cdots &A_{2,r}  \cr
A_{3,1} &A_{3,2} &\cdots &A_{3,i-1} &A_{3,i}  &A_{3,i+1} &\cdots &A_{3,r}  \cr
:       &:       &\cdots &:         &:        &:         &\cdots &:        \cr
0       &0       &\cdots &0         &A_{r,i-1}  &A_{r,i} &\cdots &A_{r,r-1}   }\right].$$
Expanding the determinants for any given $T_i$ the coefficient of any given monomial vanishes. For example,
consider the coefficient of the monomial $A_{11}A_{21}\prod_{m=3}^n A_{mm}$ coming from $T_1$. Only the first two terms
contribute and they give
$$\epsilon^{2 1 3 4 \cdots n}+\epsilon^{1 2 3 4 \cdots n}=0 .$$
Thus,
$$\sum_{k=1}^r\det (\chi^{k,b}_{(\lambda_i-i+j)})=0 ,$$
so that
\begin{eqnarray}
D\chi_{(\lambda_1,\lambda_2,\cdots ,\lambda_r)}&=&\sum_{k=1}^r\det (\chi^{k,a}_{(\lambda_i-i+j)})\cr
&=&(N+\lambda_1-1)\chi_{(\lambda_1 -1,\lambda_2,\cdots ,\lambda_r)}
  +(N+\lambda_2-2)\chi_{(\lambda_1,\lambda_2-1,\cdots ,\lambda_r)}+\cdots\cr
& &\quad  +(N+\lambda_r-r)\chi_{(\lambda_1,\lambda_2,\cdots ,\lambda_r-1)},
\nonumber
\end{eqnarray}
which is the reduction rule for the general Schur polynomial. To obtain the last equality, we needed to use the
result (\ref{MasterFormula}). In using (\ref{MasterFormula}), only terms which correspond to a valid Young diagram
give a non-zero contribution, so that only these terms should be included in the final result above.

We can give an alternative proof of exactly the same result, by writing the arbitrary Schur polynomial in terms of the
Schur polynomials corresponding to the totally antisymmetric representations. For the Schur polynomial
$\chi_{(\lambda_1,\lambda_2,\cdots ,\lambda_r)}$, let $\mu_i$ denote the number of boxes in the $i^{th}$ column. The
alternative proof uses the formula\cite{Lederman}
$$\chi_{(\lambda_1,\lambda_2,\cdots ,\lambda_r)}=\det (\chi_{(1^{\mu_i-i+j})}).$$

\subsection{Examples}

To illustrate the reduction rule, we end this appendix with two examples
$$D\chi_{\yng(3,2)}=(N+2)\chi_{\yng(2,2)}+N\chi_{\yng(3,1)},$$
$$D\chi_{\yng(3,3,1)}=(N+1)\chi_{\yng(3,2,1)}+(N-2)\chi_{\yng(3,3)}.$$

\section{Reduction Formula for Restricted Schur Polynomials with One String Attached}

In this appendix we will prove the reduction formula for restricted Schur polynomials with a single string attached.
The reduction formula we are interested in involves reduction with respect to the open string attached to the giant.
For the single restricted Schur polynomials the reduction formula follows directly from the reduction formula for
the Schur polynomials, which we proved in Appendix 1. Using very similar methods, a reduction formula for specific sums 
over multiple restricted Schur polynomials can also be obtained, with little effort, from the reduction formula for the 
Schur polynomials. We conclude this appendix with a proof of the reduction formula for the general restricted Schur 
polynomial, by employing projection operators to implement the restriction on the trace. In this appendix we will use $W$ to
denote the word describing the open string attached to the giant.

\subsection{General Comments}

Recall that for the single restricted Schur polynomial, the dimension of the representation of the Schur polynomial is equal
to the dimension of the representation of the restriction, so that the trace over $R_1$ is the same as the trace over $R$. In
this case we have
$$\chi_{R,R_1}^{(1)}(Z,W)={1\over (n-1)!}\sum_{\sigma\in S_{n}}\chi_R (\sigma )Z^{i_1}_{i_{\sigma(1)}}Z^{i_2}_{i_{\sigma(2)}}
\cdots Z^{i_{n-1}}_{i_{\sigma(n-1)}}W^{i_n}_{i_{\sigma(n)}}.$$
The reduction formula for the single restricted Schur polynomials is particularly easy to discuss because we can express the single 
restricted Schur polynomial in terms of the Schur polynomial as
\begin{equation}
\chi_{R,R_1}^{(1)}(Z,W)=W^i_j{d\over dZ^i_j}\chi_R(Z) .
\label{Sngle}
\end{equation}
Consequently, we have
$$ D_W\chi_{R,R_1}^{(1)}(Z,W) = {d\over dW^k_k} \chi_{R,R_1}^{(1)}(Z,W) =
{d\over dW^k_k}W^{i}_{j}{d\over dZ^i_j}\chi_R (Z) = {d\over dZ^i_i}\chi_R (Z).
$$
Since we are considering a single restricted Schur polynomial, only one box can be removed from $R$. The box to be removed
is the box that was associated with $W$. Thus, to compute the reduction $D_W\chi_{R,R_1}^{(1)}(Z,W)$ one simply removes the 
box associated with $W$ and multiplies by the weight of the removed box.

A similar approach to the reduction rule for multiple restricted Schur polynomials is frustrated by the fact that (\ref{Sngle})
no longer applies. The best one can do is to sum over all representations $R_\alpha$ that can be obtained from $R$ by removing a single 
box. Each term in this sum traces over a subspace $R_\alpha$; the direct sum of the subspaces is $R$, $\oplus_\alpha R_\alpha=R$ so that the sum
of these terms corresponds to tracing over $R$. Thus, in this case, we have
$$\sum_\alpha \chi_{R,R_\alpha}^{(1)}(Z,W)=W^i_j{d\over dZ^i_j}\chi_R(Z).$$
It is again straightforward to see that
$$ D_W\sum_\alpha \chi_{R,R_\alpha}^{(1)}(Z,W) = {d\over dW^k_k} \sum_\alpha \chi_{R,R_\alpha}^{(1)}(Z,W)
= {d\over dW^k_k}W^{i}_{j}{d\over dZ^i_j}\chi_R (Z) = {d\over dZ^i_i}\chi_R (Z).
$$
This result is recovered by computing the reduction $D_W\chi_{R,R_\alpha}^{(1)}$ by removing the box
associated with $W$ and multiplying by the weight of the removed box. In the remainder of this appendix we develop projection operator
methods that will allow us to prove that this is indeed the correct rule.

\subsection{Tracing over Subspaces}

The definition of the restricted Schur polynomial dual to an excited giant graviton labeled by representation $R$, with a single open string 
attached, involves a trace over one of the subspaces that can be obtained by removing a single box from $R$ to leave a valid Young diagram.
In this section our goal is to give a natural group theoretic description of tracing over this subspace.

If the representation $R$ corresponds to a Young diagram with $n$ boxes, the possible subspaces that are involved are 
associated to the irreducible representations of $S_{n-1}$ obtained by restricting
representation $R$ (a representation of $S_n$) to an $S_{n-1}\times S_1$ subgroup. Our task is to distinguish between the different
representations that can arise upon restriction. To do this, consider the operator obtained by summing all two cycles of the $S_{n-1}$
subgroup
$$ O_{S_{n-1}}(2)=\sum_{i=1}^{n-2}\sum_{j=i+1}^{n-1}(ij).$$
Since this is a sum over all elements in the conjugacy class $(2,1^{n-3})$ of $S_{n-1}$, we know that
$$ g O_{S_{n-1}}(2) g^{-1}=O_{S_{n-1}}(2)\qquad \forall g\in S_{n-1},$$
$${\rm i.e.}\qquad \left[ g,O_{S_{n-1}}(2)\right]=0.$$
Thus, by Schur's lemma, we know that $O_{S_{n-1}}(2)$ takes the form $\lambda {\bf 1}$ when acting on
each irreducible representation of $S_{n-1}$. ${\bf 1}$ denotes the identity element of $S_{n-1}$. Consider
the irreducible representations of $S_{n-1}$ labeled by Young diagram $R_a$ with $r^a_i$ boxes in the $i^{th}$ 
row of the Young diagram and $c_j^a$ boxes in the $j^{th}$ column of the Young diagram. Denote a complete
orthonormal basis of states belonging to this irreducible representation by $|R_a,i\rangle$; $i$ distinguishes 
the elements in this complete basis. We then have\cite{Chen}
$$O_{S_{n-1}}(2)|i,R_a\rangle= \left[\sum_i {r_i^a (r_i^a-1)\over 2}-\sum_j {c_j^a (c_j^a-1)\over 2}\right]
|i,R_a\rangle .$$
Clearly, the operator
$$ O^\perp_{R_a}=O_{S_{n-1}}(2)-\sum_i {r_i^a (r_i^a-1)\over 2}+\sum_j {c_j^a (c_j^a-1)\over 2} $$
projects onto the orthogonal complement of the subspace space spanned by the $|R_a,i\rangle$. By appropriate use
of the projectors, we can easily construct a projector which projects onto any desired subspace.

As an example, consider the following irreducible representation of $S_6$ 
$$R=\yng(3,2,1).$$
After restricting to $S_5\times S_1$ the following representations are subduced
$$ R_1=\yng(2,2,1),\quad R_2=\yng(3,1,1)\quad R_3=\yng(3,2).$$ 
Denoting the $O_{S_{n-1}}(2)$ eigenvalue of these irreducible representations by $\lambda_i$, $i=1,2,3$ we find
$$\lambda_1=-2,\qquad \lambda_2=0,\qquad \lambda_3=2.$$
Thus,
$$O_{S_{n-1}}(2)=\sum_{i_3} 2|R_3,i_3\rangle\langle R_3,i_3|-\sum_{i_1}2|R_1,i_1\rangle\langle R_1,i_1|.$$
Using the fact that the carrier space of $R$ is spanned by the bases of $R_1$, $R_2$ and $R_3$ we have
$$ \Gamma_R({\bf 1})=\sum_{i_1} |R_1,i_1\rangle\langle R_1,i_1|
          +\sum_{i_2} |R_2,i_2\rangle\langle R_2,i_2|
          +\sum_{i_3} |R_3,i_3\rangle\langle R_3,i_3|,$$
where $\Gamma_R({\bf 1})$ denotes the identity element of $S_6$ in representation $R$. It is now easy to verify that
the operator which projects onto $R_1$ is
$$P_{R_1}={1\over 8}O_{S_{n-1}}(2)\big[O_{S_{n-1}}(2)-2\Gamma_R({\bf 1})\big]=\sum_{i_1}|R_1,i_1\rangle\langle R_1,i_1|,$$
and consequently
$$\Tr_{R_1}(\sigma )=\Tr (P_{R_1}\sigma )={1\over 8}\Tr \left( \sum_{i=1}^{4}
\sum_{j=i+1}^{5}(ij)\left[\sum_{k=1}^{4}\sum_{l=k+1}^{5}(kl)-2\right]\sigma \right).$$

For Young diagrams with a large number of boxes it is often more convenient to consider
$$O_{S_n/S_{n-1}}(2)\equiv O_{S_{n}}(2)-O_{S_{n-1}}(2)=\sum_{j=1}^{n-1} (nj),$$
which also clearly takes a distinct eigenvalue on each irreducible representation subduced when the
representation of $S_n$ is restricted to the $S_{n-1}\times S_1$ subgroup.

Although we have only discussed how to isolate representations subduced under the restriction $S_n\to S_{n-1}\times S_1$ our
methods obviously apply to the general case of subgroup $G$, including the restriction $S_n\to S_{n-k}\times (S_1)^k$; the
discussion of the general case requires only a trivial extension of what we have described.

\subsection{Reduction Formula in General}

We will start in this section by considering the operator
$$\hat{\chi}_R (Z,W) ={1\over (n-1)!}\sum_{\sigma\in S_n}Z^{i_1}_{i_{\sigma (1)}}\cdots Z^{i_{n-1}}_{i_{\sigma (n-1)}}
W^{i_n}_{i_{\sigma (n)}}\Gamma_R(\sigma ),$$
where $\Gamma_R(\sigma )$ is the matrix representing $\sigma$ in irreducible representation $R$. There is a simple
relation between this operator and all the restricted Schur polynomials that can be obtained by restricting $R$
to $S_{n-1}\times S_1$. Denote the possible irreducible representations which arise upon restriction by $R_\alpha$. Then
$$\Tr_{R_\alpha} \left(\hat{\chi}_R (Z,W)\right)=
{1\over (n-1)!}\sum_{\sigma\in S_n}Z^{i_1}_{i_{\sigma (1)}}\cdots Z^{i_{n-1}}_{i_{\sigma (n-1)}}
W^{i_n}_{i_{\sigma (n)}}\Tr_{R_\alpha} \Gamma_R(\sigma )
=\chi_{R,R_\alpha}^{(1)} (Z,W).$$
The sum over $S_n$ can be reorganized into a sum over an $S_{n-1}$ subgroup and cosets of this subgroup as follows
(in what follows we use the cycle notation for permutations)
\begin{eqnarray}
\hat{\chi}_R (Z,W) &=&{1\over (n-1)!}\sum_{\sigma\in S_{n-1}}\left[
Z^{i_1}_{i_{\sigma (1)}}\cdots Z^{i_{n-1}}_{i_{\sigma (n-1)}} \Tr (W)\Gamma_R(\sigma )\right.\cr
&+& (WZ)^{i_1}_{i_{\sigma (1)}}\cdots Z^{i_{n-1}}_{i_{\sigma (n-1)}} \Gamma_R((1,n)\sigma )+
Z^{i_1}_{i_{\sigma (1)}}(WZ)^{i_2}_{i_{\sigma (2)}}\cdots Z^{i_{n-1}}_{i_{\sigma (n-1)}}\Gamma_R((2,n)\sigma )\cr
&+&\left. ...+ Z^{i_1}_{i_{\sigma (1)}}\cdots (WZ)^{i_{n-1}}_{i_{\sigma (n-1)}} \Gamma_R((n-1,n)\sigma )\right].
\nonumber
\end{eqnarray}
The $S_{n-1}$ subgroup is the subgroup of $S_n$ comprising of the permutations $\sigma$ that leave $n$ inert, i.e. $\sigma (n)=n$.
It is now straight forward to compute the reduction
\begin{eqnarray}
D_W\hat{\chi}_R (Z,W) &=&{d\over dW^i_i}\hat{\chi}_R (Z,W)\cr
&=&{1\over (n-1)!}\sum_{\sigma\in S_{n-1}}\left[
NZ^{i_1}_{i_{\sigma (1)}}\cdots Z^{i_{n-1}}_{i_{\sigma (n-1)}} \Gamma_R(\sigma )\right.\cr
&+& (Z)^{i_1}_{i_{\sigma (1)}}\cdots Z^{i_{n-1}}_{i_{\sigma (n-1)}} \Gamma_R((1,n)\sigma )
+ Z^{i_1}_{i_{\sigma (1)}}(Z)^{i_2}_{i_{\sigma (2)}}\cdots Z^{i_{n-1}}_{i_{\sigma (n-1)}}\Gamma_R((2,n)\sigma )\cr
&+&\left. ...+ Z^{i_1}_{i_{\sigma (1)}}\cdots (Z)^{i_{n-1}}_{i_{\sigma (n-1)}} \Gamma_R((n-1,n)\sigma )\right]\cr
&=&{1\over (n-1)!}\sum_{\sigma\in S_{n-1}}Z^{i_1}_{i_{\sigma (1)}}\cdots Z^{i_{n-1}}_{i_{\sigma (n-1)}}
\left[N+\sum_{i=1}^{n-1}\Gamma_R\left((i,n)\right)\right]\Gamma_R(\sigma )\cr
&=&{1\over (n-1)!}\sum_{\sigma\in S_{n-1}}Z^{i_1}_{i_{\sigma (1)}}\cdots Z^{i_{n-1}}_{i_{\sigma (n-1)}}
\left[N+O_{S_n/S_{n-1}}(2)\right]\Gamma_R(\sigma ) .
\nonumber
\end{eqnarray}
Tracing over the subspace of $R$ corresponding to representation $R_\alpha$ we find
$$
D_W\chi_{R,R_\alpha}^{(1)} (Z,W) =
{1\over (n-1)!}\sum_{\sigma\in S_{n-1}}Z^{i_1}_{i_{\sigma (1)}}\cdots Z^{i_{n-1}}_{i_{\sigma (n-1)}}
\Tr_{R_\alpha}\left(\left[ N + O_{S_n/S_{n-1}}(2) \right]\Gamma_R(\sigma )\right) . $$
The Young diagram labeling $R_\alpha$ is obtained from $R$ by removing a single box. Assume that the removed box
lies in th $a^{th}$ row and the $b^{th}$ column. If $R$ has $r^R_i$ boxes in the $i^{th}$ row and $c^R_j$ boxes in the 
$j^{th}$ column, then $R_\alpha$ will have 
$$r^{R_\alpha}_i=r^R_i-\delta_{ia}$$
boxes in the $i^{th}$ row and
$$c^{R_\alpha}_j=c^R_j-\delta_{jb}$$
boxes in the $j^{th}$ column. Consequently, when acting on those states of irreducible representation $R$ that span the
$R_\alpha$ subspace, we obtain
$$O_{S_n}(2)=\sum_i {r_i^R (r_i^R-1)\over 2}-\sum_j {c_j^R (c_j^R-1)\over 2},$$
$$O_{S_{n-1}}(2)=\sum_i {r_i^{R_\alpha} (r_i^{R_\alpha}-1)\over 2}-\sum_j {c_j^{R_\alpha} (c_j^{R_\alpha}-1)\over 2},$$
$$O_{S_n/S_{n-1}}(2)=O_{S_n}(2)-O_{S_{n-1}}(2)=r_a^R -c_b^R .$$
Thus,
\begin{eqnarray}
D_W\chi_{R,R_\alpha}^{(1)} (Z,W) &=&
\left[ N + r_a^R -c_b^R \right]{1\over (n-1)!}\sum_{\sigma\in S_{n-1}}Z^{i_1}_{i_{\sigma (1)}}\cdots Z^{i_{n-1}}_{i_{\sigma (n-1)}}
\Tr_{R_\alpha}\left(\Gamma_R(\sigma )\right)\cr
&=& \left[ N + r_a^R -c_b^R \right]\chi_{R_\alpha}(Z) .
\nonumber
\end{eqnarray}
Note that $\left[ N + r_a^R -c_b^R \right]$ is the weight of the box that must be removed from $R$ to obtain $R_\alpha$.
This proves that the reduction $D_W\chi_{R,R_\alpha}^{(1)}(Z,W)$ is computed by removing the box
associated with $W$ and multiplying by the weight of the removed box.

\subsection{Examples}

To illustrate the reduction rule for restricted Schur polynomials with a single string attached, we end this appendix
with two examples. Setting
$$ R=\yng(3,2),\qquad R_1=\yng(2,2),\qquad R_2=\yng(3,1),$$
we have
$$D_V\chi_{R,R_1}^{(1)}=D_V\chi^{(1)}_{\young({\,}{\,}V,{\,}{\,})}=(N+2)\chi_{\yng(2,2)}=(N+2)\chi_{R_1},$$
$$D_V\chi_{R,R_2}^{(1)}=D_V\chi^{(1)}_{\young({\,}{\,}{\,},{\,}V)}=N\chi_{\yng(3,1)}=N\chi_{R_2}.$$

\section{Reduction Formula for Restricted Schur Polynomials with Multiple Strings Attached}

In this appendix we will explain how to develop reduction formulae for giant gravitons with more than one string
attached. Our strategy is to generalize the argument of appendix B.3. For concreteness, consider the case when
two strings are attached to the system of giant gravitons. Start by introducing the operator
\begin{equation}
\hat{\chi}_R(Z,W^{(1)},W^{(2)})={1\over (n-2)!}\sum_{\sigma\in S_n}Z^{i_1}_{i_{\sigma(1)}}\cdots
Z^{i_{n-2}}_{i_{\sigma(n-2)}}(W^{(1)})^{i_{n-1}}_{i_{\sigma(n-1)}}(W^{(2)})^{i_n}_{i_{\sigma(n)}}
 \Gamma_R(\sigma ) .
\label{twoattached}
\end{equation}
There is a simple relation between this operator and all the restricted Schur polynomials that can be obtained by 
restricting $R$ to $S_{n-2}\times S_1\times S_1$. Let  $R'_\alpha$ denote the possible irreducible representations 
which arise upon restricting $R$ to $S_{n-1}\times S_1$ and let $R''_\alpha$ denote the possible irreducible 
representations which arise upon restricting $R$ to $S_{n-2}\times S_1\times S_1$. Then
\begin{eqnarray}
\chi^{(2)}_{R,R''_\alpha}(Z,W^{(1)},W^{(2)})&=&\Tr_{R''_\alpha}(\hat{\chi}_R(Z,W^{(1)},W^{(2)}))\cr
&=&{1\over (n-2)!}\sum_{\sigma\in S_n}Z^{i_1}_{i_{\sigma(1)}}\cdots
Z^{i_{n-2}}_{i_{\sigma(n-2)}}(W^{(1)})^{i_{n-1}}_{i_{\sigma(n-1)}}(W^{(2)})^{i_n}_{i_{\sigma(n)}}
\Tr_{R_\alpha''}\left( \Gamma_R(\sigma )\right) .
\nonumber
\end{eqnarray}
Consider (\ref{twoattached}); we will now write the sum over $S_n$ as a sum over $S_{n-2}$ and its cosets. It is not
hard to see that
\begin{eqnarray}
\hat{\chi}_R(Z,W^{(1)},W^{(2)})&=&{1\over (n-2)!}\sum_{\sigma\in S_{n-2}}\left[
Z^{i_1}_{i_{\sigma(1)}}\cdots
Z^{i_{n-2}}_{i_{\sigma(n-2)}}\Tr (W^{(1)}W^{(2)})\Gamma_R((n,n-1)\sigma )\right. \cr
+Z^{i_1}_{i_{\sigma(1)}}&\cdots&
Z^{i_{n-2}}_{i_{\sigma(n-2)}}\Tr (W^{(1)})\Tr (W^{(2)})\Gamma_R(\sigma )\cr
+\sum_{b=1}^{n-2}Z^{i_1}_{i_{\sigma(1)}}&\cdots& (W^{(2)}Z)^{i_b}_{i_{\sigma (b)}}\cdots
Z^{i_{n-2}}_{i_{\sigma(n-2)}}\Tr (W^{(1)}) \Gamma_R((b,n)\sigma )\cr
+\sum_{b=1}^{n-2}Z^{i_1}_{i_{\sigma(1)}}&\cdots& (W^{(1)}Z)^{i_b}_{i_{\sigma (b)}}\cdots
Z^{i_{n-2}}_{i_{\sigma(n-2)}}\Tr (W^{(2)}) \Gamma_R((b,n-1)\sigma ) \cr
+\sum_{b=1}^{n-2}Z^{i_1}_{i_{\sigma(1)}}&\cdots& (W^{(1)}W^{(2)}Z)^{i_b}_{i_{\sigma (b)}}\cdots
Z^{i_{n-2}}_{i_{\sigma(n-2)}} \Gamma_R((b,n-1)(n-1,n)\sigma ) \cr
+\sum_{b=1}^{n-2}Z^{i_1}_{i_{\sigma(1)}}&\cdots& (W^{(2)}W^{(1)}Z)^{i_b}_{i_{\sigma (b)}}\cdots
Z^{i_{n-2}}_{i_{\sigma(n-2)}} \Gamma_R((b,n-1)(b,n)\sigma ) \cr
+ \sum_{a,b=1\, a\ne b}^{n-2}Z^{i_1}_{i_{\sigma(1)}}&\cdots& \left. (W^{(1)}Z)^{i_a}_{i_{\sigma (a)}}
\cdots (W^{(2)}Z)^{i_b}_{i_{\sigma (b)}}\cdots
Z^{i_{n-2}}_{i_{\sigma(n-2)}} \Gamma_R((a,n-1)(b,n)\sigma )\right].
\nonumber
\end{eqnarray}
The subgroup $S_{n-2}$ consists of those permutations $\sigma\in S_n$ that leave $n$ and $n-1$ inert: $\sigma(n-1)=n-1$
and $\sigma(n)=n$. The reduction of this expression, with respect to $W^{(1)}$ is
\begin{eqnarray}
D_{W^{(1)}}\hat{\chi}_R(Z,W^{(1)},W^{(2)})&=&{1\over (n-2)!}\sum_{\sigma\in S_{n-2}}\left[
Z^{i_1}_{i_{\sigma(1)}}\cdots
Z^{i_{n-2}}_{i_{\sigma(n-2)}}\Tr (W^{(2)})\Gamma_R((n,n-1)\sigma )\right. \cr
+N Z^{i_1}_{i_{\sigma(1)}}&\cdots&
Z^{i_{n-2}}_{i_{\sigma(n-2)}}\Tr (W^{(2)})\Gamma_R(\sigma )\cr
+N\sum_{b=1}^{n-2}Z^{i_1}_{i_{\sigma(1)}}&\cdots& (W^{(2)}Z)^{i_b}_{i_{\sigma (b)}}\cdots
Z^{i_{n-2}}_{i_{\sigma(n-2)}} \Gamma_R((b,n)\sigma )\cr
+\sum_{b=1}^{n-2}Z^{i_1}_{i_{\sigma(1)}}&\cdots& 
Z^{i_{n-2}}_{i_{\sigma(n-2)}}\Tr (W^{(2)}) \Gamma_R((b,n-1)\sigma ) \cr
+\sum_{b=1}^{n-2}Z^{i_1}_{i_{\sigma(1)}}&\cdots& (W^{(2)}Z)^{i_b}_{i_{\sigma (b)}}\cdots
Z^{i_{n-2}}_{i_{\sigma(n-2)}} \Gamma_R((b,n-1)(n-1,n)\sigma ) \cr
+\sum_{b=1}^{n-2}Z^{i_1}_{i_{\sigma(1)}}&\cdots& (W^{(2)}Z)^{i_b}_{i_{\sigma (b)}}\cdots
Z^{i_{n-2}}_{i_{\sigma(n-2)}} \Gamma_R((b,n-1)(b,n)\sigma ) \cr
+ \sum_{a,b=1\, a\ne b}^{n-2}Z^{i_1}_{i_{\sigma(1)}}&\cdots&\left. 
(W^{(2)}Z)^{i_b}_{i_{\sigma (b)}}\cdots
Z^{i_{n-2}}_{i_{\sigma(n-2)}} \Gamma_R((a,n-1)(b,n)\sigma )\right].
\nonumber
\end{eqnarray}
Collecting the second and third terms, we obtain
$${N\over (n-2)!}\sum_{\sigma\in S_{n-2}}\left[
Z^{i_1}_{i_{\sigma(1)}}\cdots Z^{i_{n-2}}_{i_{\sigma(n-2)}}\Tr (W^{(2)})\Gamma_R(\sigma )\right.$$
$$\left. +\sum_{b=1}^{n-2}Z^{i_1}_{i_{\sigma(1)}}\cdots (W^{(2)}Z)^{i_b}_{i_{\sigma (b)}}\cdots
Z^{i_{n-2}}_{i_{\sigma(n-2)}} \Gamma_R((b,n)\sigma )\right]
=N\oplus_{R'_\alpha} \hat{\chi}_{R'_\alpha}(Z,W^{(2)}).$$
Collect the fifth, sixth and seventh terms. After using $(b,n-1)(n-1,n)=(n-1,n)(b,n)$ in the fifth term
we obtain
$${1\over (n-2)!}\sum_{\sigma\in S_{n-2}}
\hat{O}_{S_n/S_{n-1}}(2)\sum_{b=1}^{n-2}Z^{i_1}_{i_{\sigma(1)}} \cdots 
(W^{(2)}Z)^{i_b}_{i_{\sigma (b)}}\cdots
Z^{i_{n-2}}_{i_{\sigma(n-2)}} \Gamma_R((b,n)\sigma )$$
$$=\hat{O}_{S_n/S_{n-1}}(2)\big[ \sum_{R'_\alpha} \hat{\chi}_{R'_\alpha}(Z,W^{(2)})
-{1\over (n-2)!}\sum_{\sigma\in S_{n-2}}
Z^{i_1}_{i_{\sigma(1)}}\cdots Z^{i_{n-2}}_{i_{\sigma(n-2)}}\Tr (W^{(2)})\Gamma_R(\sigma )\big],$$
where (i.e. here $S_{n-1}$ is the subgroup that leaves the index of $W^{(1)}$ inert, $\sigma (n-1)=n-1$)
$$\hat{O}_{S_n/S_{n-1}}(2)=\sum_{a=1}^{n-2}\Gamma_R\left((a,n-1)\right)+\Gamma_R\left((n,n-1)\right).$$
Finally, collecting the first and fourth terms we have
$${1\over (n-2)!}\sum_{\sigma\in S_{n-2}}
Z^{i_1}_{i_{\sigma(1)}}\cdots Z^{i_{n-2}}_{i_{\sigma(n-2)}}\Tr (W^{(2)})
\hat{O}_{S_{n}/S_{n-1}}(2)\Gamma_R(\sigma ).$$
Thus, we have the remarkably simple result
\begin{equation}
D_{W^{(1)}}\hat{\chi}_R(Z,W^{(1)},W^{(2)})=\left(N+\hat{O}_{S_n/S_{n-1}}(2)\right)
\oplus_{R'_\alpha} \hat{\chi}_{R'_\alpha}(Z,W^{(2)}).
\label{multiop}
\end{equation}
When acting on the subspace $R_\alpha'$, $\big[ N+\hat{O}_{S_n/S_{n-1}}(2)\big]$ is equal to the weight of the
box that must be removed to get $R_\alpha'$ from $R$ - we argued this in the previous appendix.
To obtain the reduction formula for the restricted Schur polynomial, we now simply need to trace both sides over
the representation of the restriction. There are two cases to consider: 

\noindent
{\it Each string starts and ends on a specific brane:}
In this case we are tracing over an ``on the diagonal block". The result is
$$ D_{W^{(1)}}\chi^{(2)}_{R,R''}(Z,W^{(1)},W^{(2)})=c_{R,R'}\chi^{(1)}_{R',R''}(Z,W^{(2)}),$$
where $c_{R,R'}$ is the weight of the box that must be removed from $R$ to obtain $R'$.

\noindent
{\it The strings stretch between two branes:}
In this case we are tracing over an ``off the diagonal block". The off diagonal blocks have row and column indices from different 
$R_\alpha'$ irreducible representations. Our operator (\ref{multiop}) is diagonal in $R_\alpha'$ indices so that the trace over
the off diagonal blocks vanish. Thus, in this case
$$ D_{W^{(1)}}\chi^{(2)}_{R,R''}(Z,W^{(1)},W^{(2)})=0.$$
This is not unexpected. Indeed, reducing a restricted Schur polynomial removes the open string with respect to which we are reducing.
For the states with strings stretched between the two branes it is not possible to remove a single string and still have a state
consistent with the Gauss law.

\subsection{An Example}

We will consider
$$R=\yng(3,1).$$
By removing one box from $R$, the following two irreducible representations are subduced
$$ R_1'=\yng(2,1)\qquad R_2'=\yng(3).$$
By removing a box from $R_1'$ we subduce
$$R_1''=\yng(1,1)\qquad R_2''=\yng(2),$$
and by removing a box from $R_2'$ we subduce
$$ R_3''=\yng(2).$$
$R_1''$, $R_2''$ and $R_3''$ are all one dimensional irreducible representations. We need to make some choices
when specifying the subgroups with respect to which we restrict. Once one has decided in what order the reductions
of the restricted Schur polynomial will be computed, there is a natural choice for these subgroups. In this example,
we will reduce first with respect to $W^{(1)}$ and then with respect to $W^{(2)}$. The choice of the order in which
we restrict is arbitrary - it corresponds to choosing the order in which we will Wick contract the open string words
attached to the giant. In view of our results from the previous subsection, it is simplest to take
$$ S_3=\{ 1,(12),(14),(24),(124),(142)\},$$
and
$$S_2=\{ 1,(12)\}.$$
States belonging to the carrier spaces of $R_1'$ and $R_2'$ are eigenvectors of the operator
$$\hat{O}(2)=(12)+(14)+(24),$$
with eigenvalue
$$\lambda_{R_1'}=0,\qquad \lambda_{R_2'}=3.$$
Thus,
$$P_{R,R_1'}=1-{1\over 3}\hat{O}(2),$$
projects from $R$ onto $R_1'$ and
$$P_{R,R_2'}={1\over 3}\hat{O}(2),$$
projects from $R$ onto $R_2'$. In a similar way, we can argue that
$$P_{R_1',R_2''}={1+(12)\over 2},$$
projects from $R_1'$ onto $R_2''$ and
$$P_{R_1',R_1''}={1-(12)\over 2},$$
projects from $R_1'$ onto $R_1''$. Composing these projections we easily find that $P_{A,B}$ projects from $A$
onto $B$ where
$$ P_{R,R_1''}={1-(12)\over 2}\left[1-{1\over 3}\hat{O}(2)\right],$$
$$ P_{R,R_2''}={1+(12)\over 2}\left[1-{1\over 3}\hat{O}(2)\right],$$
$$ P_{R,R_3''}={1\over 3}\hat{O}(2).$$
Computing
$$\chi_{R,R_i''}(Z,W^{(1)},W^{(2)})={1\over 2}\sum_{\sigma\in S_4}Z^{i_1}_{i_{\sigma(1)}}Z^{i_2}_{i_{\sigma(2)}}
(W^{(1)})^{i_3}_{i_{\sigma(3)}}(W^{(2)})^{i_4}_{i_{\sigma(4)}}\Tr (P_{R,R_i''}\Gamma_R(\sigma) ),$$
we find
\begin{eqnarray}
& &\chi_{R,R_1''}(Z,W^{(1)},W^{(2)})={1\over 2}\times\cr
& &\left[
\Tr (Z)^2\Tr (W^{(1)})\Tr (W^{(2)})-\Tr (Z^2)\Tr (W^{(1)})\Tr (W^{(2)})+\Tr (ZW^{(1)})\Tr (Z)\Tr (W^{(2)})\right.\cr
& &+\Tr (ZW^{(2)})\Tr (Z)\Tr (W^{(1)})+\Tr (Z)^2\Tr (W^{(1)}W^{(2)})-\Tr (Z^2)\Tr (W^{(1)}W^{(2)})\cr
& &-\Tr (Z^2W^{(1)})\Tr (W^{(2)})-\Tr (Z^2W^{(2)})\Tr (W^{(1)})+\Tr (ZW^{(1)}W^{(2)})\Tr (Z)\cr
& &+\left.\Tr (Z W^{(2)}W^{(1)})\Tr (Z)-\Tr (Z^2W^{(1)}W^{(2)})-\Tr (Z^2 W^{(2)}W^{(1)})\right],
\nonumber
\end{eqnarray}
\begin{eqnarray}
& &\chi_{R,R_2''}(Z,W^{(1)},W^{(2)})={1\over 2}\times\cr
& &\left[
\Tr (Z)^2\Tr (W^{(1)})\Tr (W^{(2)})+\Tr (Z^2)\Tr (W^{(1)})\Tr (W^{(2)})+{5/3}\Tr (ZW^{(1)})\Tr (Z)\Tr (W^{(2)})\right.\cr
& &-\Tr (Z)\Tr (W^{(1)})\Tr (ZW^{(2)})+{1\over 3}\Tr (W^{(2)}W^{(1)})\Tr (Z)^2+{1\over 3}\Tr (Z^2)\Tr( W^{(1)}W^{(2)})\cr
& &-{4\over 3}\Tr (ZW^{(1)})\Tr (ZW^{(2)})+{5\over 3}\Tr (Z^2 W^{(1)})\Tr (W^{(2)})-\Tr (W^{(1)})\Tr (Z^2 W^{(2)})\cr
& &-{1\over 3}\Tr (Z)\Tr (ZW^{(1)}W^{(2)})-{1\over 3}\Tr (Z)\Tr (ZW^{(2)}W^{(1)})-{1\over 3}\Tr (Z^2 W^{(1)}W^{(2)})\cr
& &\left.-{1\over 3}\Tr (Z^2W^{(2)}W^{(1)})-{4\over 3}\Tr (ZW^{(1)}ZW^{(2)})\right],
\nonumber
\end{eqnarray}
\begin{eqnarray}
& &\chi_{R,R_3''}(Z,W^{(1)},W^{(2)})={1\over 2}\times\cr
& &\left[
\Tr (Z)^2\Tr (W^{(1)})\Tr (W^{(2)})+\Tr (Z^2)\Tr (W^{(1)})\Tr (W^{(2)})-{2/3}\Tr (ZW^{(1)})\Tr (Z)\Tr (W^{(2)})\right.\cr
& &+2\Tr (Z)\Tr (W^{(1)})\Tr (ZW^{(2)})-{1\over 3}\Tr (W^{(2)}W^{(1)})\Tr (Z)^2-{1\over 3}\Tr (Z^2)\Tr( W^{(1)}W^{(2)})\cr
& &-{2\over 3}\Tr (ZW^{(1)})\Tr (ZW^{(2)})-{2\over 3}\Tr (Z^2 W^{(1)})\Tr (W^{(2)})+2\Tr (W^{(1)})\Tr (Z^2 W^{(2)})\cr
& &-{2\over 3}\Tr (Z)\Tr (ZW^{(1)}W^{(2)})-{2\over 3}\Tr (Z)\Tr (ZW^{(2)}W^{(1)})-{2\over 3}\Tr (Z^2 W^{(1)}W^{(2)})\cr
& &\left.-{2\over 3}\Tr (Z^2W^{(2)}W^{(1)})-{2\over 3}\Tr (ZW^{(1)}ZW^{(2)})\right].
\nonumber
\end{eqnarray}
It is now a simple task to verify that
\begin{eqnarray}
& &D_{W^{(1)}}\chi_{R,R_1''}(Z,W^{(1)},W^{(2)})=(N+2)\chi_{R_1',R_1''}(Z,W^{(2)})\cr
&=&{N+2\over 2}\left[\Tr (Z)^2\Tr (W^{(2)})-\Tr (Z^2)\Tr (W^{(2)})+\Tr (ZW^{(2)})\Tr (Z)-\Tr (Z^2 W^{(2)})\right],
\nonumber
\end{eqnarray}
\begin{eqnarray}
& &D_{W^{(1)}}\chi_{R,R_2''}(Z,W^{(1)},W^{(2)})=(N+2)\chi_{R_1',R_2''}(Z,W^{(2)})\cr
&=&{N+2\over 2}\left[\Tr (Z)^2\Tr (W^{(2)})+\Tr (Z^2)\Tr (W^{(2)})-\Tr (ZW^{(2)})\Tr (Z)-\Tr (Z^2 W^{(2)})\right],
\nonumber
\end{eqnarray}
\begin{eqnarray}
& &D_{W^{(1)}}\chi_{R,R_3''}(Z,W^{(1)},W^{(2)})=(N-1)\chi_{R_2',R_3''}(Z,W^{(2)})\cr
&=&{N-1\over 2}\left[\Tr (Z)^2\Tr (W^{(2)})+\Tr (Z^2)\Tr (W^{(2)})+2\Tr (ZW^{(2)})\Tr (Z)+2\Tr (Z^2 W^{(2)})\right],
\nonumber
\end{eqnarray}
in perfect agreement with the previous subsection. To construct the restricted Schur polynomials dual
to strings stretched between 2 branes, we need the intertwiners $A$ and $B$ which satisfy
$$ P_{R,R_2''}A=AP_{R,R_3''},\qquad BP_{R,R_2''}=P_{R,R_3''}B. $$
In this case, it is a simple matter to verify that suitable intertwiners are
$$ A={\cal M}P_{R,R_3''}(34)P_{R,R_2''},\qquad B={\cal M}P_{R,R_2''}(34)P_{R,R_3''},$$
$$ {\cal M}=\sqrt{d_{R_3''}\over\Tr (P_{R,R_3''}(34)P_{R,R_2''}(34))}.$$
In terms of these two intertwiners, the remaining two restricted Schur polynomials we can build are
$$\chi_{R,A}(Z,W^{(1)},W^{(2)})={1\over 2}\sum_{\sigma\in S_4}Z^{i_1}_{i_{\sigma(1)}}Z^{i_2}_{i_{\sigma(2)}}
(W^{(1)})^{i_3}_{i_{\sigma(3)}}(W^{(2)})^{i_4}_{i_{\sigma(4)}}\Tr (A\Gamma_R(\sigma) ),$$
$$\chi_{R,B}(Z,W^{(1)},W^{(2)})={1\over 2}\sum_{\sigma\in S_4}Z^{i_1}_{i_{\sigma(1)}}Z^{i_2}_{i_{\sigma(2)}}
(W^{(1)})^{i_3}_{i_{\sigma(3)}}(W^{(2)})^{i_4}_{i_{\sigma(4)}}\Tr (B\Gamma_R(\sigma) ).$$
We find
\begin{eqnarray}
& &\chi_{R,A}(Z,W^{(1)},W^{(2)})={1\over 2}\times\cr
& &\left[
-\Tr (ZW^{(1)})\Tr (Z)\Tr (W^{(2)})+\Tr (W^{(2)}W^{(1)})\Tr (Z)^2+\Tr (Z^2)\Tr( W^{(1)}W^{(2)})\right.\cr
& &-\Tr (ZW^{(1)})\Tr (ZW^{(2)})-\Tr (Z^2 W^{(1)})\Tr (W^{(2)})+2\Tr (Z)\Tr (ZW^{(1)}W^{(2)})\cr
& &-\Tr (Z)\Tr (ZW^{(2)}W^{(1)})+2\Tr (Z^2 W^{(1)}W^{(2)})-\Tr (Z^2W^{(2)}W^{(1)})\cr
& &-\left. \Tr (ZW^{(1)}ZW^{(2)})\right],
\nonumber
\end{eqnarray}
\begin{eqnarray}
& &\chi_{R,B}(Z,W^{(1)},W^{(2)})={1\over 2}\times\cr
& &\left[
-\Tr (ZW^{(1)})\Tr (Z)\Tr (W^{(2)})+\Tr (W^{(2)}W^{(1)})\Tr (Z)^2+\Tr (Z^2)\Tr( W^{(1)}W^{(2)})\right.\cr
& &-\Tr (ZW^{(1)})\Tr (ZW^{(2)})-\Tr (Z^2 W^{(1)})\Tr (W^{(2)})-\Tr (Z)\Tr (ZW^{(1)}W^{(2)})\cr
& &+2\Tr (Z)\Tr (ZW^{(2)}W^{(1)})-\Tr (Z^2 W^{(1)}W^{(2)})+2\Tr (Z^2W^{(2)}W^{(1)})\cr
& &-\left. \Tr (ZW^{(1)}ZW^{(2)})\right].
\nonumber
\end{eqnarray}
It is now a simple task to verify that
$$ D_{W^{(1)}}\chi_{R,A}(Z,W^{(1)},W^{(2)})=0= D_{W^{(1)}}\chi_{R,B}(Z,W^{(1)},W^{(2)}),$$
in perfect agreement with the previous subsection.

\subsection{Many Strings}

The reduction formula for operators with many strings attached can be derived following the same basic procedure
used in the two string example. Imagine we are going to reduce a $k$ string word with respect to $W^{(k)}$.
One breaks the summation over the group into a sum over a subgroup and its cosets
\begin{eqnarray}
\hat{\chi}_R^{(k)} &(&Z,W^{(1)},...,W^{(k)}) ={1\over (n-k)!}\times\cr
& &\sum_{\sigma\in S_{n-1}}\left[
Z^{i_1}_{i_{\sigma (1)}}\cdots Z^{i_{n-k}}_{i_{\sigma (n-k)}} 
(W^{(1)})^{i_{n-k+1}}_{i_{\sigma (n-k+1)}}\cdots (W^{(k-1)})^{i_{n-1}}_{i_{\sigma (n-1)}}
\Tr (W^{(k)})\Gamma_R(\sigma )\right.\cr
&+& (W^{(k)}Z)^{i_1}_{i_{\sigma (1)}}\cdots Z^{i_{n-k}}_{i_{\sigma (n-k)}} 
(W^{(1)})^{i_{n-k+1}}_{i_{\sigma (n-k+1)}}\cdots (W^{(k-1)})^{i_{n-1}}_{i_{\sigma (n-1)}}
\Gamma_R((1,n)\sigma )\cr
&+& Z^{i_1}_{i_{\sigma(1)}}(W^{(k)}Z)^{i_2}_{i_{\sigma (2)}}\cdots Z^{i_{n-k}}_{i_{\sigma (n-k)}} 
(W^{(1)})^{i_{n-k+1}}_{i_{\sigma (n-k+1)}}\cdots (W^{(k-1)})^{i_{n-1}}_{i_{\sigma (n-1)}}
\Gamma_R((2,n)\sigma )\cr
&+&\left.
...+ Z^{i_1}_{i_{\sigma(1)}}\cdots Z^{i_{n-k}}_{i_{\sigma (n-k)}} 
(W^{(1)})^{i_{n-k+1}}_{i_{\sigma (n-k+1)}}\cdots (W^{(k)}W^{(k-1)})^{i_{n-1}}_{i_{\sigma (n-1)}}
\Gamma_R((n-1,n)\sigma )\right].
\nonumber
\end{eqnarray}
The $S_{n-1}$ subgroup is the subgroup of $S_n$ comprising of the permutations $\sigma$ that leave $n$ inert, i.e. $\sigma (n)=n$.
It is straight forward to compute the reduction
$$
D_{W^{(k)}}\hat{\chi}_R^{(k)}(Z,W^{(1)},...,W^{(k)}) 
={1\over (n-k)!}\times
$$
$$
\sum_{\sigma\in S_{n-1}}Z^{i_1}_{i_{\sigma (1)}}\cdots Z^{i_{n-k}}_{i_{\sigma (n-k)}} 
(W^{(1)})^{i_{n-k+1}}_{i_{\sigma (n-k+1)}}\cdots (W^{(k-1)})^{i_{n-1}}_{i_{\sigma (n-1)}}
\left[N+O_{S_n/S_{n-1}}(2)\right]\Gamma_R(\sigma ) .
$$
Taking traces of this result gives the reduction rule. The result of this straightforward analysis is 
that the reduction of an excited giant graviton is equal to the same operator with the open string and its box
removed, multiplied by the weight of the removed box. {\it To get this simple result, we need to choose a sequence of 
reductions and use this to define the subgroups used in the restrictions, exactly as in the two string example we 
discussed in the previous subsection.} We will not provide the explicit details of this analysis; by closely
following our two string example, the diligent reader should be able construct this proof herself.

\subsection{Graphical Notation}

We have seen that, in general, an operator dual to an excited giant graviton takes the form  
$$
\chi_{R,R_1}^{(k)}(Z,W^{(1)},...,W^{(k)})
={1\over (n-k)!}\sum_{\sigma\in S_n}\Tr (\Pi \Gamma_R(\sigma))\Tr(\sigma Z^{\otimes n-k}W^{(1)}\cdots W^{(k)}),
$$
where $\Pi$ is a product of projection operators and/or intertwiners. To define $\Pi$ we need to specify
the sequence of irreducible representations used to subduce $R_1$ from $R$, as well as the chain of subgroups
to which these representations belong. Further, since in general the row and column indices of the block that we trace over
(denoted by $R_1$ in the above formula) need not coincide, we need to specify this data separately for both indices. In this subsection we will 
explain a graphical notation that summarizes this information.

For the case that we have $k$ strings, we label the words describing the open strings $1,2,...,k$. Denote the chain of 
subgroups involved in the reduction by ${\cal G}_k\subset {\cal G}_{k-1}\subset\cdots\subset {\cal G}_2\subset {\cal G}_1\subset S_n$.
${\cal G}_m$ is obtained by taking all elements $S_n$ that leave the indices of the strings $W^{(i)}$ with $i\le m$ inert.

To specify the sequence of irreducible representations employed in subducing $R_1$, place a pair of labels into each box,
a lower label and an upper label.
The representations needed to subduce the row label of $R_1$ are obtained by starting with $R$. The second representation
is obtained by dropping the box with upper label equal to 1; the third representation is obtained from the
second by dropping the box with upper label equal to 2 and so on until the box with label $k$ is dropped. The 
representations needed to subduce the column label are obtained in exactly the same way except that instead of using the
upper label, we now use the lower label.

See sections 2.2 and D.2 where this rule is illustrated graphically.

\subsection{Examples}

To illustrate our reduction rule we end this appendix by showing the sequence of reductions that can be performed in two examples:
$$ D_{W^{(1)}}\chi_{\young({\,}{\,}{\,}{\,},{\,}{\,}{\,}{\,},{\,}{\,}{1},{3}{2})}=
N \chi_{\young({\,}{\,}{\,}{\,},{\,}{\,}{\,}{\,},{\,}{\,},{2}{3})},$$
$$D_{W^{(2)}}\chi_{\young({\,}{\,}{\,}{\,},{\,}{\,}{\,}{\,},{\,}{\,},{3}{2})}=
(N-2) \chi_{\young({\,}{\,}{\,}{\,},{\,}{\,}{\,}{\,},{\,}{\,},{3})},$$

$$ D_{W^{(3)}}\chi_{\young({\,}{\,}{\,}{\,},{\,}{\,}{\,}{\,},{\,}{\,},{3})}=
(N-3) \chi_{\young({\,}{\,}{\,}{\,},{\,}{\,}{\,}{\,},{\,}{\,})}.$$

$$ D_{W^{(1)}}\chi_{\young({\,}{\,}{\,}{\,},{\,}{\,}{\,}{\,},{\,}{\,}{1},{\,}{\threetwo},{\twothree})}=
N \chi_{\young({\,}{\,}{\,}{\,},{\,}{\,}{\,}{\,},{\,}{\,},{\,}{\threetwo},{\twothree})},$$
$$D_{W^{(2)}}\chi_{\young({\,}{\,}{\,}{\,},{\,}{\,}{\,}{\,},{\,}{\,},{\,}{\threetwo},{\twothree})}=0.$$

\section{Subgroup Swap Rule}

To define the restricted Schur polynomials, we had to choose a sequence of subgroups that are used to perform the restrictions. 
{\it Choosing different sequences of subgroups leads to different polynomials.} Consider for example
$$
\chi_{R,R_1}^{(k)}(Z,W^{(1)},...,W^{(k)})={1\over (n-k)!}
\sum_{\sigma\in S_n}\Tr_{R_1} (\Gamma_R(\sigma))\Tr(\sigma Z^{\otimes n-k}W^{(1)}\cdots W^{(k)}),
$$
If we first restrict with respect to the subgroup that leaves the index of $W^{(1)}$ inert and second with respect to the
subgroup that leaves the index of $W^{(2)}$ inert, then in general, we'll get a different polynomial to what we'd get if
we first restrict with respect to the subgroup that leaves the index of $W^{(2)}$ inert and then with respect to the
subgroup that leaves the index of $W^{(1)}$ inert. In this appendix we will derive a relation between these two sets of
polynomials, which we call the ``subgroup swap rule". This rule is useful in the computation of two point functions.

\subsection{When stretched strings states can be ignored}

For the sake of clarity, we will develop our rule using a specific example. Consider 
$$
\chi_{R,R_2''}^{(2)}(Z,W^{(1)},W^{(2)})={1\over 2!}\sum_{\sigma\in S_4}
\Tr_{R_2''} (\Gamma_R(\sigma))\Tr(\sigma Z^{\otimes 2}W^{(1)}W^{(2)})
$$
$$={1\over 2!}\sum_{\sigma\in S_4}
\Tr_{R_2''} (\Gamma_R(\sigma))Z^{i_1}_{i_{\sigma(1)}}Z^{i_2}_{i_{\sigma(2)}}
(W^{(1)})^{i_3}_{i_{\sigma(3)}}(W^{(2)})^{i_4}_{i_{\sigma(4)}},$$
with the representation
$$ R=\yng(3,1),$$
of $S_4$. Restricting to an $S_3$ subgroup, we can obtain 
$$ R_1'=\yng(2,1)\quad {\rm or}\quad R_2'=\yng(3).$$
Restricting further to $S_2$, $R_1'$ subduces
$$ R_1''=\yng(1,1)\quad {\rm or} \quad R_2''=\yng(2),$$
while $R_2'$ subduces
$$ R_3''=\yng(2).$$
The two possible $S_3$ subgroups we will consider are
$$ S_3=\{ 1,(12),(13),(23),(123),(132)\},$$
which corresponds to leaving the index of $W^{(2)}$ inert, and
$$ S_3=\{ 1,(12),(14),(24),(124),(142)\},$$
which corresponds to leaving the index of $W^{(1)}$ inert. We denote the restricted Schur polynomial obtained holding
$W^{(1)}$ inert by $\chi_{R,R_i''}^{(2)}\Big|_{1}$ and the Schur polynomial obtained by holding $W^{(2)}$ inert by
$\chi_{R,R_i''}^{(2)}\Big|_{2}$. Regardless of which $S_3$ subgroup we use, the $S_2$ subgroup we consider is
$$ S_2=\{ 1,(12)\},$$
The representation $\yng(1,1)$ appears only once under the restriction of $S_4$ to $S_2$ so that
we must have
$$\chi_{R,R_1''}^{(2)}\Big|_{1}=\chi_{R,R_1''}^{(2)}\Big|_{2}.$$
$\yng(2)$ appears with multiplicity 2 under the restriction of $S_4$ to $S_2$. In general, the subspaces corresponding to
irreducible representations $R_2''$ and $R_3''$ will depend on the particular subgroups employed in the restrictions.
Denote the projection operators which project from $R$ to $R_2''$ or $R_3''$ using the subgroup which keeps $W^{(1)}$ inert by
$P_{R,R_2''}|_1$ and $P_{R,R_3''}|_1$, and denote the projection operators which keep $W^{(2)}$ inert by 
$P_{R,R_2''}|_2$ and $P_{R,R_3''}|_2$. We have
$$ P_{R,R_2''}\big|_1\ne P_{R,R_2''}\big|_2,\qquad P_{R,R_3''}\big|_1 \ne P_{R,R_3''}\big|_2 ,$$
$$ P_{R,R_2''}\big|_1 + P_{R,R_3''}\big|_1 = P_{R,R_2''}\big|_2 + P_{R,R_3''}\big|_2 .$$
Further, since $P_{R,R_a''}|_i P_{R,R_b''}|_j$ commutes with every element of $S_2$, we know by Schur's Lemma that,
when acting on the $R_a''$ subspace,
(unless specified otherwise, all traces in this section are over $R$; ${\bf 1}$ in the next equation is the identity
operator acting on the $R_a''$ subspace)
$$P_{R,R_a''}|_i P_{R,R_b''}|_j=\lambda {\bf 1},\qquad \lambda = {\Tr (P_{R,R_a''}|_i P_{R,R_b''}|_j )\over d_{R_a''}}.$$
It is now straight forward to obtain
\begin{eqnarray}
\chi_{R,R_2''}^{(2)}\Big|_{1} &=&
{1\over 2!}\sum_{\sigma\in S_4}\Tr (P_{R,R_2''}\Big|_1\Gamma_R(\sigma))\Tr(\sigma Z^{\otimes 2}W^{(1)}W^{(2)})\cr
&=& {1\over 2!}\sum_{\sigma\in S_4}\Tr 
(P_{R,R_2''}\Big|_1 (P_{R,R_2''}\Big|_1 +P_{R,R_3''}\Big|_1)\Gamma_R(\sigma)
(P_{R,R_2''}\Big|_1 +P_{R,R_3''}\Big|_1))\Tr(\sigma Z^{\otimes 2}W^{(1)}W^{(2)})\cr
&=& {1\over 2!}\sum_{\sigma\in S_4}\Tr 
(P_{R,R_2''}\Big|_1 (P_{R,R_2''}\Big|_2 +P_{R,R_3''}\Big|_2)\Gamma_R(\sigma)
(P_{R,R_2''}\Big|_2 +P_{R,R_3''}\Big|_2))\Tr(\sigma Z^{\otimes 2}W^{(1)}W^{(2)})\cr
&=& {1\over 2!}\sum_{\sigma\in S_4}\Tr 
(P_{R,R_2''}\Big|_1 P_{R,R_2''}\Big|_2 P_{R,R_2''}\Big|_2\Gamma_R(\sigma)
P_{R,R_2''}\Big|_2)\Tr(\sigma Z^{\otimes 2}W^{(1)}W^{(2)})\cr
& &+ {1\over 2!}\sum_{\sigma\in S_4}\Tr 
(P_{R,R_2''}\Big|_1 P_{R,R_3''}\Big|_2 P_{R,R_3''}\Big|_2\Gamma_R(\sigma)
P_{R,R_3''}\Big|_2)\Tr(\sigma Z^{\otimes 2}W^{(1)}W^{(2)})\cr
& &+{\rm \,\,\, stretched\,\,\, string\,\,\, states}\cr
&=& a{1\over 2!}\sum_{\sigma\in S_4}\Tr 
(P_{R,R_2''}\Big|_2\Gamma_R(\sigma))\Tr(\sigma Z^{\otimes 2}W^{(1)}W^{(2)})\cr
& &+ b{1\over 2!}\sum_{\sigma\in S_4}\Tr 
(P_{R,R_3''}\Big|_2\Gamma_R(\sigma))\Tr(\sigma Z^{\otimes 2}W^{(1)}W^{(2)})\cr
& &+{\rm \,\,\, stretched\,\,\, string\,\,\, states}\cr
&=&a\chi_{R,R_2''}^{(2)}\Big|_{2}+b\chi_{R,R_3''}^{(2)}\Big|_{2}
+{\rm \,\,\, stretched\,\,\, string\,\,\, states},
\nonumber
\end{eqnarray}
where the ``stretched string states" terms are obtained by tracing over off the diagonal blocks and
$$ a={\Tr (P_{R,R_2''}\Big|_1 P_{R,R_2''}\Big|_2)\over d_{R_2''}},\qquad 
   b={\Tr (P_{R,R_2''}\Big|_1 P_{R,R_3''}\Big|_2)\over d_{R_2''}}.$$
After inserting the explicit expressions for the projection operators, it is
easy to see that the computation of $a$ and $b$ reduce to a sum of characters. Performing this simple computation
we obtain
$$ a={1\over 9},\qquad b={8\over 9}.$$

Using the subgroup swap rule we can compute the reduction of a restricted Schur polynomial, 
with one or two strings attached, with respect to
any of the open strings attached to the giant. Of course, the ``stretched string states" terms do not contribute
and it is for this reason that we have not computed their explicit form.
For the example we used in this appendix,
$$ D_{W^{(1)}} \chi_{R,R_2''}^{(2)}\Big|_{1} (Z,W^{(1)},W^{(2)})=D_{W^{(1)}}\chi_{\young({\,}{\,}1,2)}\Big|_{1}
=(N+2)\chi_{\young({\,}{\,},2)}=(N+2) \chi_{R_1',R_2''}^{(1)}(Z,W^{(2)}),$$
\begin{eqnarray}
D_{W^{(2)}} \chi_{R,R_2''}^{(2)}\Big|_{1}(Z,W^{(1)},W^{(2)}) &=&
D_{W^{(2)}}\left({1\over 9}\chi_{R,R_2''}^{(2)}\Big|_{2}
                +{8\over 9}\chi_{R,R_3''}^{(2)}\Big|_{2}\right)\cr 
&=&D_{W^{(2)}}\left({1\over 9}\chi_{\young({\,}{\,}2,1)}\Big|_{2}
                +{8\over 9}\chi_{\young({\,}{\,}1,2)}\Big|_{2}\right)\cr 
&=&{N+2\over 9}\chi_{\young({\,}{\,},1)}
                +{8(N-1)\over 9}\chi_{\young({\,}{\,}1)}\cr 
&=&{N+2\over 9}\chi_{R_1',R_2''}^{(1)}(Z,W^{(1)})+{8(N-1)\over 9}\chi_{R_2',R_3''}^{(1)}(Z,W^{(1)}).
\nonumber
\end{eqnarray}

We conclude this appendix with an efficient method to compute $\Tr (P_{R,R_a''}|_n P_{R,R_a''}|_{n-1} )$ which
appears in the application of the subgroup swap rule. We assume, in the following discussion, that the string label 
matches the index label, i.e. $P_{R,R_a''}|_n =$ the projector obtained using the subgroup which holds the index of
string $W^{(n)}$ inert $=$ the projector obtained using the subgroup which holds the index $n$ inert. Start by noting that
$$ P_{R,R_a''}|_n =\Gamma_R\left( (n,n-1)\right) P_{R,R_a''}|_{n-1} \Gamma_R\left((n,n-1)\right),$$
so that
$$\Tr (P_{R,R_a''}|_n P_{R,R_a''}|_{n-1})=\Tr \left(P_{R,R_a''}|_n \Gamma_R\left((n,n-1)\right) P_{R,R_a''}|_n 
\Gamma_R\left((n,n-1)\right) \right)\equiv\Tr (M^2),$$
where
$$ M_{\alpha\beta}\equiv \langle R_a'',\alpha|\Gamma_R\left((n,n-1)\right)|R_a'',\beta\rangle,\qquad \alpha,\beta =1,...,d_{R_a''} .$$
Recall that $R_a''$ is a representation of the subgroup of $S_n$ which includes all elements which leave $n$ and $n-1$
inert. Clearly, $(n,n-1)$ commutes with every element of the subgroup and hence, by Schur's lemma
$$ M_{\alpha\beta}={\Tr (M)\over d_{R_a''}}\delta_{\alpha\beta}.$$
It is now a simple matter to obtain
$$\Tr (P_{R,R_a''}|_n P_{R,R_a''}|_{n-1} )={(\Tr (M))^2 \over d_{R_a''}}=
{\left( \Tr_{R_a''}\left( \Gamma_R\left((n,n-1)\right) \right)\right)^2\over d_{R_a''}}.$$
Thus, the computation of the trace of the projectors has been reduced to the computation of 
$\Tr_{R_a''}\left( \Gamma_R\left((n,n-1)\right) \right)$. To derive a simple formula for 
$\Tr_{R_a''}\left( \Gamma_R\left((n,n-1)\right) \right)$, note that
\begin{eqnarray}
\Tr_{R_a''}\left(\Gamma_R\left((n,n-1)(n,i)\right) \right)
&=& \Tr \left(P_{R,R_a''}|_{n}\Gamma_R\left((n,n-1)\right)\Gamma_R\left((n,i)\right)\right)\cr
&=& \Tr \left(P_{R,R_a''}|_{n-1}\Gamma_R\left((n-1,n)\right)\Gamma_R\left((n-1,i)\right)\right)\cr
&=& \Tr \left(\Gamma_R\left((n-1,n)\right)P_{R,R_a''}|_{n}\Gamma_R\left((n-1,i)\right)\right)\cr
&=& \Tr \left(P_{R,R_a''}|_{n}\Gamma_R\left((n-1,i)\right)\Gamma_R\left((n-1,n)\right)\right)\cr
&=&\Tr_{R_a''}\left(\Gamma_R\left((n-1,i)(n,n-1)\right) \right).
\nonumber
\end{eqnarray}
Now consider (${\bf 1}$ is the identity matrix)
\begin{eqnarray}
d_{R_a''}&=&\Tr_{R_a''}({\bf 1})\cr
&=&\Tr_{R_a''}\left( \Gamma_R\left((n,n-1)(n,n-1)\right)\right)\cr
&=&\Tr_{R_a''}\left( \Gamma_R\left((n,n-1)(n,n-1)\right)\right)\cr
&&\quad +\sum_{i=1}^{n-2}\left[
\Tr_{R_a''}\left(\Gamma_R\left((n,n-1)(n,i)\right) \right)
-\Tr_{R_a''}\left(\Gamma_R\left((n-1,i)(n,n-1)\right) \right)\right]\cr
&=&\sum_{i=1}^{n-1}\Tr_{R_a''}\left(\Gamma_R\left((n,n-1)(n,i)\right) \right)
-\sum_{j=1}^{n-2}\Tr_{R_a''}\left(\Gamma_R\left((n-1,j)(n,n-1)\right) \right)\cr
&=&(c_{R,R'}-c_{R',R_a''})\Tr_{R_a''}\left(\Gamma_R\left((n,n-1)\right)\right).
\nonumber
\end{eqnarray}
Thus, we obtain
$$ \Tr_{R_a''}\left(\Gamma_R\left((n,n-1)\right)\right)={d_{R_a''}\over c_{R,R'}-c_{R',R_a''}},$$
$$\Tr (P_{R,R_a''}|_n P_{R,R_a''}|_{n-1} )={d_{R_a''}\over (c_{R,R'}-c_{R',R_a''})^2}.$$

\subsection{Stretched String Contributions}

In the previous subsection, we have written down the result of performing the subgroup swap, up to the stretched
string contributions which we dropped. This is all that is needed to compute two point functions of operators that
have one or two strings attached. This follows because one reduces the result of the swap, and the reduction of 
twisted strings vanishes. If however, one considers more than two strings, the stretched string states need to be 
accounted for. In this section we will obtain the explicit form of the stretched string states.

To make our discussion concrete, we will frame it in a specific example. Consider the restricted Schur
polynomial
$$\chi_{\young({\,}{\,}{1},{\,}{2},{3})}|_1|_2|_3.$$
The chain of representations involved in the restrictions are
$$\young({\,}{\,}{\,},{\,}{\,},{\,})\to\young({\,}{\,},{\,}{\,},{\,})\to\young({\,}{\,},{\,},{\,})\to\young({\,}{\,},{\,}).$$
The first subgroup is obtained as the set of all elements of 
$S_6$ that leave the index of string $1$ inert;
the second subgroup is obtained as the set of all elements of $S_6$ 
that leave the indices of strings $1$ and $2$ inert and
the third subgroup is obtained as the set of all elements of 
$S_6$ that leave the indices of strings $1$, $2$ and $3$ inert. We have indicated
this in our operator with the symbol $|_1|_2|_3$.
We will denote the states that span the carrier space of the 
representation $\young({\,}{\,},{\,})$ obtained by the
above restriction by $|(1,2,3);i\rangle^A$. The $(1,2,3)$ labels specify the subgroups used in the restriction, $i=1,2$
labels the vectors in the basis and $A$ labels the space obtained. We can always arrange things so that we are working with a unitary
representation. In this case, the operator projecting onto this subspace is given by
$$ P_A(1,2,3)=\sum_i |(1,2,3);i\rangle^A{}^A\langle (1,2,3);i|.$$

We are interested in swapping the $2\leftrightarrow 3$ subgroups. The states spanning the 
carrier space of the representations $\young({\,}{\,},{\,})$ defined
by the restrictions used to construct
$$\chi_{\young({\,}{\,}{1},{\,}{2},{3})}|_1|_3|_2\qquad {\rm and}\qquad \chi_{\young({\,}{\,}{1},{\,}{3},{2})}|_1|_3|_2 ,$$
are denoted by $|(1,3,2);i\rangle^A$ and $|(1,3,2);i\rangle^B$ respectively. The corresponding projection operators are
$$P_A(1,3,2)=\sum_i |(1,3,2);i\rangle^A{}^A\langle (1,3,2);i|, \qquad
  P_B(1,3,2)=\sum_i |(1,3,2);i\rangle^B{}^B\langle (1,3,2);i|.$$

From the previous subsection, it is clear that the direct sum of the subspaces with bases $\{ |(1,3,2);i\rangle^A\}$
and $\{ |(1,3,2);i\rangle^B\}$ is the same space as the direct sum of the subspaces with bases
$\{|(1,2,3);i\rangle^A\}$ and $\{|(1,2,3);i\rangle^B\}$. Thus, we can write
$$|(1,2,3);i\rangle^A =\sum_j\alpha_{ij}|(1,3,2);j\rangle^A +\sum_j\beta_{ij}|(1,3,2);j\rangle^B .$$
Using this expression, it is clear that
$$ P_A(1,2,3)_{il}=\sum_{j,k}\big[
\alpha_{ij}^*|(1,3,2);j\rangle^A{}^A\langle (1,3,2);k|\alpha_{kl}+
\alpha_{ij}^*|(1,3,2);j\rangle^A{}^B\langle (1,3,2);k|\beta_{kl}$$
$$+\beta_{ij}^*|(1,3,2);j\rangle^B{}^A\langle (1,3,2);k|\alpha_{kl}+
\beta_{ij}^*|(1,3,2);j\rangle^B{}^B\langle (1,3,2);k|\beta_{kl}\big].$$
The subgroup swap rule tells us that the piece of this operator that acts in the 
$|(1,3,2);j\rangle^A$ subspace (the first of the four terms) is proportional to $P_A(1,3,2)$ and hence is full rank.
This implies that $\alpha_{ij}$ is full rank. Thus, we can now use $\alpha_{ij}$ to define an
equivalent representation, with a new basis  $\sim \sum_j(\alpha^{-1})_{ij} |(1,3,2);j\rangle^A$.
This shows that we can always take $\alpha_{ij}$ proportional to the identity. In a similar way, we can always ensure that 
$\beta_{ij}$ is proportional to the identity, so that
$$|(1,2,3);i\rangle^A =\alpha|(1,3,2);i\rangle^A + \beta |(1,3,2);i\rangle^B .$$
Consistency with the subgroup swap rule of the previous section now implies that
$$\alpha^2 = {3\over 4},\qquad \beta^2={1\over 4}.$$
By a convenient choice of phases, we can now write
$$|(1,2,3);i\rangle^A ={\sqrt{3}\over 2}|(1,3,2);i\rangle^A + e^{i\phi}{1\over 2}|(1,3,2);i\rangle^B .$$
The remaining phase $\phi$ is convention dependent. With the conventions we use here, if one is swapping
indices $n$ and $m$ where the subgroups were defined by first holding $n$ fixed and then holding $m$ 
fixed, and if $c_n$ and $c_m$ are the weights of the associated boxes in the Young diagram,
we have
$$ e^{i\phi}= {c_n-c_m \over |c_n-c_m|}.$$
It is now clear that
$${}^A\langle (1,2,3);i|\Gamma_R (\sigma )|(1,2,3);i\rangle^A=
{3\over 4}{}^A\langle (1,3,2);i|\Gamma_R (\sigma )|(1,3,2);i\rangle^A +
{1\over 4}{}^B\langle (1,3,2);i|\Gamma_R (\sigma )|(1,3,2);i\rangle^B $$ 
$$ + {\sqrt{3}\over 4}{}^A\langle (1,3,2);i|\Gamma_R (\sigma )|(1,3,2);i\rangle^B 
   + {\sqrt{3}\over 4}{}^B\langle (1,3,2);i|\Gamma_R (\sigma )|(1,3,2);i\rangle^A ,$$
and hence
$$\chi_{\young({\,}{\,}{1},{\,}{2},{3})}|_1|_2|_3 =
{3\over 4} \chi_{\young({\,}{\,}{1},{\,}{2},{3})}|_1|_3|_2
+ {1\over 4}\chi_{\young({\,}{\,}{1},{\,}{3},{2})}|_1|_3|_2
+{\sqrt{3}\over 4}\chi_{\young({\,}{\,}{1},{\,}{\twothree},{\threetwo})}|_1|_3|_2
+{\sqrt{3}\over 4}\chi_{\young({\,}{\,}{1},{\,}{\threetwo},{\twothree})}|_1|_3|_2 .$$
Our conventions for the graphical notation is spelled out by indicating the ``trace" being used to define the operator
$$\chi_{\young({\,}{\,}{1},{\,}{\twothree},{\threetwo})}|_1|_3|_2\leftrightarrow
\sum_i {}^A\langle (1,3,2);i|\Gamma_R (\sigma )|(1,3,2);i\rangle^B =\Tr (P_{AB}\Gamma_R(\sigma )),$$
$$ P_{AB} = N \, P_A(1,3,2)\Gamma_R\left( (23)\right) P_B(1,3,2),$$
$$N=\sqrt{d_{\yng(2,1)}\over\Tr (P_A(1,3,2)\Gamma_R\left( (23)\right) P_B(1,3,2)\Gamma_R\left( (23)\right))},$$
where $d_{\yng(2,1)}$ is the dimension of the $\yng(2,1)$ rep of $S_3$, and
$$\chi_{\young({\,}{\,}{1},{\,}{\threetwo},{\twothree})}|_1|_3|_2\leftrightarrow
\sum_i {}^B\langle (1,3,2);i|\Gamma_R (\sigma )|(1,3,2);i\rangle^A =\Tr (P_{BA}\Gamma_R(\sigma )),$$
$$ P_{BA} = N\, P_B(1,3,2)\Gamma_R\left( (23)\right) P_A(1,3,2).$$
Our graphical notation summarizes how the operator in constructed. In computing the trace for the operator above,
the bra and kets that are summed are from different carrier spaces. Each carrier space is naturally associated
with a restricted Schur polynomial with no stretched strings, in the sense that the restriction used to define
the polynomial is the restriction used to define the carrier space. 
In labelling the boxes, if the label in a box of the restricted Schur polynomial
associated with the bra is not the same as the label in the corresponding box in the 
ket, then the bra label is written above the ket label. 

The same rules can be used to swap subgroups for twisted string states. To illustrate this, imagine that we
swap $3\leftrightarrow 1$. Now, four carrier spaces participate. Their bases are denoted $\{ |(3,2,1);i\rangle^A\}$,
$\{ |(3,2,1);i\rangle^B\}$, $\{ |(3,2,1);i\rangle^C\}$ and $\{ |(3,2,1);i\rangle^D\}$. The restrictions needed to
define these bases are specified by the following four operators
$$\chi_{\young({\,}{\,}{1},{\,}{2},{3})}|_3|_1|_2 ,\quad
\chi_{\young({\,}{\,}{1},{\,}{3},{2})}|_3|_1|_2 ,\quad
\chi_{\young({\,}{\,}{3},{\,}{2},{1})}|_3|_1|_2 ,\quad
\chi_{\young({\,}{\,}{3},{\,}{1},{2})}|_3|_1|_2 ,$$
respectively. Arguing exactly as above, we have
$$ |(1,3,2);i\rangle^A ={\sqrt{15}\over 4}|(3,1,2);i\rangle^A +{1\over 4}|(3,1,2);i\rangle^C ,$$
$$ |(1,3,2);i\rangle^B ={\sqrt{3}\over 2}|(3,1,2);i\rangle^B+{1\over 2}|(3,1,2);i\rangle^D .$$
Thus,
$$ \sum_i {}^A\langle (1,3,2);i|\Gamma_R (\sigma )|(1,3,2);i\rangle^B
=$$
$${3\sqrt{5}\over 8}\sum_i {}^A\langle (3,1,2);i|\Gamma_R (\sigma )|(3,1,2);i\rangle^B+
{\sqrt{15}\over 8}\sum_i {}^A\langle (3,1,2);i|\Gamma_R (\sigma )|(3,1,2);i\rangle^D$$
$$ +
{\sqrt{3}\over 8}\sum_i {}^C\langle (3,1,2);i|\Gamma_R (\sigma )|(3,1,2);i\rangle^B+
{1\over 8}\sum_i {}^C\langle (3,1,2);i|\Gamma_R (\sigma )|(3,1,2);i\rangle^D
$$
and consequently
$$ \chi_{\young({\,}{\,}{1},{\,}{\twothree},{\threetwo})}|_1|_3|_2 =
{3\sqrt{5}\over 8}\chi_{\young({\,}{\,}{1},{\,}{\twothree},{\threetwo})}|_3|_1|_2
+{\sqrt{15}\over 8}\chi_{\young({\,}{\,}{\onethree},{\,}{\twoone},{\threetwo})}|_3|_1|_2
+{\sqrt{3}\over 8}\chi_{\young({\,}{\,}{\threeone},{\,}{\twothree},{\onetwo})}|_3|_1|_2
+{1\over 8}\chi_{\young({\,}{\,}{3},{\,}{\twoone},{\onetwo})}|_3|_1|_2 .$$

\section{Constrained Two Point Functions}

In the computation of two point functions of restricted Schur polynomials, we have seen that we need to evaluate two 
point functions of restricted Schur polynomials
in which we do not sum over all possible contractions, but rather, we contract the $i^{th}$ field in the first
Schur polynomial with the $i^{th}$ field in the second Schur polynomial, with $i=n,n-1,...,n-k$. Recall that we
denote this correlator by
$$\langle\chi_{R,R'}(Z)(\chi_{S,S'}(Z))^\dagger \rangle\Big|_{n,n-1,...,n-k} .$$
$R$ is a representation of $S_n$; $R'$ is a representation of $S_{n-k-1}$. Thus, $k+1$ boxes must be removed from $R$
to obtain $R'$. In this appendix we will evaluate these constrained two point functions.

The computation is straightforward. We use the methods of \cite{Corley:2001zk}. See also
\cite{Balasubramanian:2004nb} where the correlator 
$$\langle\chi_{R,R'}(Z)(\chi_{S,S'}(Z))^\dagger\rangle\Big|_{n} ,$$
was already studied. We find (we assume that $R$ and $S$ are unitary representations with 
$n$ boxes, without a loss of generality)
\begin{eqnarray}
G_2&\equiv& (n!)^2 \langle\chi_{R,R'}(Z)(\chi_{S,S'}(Z))^\dagger\rangle\Big|_{n,n-1,...,n-k} \cr
&=& \sum_{i,j} \sum_{\sigma,\tau\in S_n} \Tr_{R'}(\Gamma_R(\sigma)) \Tr_{S'}(\Gamma_S(\tau))^*
\langle \prod_{a=1}^{n-k-1} Z^{i_a}_{i_{\sigma(a)}}(Z^{\dagger})_{j_a}^{j_{\tau(a)}}\rangle 
\prod_{b=n-k}^n \langle Z^{i_b}_{i_{\sigma(b)}}(Z^{\dagger})_{j_b}^{j_{\tau(b)}}\rangle \cr
&=& \sum_{i,j} \sum_{\sigma,\tau\in S_n} \Tr_{R'}(\Gamma_R(\sigma)) \Tr_{S'}(\Gamma_S(\tau))^*
\sum_{\alpha \in S_{n-k-1}} \prod_{a=1}^{n-k-1}
\langle  Z^{i_a}_{i_{\sigma(a)}}(Z^{\dagger})_{j_\alpha( a)}^{j_{\tau(\alpha (a))}}\rangle 
\prod_{b=n-k}^n \langle Z^{i_b}_{i_{\sigma(b)}}(Z^{\dagger})_{j_b}^{j_{\tau(b)}}\rangle 
\nonumber
\end{eqnarray}
The sum $\sum_{\alpha\in S_{n-k-1}}$ is needed to include all possible contractions. Again ignoring
spacetime dependence, the two point function we use is
$$
\langle Z^{i}_{j} (Z^\dagger)^{k}_l\rangle =\delta^{i}_{l}\delta^{k}_j .$$
Using this two point function, we obtain
\begin{eqnarray}
G_2 &=& \sum_{i,j} \sum_{\sigma,\tau\in S_n} 
\Tr_{R'}(\Gamma_R(\sigma)) \Tr_{S'}(\Gamma_S(\tau))^*
\sum_{\alpha \in S_{n-k-1}}\left[\prod_{l=1}^{n-k-1}
\delta^{i_l}_{j_{\alpha(l)}} \delta^{j_{\tau(\alpha(l))}}_{i_{\sigma(l)}} \right]
\prod_{l=n-k}^n\delta^{i_l}_{j_l}\delta^{j_{\tau(l)}}_{i_{\sigma(l)}} \cr
&=& \sum_{i,j} \sum_{\sigma,\tau\in S_n} 
\Tr_{R'}(\Gamma_R(\sigma)) \Tr_{S'}(\Gamma_S(\tau))^*
\sum_{\alpha\in S_{n}|_{\alpha(n)=n,...,\alpha(n-k)=n-k}}\left[\prod_{l=1}^{n}
\delta^{i_l}_{j_{\alpha(l)}} \delta^{j_{\tau(\alpha(l))}}_{i_{\sigma(l)}} \right]\cr
&=& \sum_{i,j} \sum_{\sigma,\tau\in S_n} 
\Tr_{R'}(\Gamma_R(\sigma)) \Tr_{S'}(\Gamma_S(\tau))^* 
\sum_{\alpha\in S_{n}|_{\alpha(n)=n,...,\alpha(n-k)=n-k}}\left[\prod_{l=1}^{n}
\delta^{i_{\sigma(l)}}_{j_{\alpha (\sigma(l))}} \delta^{j_{\tau(\alpha (l))}}_{i_{\sigma(l)}} \right]\cr
&=& \sum_{j} \sum_{\sigma,\tau\in S_n} 
\Tr_{R'}(\Gamma_R(\sigma)) \Tr_{S'}(\Gamma_S(\tau))^* 
\sum_{\alpha\in S_{n}|_{\alpha(n)=n,...,\alpha(n-k)=n-k}}
\left[\prod_{l=1}^{n} \delta^{j_{\tau(\alpha(l))}}_{j_{\alpha(\sigma(l))}}\right] \cr
&=&  \sum_{\sigma,\tau\in S_n} 
\Tr_{R'}(\Gamma_R(\sigma)) \Tr_{S'}(\Gamma_S(\tau))^* 
\sum_{\alpha \in S_{n}|_{\alpha(n)=n,...,\alpha(n-k)=n-k}}
N^{C(\alpha^{-1}\cdot \tau^{-1} \cdot \alpha \cdot \sigma)}
\nonumber
\end{eqnarray}
where $C(\alpha^{-1}\cdot \tau^{-1} \cdot \alpha \cdot \sigma)$ is the number of cycles in the permutation
$\alpha^{-1}\cdot \tau^{-1} \cdot \alpha \cdot \sigma $ and $S_{n}|_{\alpha(n)=n,...,\alpha(n-k)=n-k}$ are those
elements $\alpha$ of $S_n$ that have $\alpha(n)=n,...,\alpha(n-k)=n-k$. Inserting the delta-function 
we find
\begin{eqnarray}
G_2 & = & \sum_{\sigma,\tau\in S_n} 
\Tr_{R'}(\Gamma_R(\sigma)) \Tr_{S'}(\Gamma_S(\tau))^* 
\sum_{\alpha\in S_{n}|_{\alpha(n)=n,...,\alpha(n-k)=n-k}}
\sum_{\beta\in S_n} N^{C(\beta)}\delta(\beta^{-1} \cdot \alpha^{-1}\cdot \tau^{-1} \cdot \alpha \cdot \sigma) \cr
&=& \sum_{\tau\in S_n}\sum_{\alpha\in S_{n}|_{\alpha(n)=n,...,\alpha(n-k)=n-k}}
\sum_{\beta \in S_n}\Tr_{R'}(\Gamma_R(\alpha^{-1}\tau\alpha\beta )) \Tr_{S'}(\Gamma_S(\tau))^* N^{C(\beta)}.
\nonumber 
\end{eqnarray}
We now compute
$$\sum_{\tau\in S_n}\Tr_{R'}(\Gamma_R(\alpha^{-1}\tau\alpha\beta )) \Tr_{S'}(\Gamma_S(\tau))^* .$$
We need to consider two possibilities: (I) the strings of the original operator are attached to a specific giant 
so that the traces are over ``on the diagonal blocks," and (II) some strings of the original operator stretch 
between the giants so that the traces are over ``off the diagonal blocks." For case (I) we can write
(unless specified otherwise, all traces are over $R$)
\begin{eqnarray}
\sum_{\tau\in S_n}\Tr_{R'}&&(\Gamma_R(\alpha^{-1}\tau\alpha\beta )) \Tr_{S'}(\Gamma_S(\tau))^*
= \sum_{\tau\in S_n}\Tr(P_{R,R'}\Gamma_R(\alpha^{-1}\tau\alpha\beta )) \Tr(P_{S,S'}\Gamma_S(\tau))^*\cr
&=& \sum_{\tau\in S_n}\sum_{i,j,k,l=1}^{d_R}\left(\Gamma_R(\alpha\beta )P_{R,R'}\Gamma_R(\alpha^{-1})\right)_{ij}
(P_{S,S'}^*)_{kl}\Gamma_R(\tau)_{ji} \Gamma_S(\tau)^*_{lk}.\nonumber
\end{eqnarray}
where $P_{R,R'}$ and $P_{S,S'}$ are projection operators and $d_R$ is the dimension of the irreducible representation
$R$ of $S_n$. Using the orthogonality relation
$$
\sum_{\sigma\in G} \left(\Gamma_{R}(\sigma)\right)_{il} \left(\Gamma_{S}(\sigma)^*\right)_{sm}
= {g\over d_{R}} \delta_{is}\delta_{lm} \delta_{RS}
$$
with $g=|G|$, and the fact that the projection operators are hermittian ($(P_{S,S'}^*)_{kl}=(P_{S,S'})_{lk}$)we obtain
\begin{eqnarray}
\sum_{\tau\in S_n}\Tr_{R'}(\Gamma_R(\alpha^{-1}\tau\alpha\beta )) \Tr_{S'}(\Gamma_S(\tau))^*
&=&\delta_{RS}\Tr (\Gamma_R(\alpha\beta )P_{R,R'}\Gamma_R(\alpha^{-1})P_{R,S'}){n!\over d_R}\cr
&=&\delta_{RS}\delta_{R'S'}\Tr (\Gamma_R(\alpha\beta )P_{R,R'}\Gamma_R(\alpha^{-1})){n!\over d_R}\cr
&=&\delta_{RS}\delta_{R'S'}\Tr (\Gamma_R(\beta )P_{R,R'}){n!\over d_R}\cr
&=&\delta_{RS}\delta_{R'S'}\Tr_{R'} (\Gamma_R(\beta )){n!\over d_R}.
\nonumber
\end{eqnarray}
To obtain this result, we used the fact that since $\alpha\in S_{n-k-1}$ we have
$$P_{R,R'}\Gamma_R(\alpha^{-1})P_{R,S'}=\delta_{R'S'}P_{R,R'}\Gamma_R(\alpha^{-1})P_{R,R'}
=\delta_{R'S'}P_{R,R'}\Gamma_R(\alpha^{-1}).$$
Thus,
\begin{eqnarray}
G_2 &=& \sum_{\alpha\in S_{n}|_{\alpha(n)=n,...,\alpha (n-k)=n-k}} \sum_{\beta\in S_n} N^{C(\beta)} 
\Tr_{R'}(\Gamma_R(\beta)){n!\over d_R} \delta_{RS}\delta_{R'S'}\cr
&=&{n!(n-k-1)!\over d_R}\delta_{RS}\delta_{R'S'}\sum_{\beta\in S_n} N^{C(\beta)} \Tr_{R'} ( \Gamma_R(\beta )).\nonumber
\end{eqnarray}
Our result for case (I) is therefore
$$\langle\chi_{R,R'}(Z)\chi_{S,S'}(Z)^\dagger\rangle\Big|_{n,n-1,...,n-k}=
{(n-k-1)!\over n! d_R}\delta_{R\to R',S\to S'}\sum_{\beta\in S_n} N^{C(\beta)} \Tr_{R'} ( \Gamma_R(\beta )).$$
The delta function $\delta_{R\to R',S\to S'}$ is only non-zero if the complete chain of representations used to restrict
$R$ to $R'$ match with the representations used to restrict $S$ to $S'$.

For case (II) we can write
$$
\sum_{\tau\in S_n}\Tr_{T',R'}(\Gamma_R(\alpha^{-1}\tau\alpha\beta )) \Tr_{U',S'}(\Gamma_S(\tau))^*$$
$$= \sum_{\tau\in S_n}\Tr(A_{T'R'}P_{R,R'}\Gamma_R(\alpha^{-1}\tau\alpha\beta )) \Tr(A_{U'S'}P_{S,S'}\Gamma_S(\tau))^* $$
$$= \sum_{\tau\in S_n}\sum_{i,j,k,l=1}^{d_R}\left(\Gamma_R(\alpha\beta )A_{T'R'}P_{R,R'}\Gamma_R(\alpha^{-1})\right)_{ij}
(A_{U'S'}^* P_{S,S'}^*)_{kl}\Gamma_R(\tau)_{ji} \Gamma_S(\tau)^*_{lk}.$$
where $P_{R,R'}$ and $P_{S,S'}$ are projection operators and $A_{T'R'}$ and $A_{U'S'}$ are intertwiners defined by
$$ A_{T'R'}P_{R,R'}=P_{R,T'}A_{T'R'},\qquad A_{U'S'}P_{S,S'}=P_{S,U'}A_{U'S'}.$$
Now, after using the orthogonality relation, the hermitticity of the projection operator,  
$A^\dagger_{S'U'} =A_{U'S'}$ and the fact that $\alpha^{-1}\in S_{n-k-1}$ we have
\begin{eqnarray}
\sum_{\tau\in S_n}\Tr_{T',R'}(\Gamma_R(\alpha^{-1}\tau\alpha\beta )) \Tr_{U',S'}(\Gamma_S(\tau))^*
&=&\delta_{RS}\Tr (\Gamma_R(\alpha\beta )A_{T'R'}P_{R,R'}\Gamma_R(\alpha^{-1})P_{R,S'}A_{S'U'}){n!\over d_R}\cr
&=&\delta_{RS}\Tr (\Gamma_R(\alpha\beta )A_{T'R'}\Gamma_R(\alpha^{-1})A_{S'U'}){n!\over d_R}\cr
&=&\delta_{RS}\delta_{T'U'}\delta_{R'S'}\Tr (\Gamma_R(\alpha\beta )A_{T'R'}\Gamma_R(\alpha^{-1})A_{R'T'}){n!\over d_R}.
\nonumber
\end{eqnarray}
Notice that the subgroup labels have contracted exactly as if they were Chan-Paton indices. In the last line above, we
are multiplying group elements that come from different ``on the diagonal blocks". If we now assume that we have chosen
our bases of these different on the diagonal blocks so that different blocks represent the same group element with the 
same matrix, we obtain 
$$
\sum_{\tau\in S_n}\Tr_{T',R'}(\Gamma_R(\alpha^{-1}\tau\alpha\beta )) \Tr_{U',S'}(\Gamma_S(\tau))^*
=\delta_{RS}\delta_{T'U'}\delta_{R'S'}\Tr_{T'} (\Gamma_R(\beta )){n!\over d_R}.
$$
Following what we did for case (I), we obtain
$$\langle\chi_{R,R'T'}(Z)(\chi_{S,U'S'}(Z))^\dagger\rangle\Big|_{n,n-1,...,n-k}=
{(n-k-1)!\over n! d_R}\delta_{R\to R'T',S\to U'S'}\sum_{\beta\in S_n} N^{C(\beta)} \Tr_{R'} ( \Gamma_R(\beta )).$$
for case (II). In this last formula we have explicitly displayed the indices that are twisted.
The delta function $\delta_{R\to R'T',S\to U'S'}$ is only non-zero if the complete chain of representations used to restrict
$R$ to $R'T'$ match with the representations used to restrict $S$ to $U'S'$.

\section{An Identity}

In this appendix we will evaluate
$$ I = \sum_{\beta\in S_n} N^{C(\beta)} \Tr_{R'} ( \Gamma_R(\beta )).$$
The symmetric group $S_n$ can be given an action
$$\beta |i_1,i_2,...,i_n\rangle =|i_{\beta (1)},i_{\beta (2)},...,i_{\beta (n)}\rangle,\qquad \beta\in S_n $$
on the space $V^{\otimes n}$ with $V$ the fundamental representation of $U(N)$. Using the formula
$$\langle j_1,j_2,...,j_n |\sigma |i_1,i_2,...,i_n\rangle =\delta^{j_1}_{i_{\sigma (1)}}
\delta^{j_2}_{i_{\sigma (2)}}...\delta^{j_n}_{i_{\sigma (n)}},$$
we easily obtain
$$ N^{C(\beta )}=\delta^{i_1}_{i_{\beta (1)}}\delta^{i_2}_{i_{\beta (2)}}\cdots \delta^{i_n}_{i_{\beta (n)}}=\Tr_n (\beta ),$$
where $\Tr_n$ is a trace over $V^{\otimes n}$. Thus, $I$ can be rewritten as
$$ I = \sum_{\beta\in S_n} \Tr_{R'} ( \Gamma_R(\beta )) \Tr_n (\beta ) .$$
In this appendix we will make use of the result, that the dimension of the irreducible representation $R$ of $SU(N)$ is
given by
$$ Dim_N(R)={1\over n!}\sum_{\sigma\in S_n}\chi_R (\sigma )N^{C(\sigma )}. $$
This follows as a consequence of the fact that the Schur polynomials compute characters of the unitary group. See 
\cite{Corley:2001zk},\cite{Fulton} for a nice discussion.

Consider first the case that $R'$ is obtained from $R$ by removing a single box, i.e. $R'$ is an irreducible representation
of $S_{n-1}$. In this case, $I$ can be written as
$$ 
I = \sum_{\beta\in S_{n}|_{\beta (n)=n}}
     \left( \Tr_{R'}( \Gamma_R(\beta ))\Tr_n (\beta )+\sum_{i=1}^{n-1} \Tr_{R'} ( \Gamma_R((i,n)\beta ))\Tr_n ((i,n)\beta )\right) .$$
It is a simple matter to see that
$$ \Tr_n (\beta )=N\Tr_n ((i,n)\beta ),$$
so that
\begin{eqnarray}
I &=& {1\over N}\sum_{\beta\in S_{n}|_{\beta (n)=n}}
     \Tr_{R'} \left( \left(N +\sum_{i=1}^{n-1}\Gamma_R((i,n))\right)\Gamma_R(\beta )\right)\Tr_n (\beta )\cr
&=&  {1\over N}\sum_{\beta\in S_{n}|_{\beta (n)=n}}c_{R,R'}
     \Tr_{R'} \left(\Gamma_R(\beta )\right)\Tr_n (\beta ),
\end{eqnarray}
where $c_{R,R'}$ is the weight of the box that must be removed to obtain $R'$ from $R$. Writing this in terms
of the $S_{n-1}$ subgroup we have
$$ I= c_{R,R'}\sum_{\beta\in S_{n-1}} \chi_{R'}(\beta )\Tr_{n-1} (\beta )=(n-1)!c_{R,R'}Dim_N (R'). $$

Next consider the case that $R'$ is obtained from $R$ by removing two boxes, i.e. $R'$ is an irreducible representation
of $S_{n-2}$. Removing one box $R\to \bar{R}$; removing a second box, $\bar{R}\to R'$. Start by writing $I$ as
\begin{eqnarray}
I &=& \sum_{\beta\in S_{n}|_{\beta (n)=n}}
     \left( \Tr_{R'}( \Gamma_R(\beta ))\Tr_n (\beta )+\sum_{i=1}^{n-1} \Tr_{R'} ( \Gamma_R((i,n)\beta ))\Tr_n ((i,n)\beta )\right)\cr
&=& {1\over N}\sum_{\beta\in S_{n}|_{\beta (n)=n}}
     \Tr_{R'} \left( \left(N +\sum_{i=1}^{n-1}\Gamma_R((i,n))\right)\Gamma_R(\beta )\right)\Tr_n (\beta )\cr
&=&  {1\over N}\sum_{\beta\in S_{n}|_{\beta (n)=n}}c_{R,\bar{R}}
     \Tr_{R'} \left(\Gamma_R(\beta )\right)\Tr_n (\beta ),
\end{eqnarray}
where $c_{R,\bar{R}}$ is the weight of the box that must be removed to obtain $\bar{R}$ from $R$. We can rewrite this again
\begin{eqnarray}
I &=& {1\over N}\sum_{\beta\in S_{n}|_{\beta (n)=n,\beta(n-1)=n-1}}c_{R,\bar{R}}
     \left( \Tr_{R'}( \Gamma_R(\beta ))\Tr_n (\beta )+\sum_{i=1}^{n-2} \Tr_{R'} ( \Gamma_R((i,n-1)\beta ))\Tr_n ((i,n-1)\beta )\right)\cr
&=& {1\over N^2}\sum_{\beta\in S_{n}|_{\beta (n)=n,\beta (n-1)=n-1}}c_{R,\bar{R}}
     \Tr_{R'} \left( \left( N +\sum_{i=1}^{n-2}\Gamma_R((i,n-1))\right)\Gamma_R(\beta )\right)\Tr_n (\beta )\cr
&=&  {1\over N^2 }\sum_{\beta\in S_{n}|_{\beta (n)=n,\beta (n-1)=n-1}}c_{R,\bar{R}}c_{\bar{R},R'}
     \Tr_{R'} \left(\Gamma_R(\beta )\right)\Tr_n (\beta ),
\end{eqnarray}
where $c_{\bar{R},R'}$ is the weight of the box that must be removed from $\bar{R}$ to obtain $R'$. Writing this in terms
of the $S_{n-2}$ subgroup we have
$$ I= c_{R,R'}c_{\bar{R},R'}\sum_{\beta\in S_{n-2}} \chi_{R'}(\beta )\Tr_{n-2} (\beta )=(n-2)!c_{R,R'}c_{\bar{R},R'}Dim_N (R'). $$

These arguments can be extended to the general case where $k$ boxes need to be removed from $R$ to obtain $R'$. The result is
$$ \sum_{\beta\in S_n} N^{C(\beta)} \Tr_{R'} ( \Gamma_R(\beta )) = (n-k)!\left( \prod_i c_i \right) Dim_N (R'), $$
where $\left( \prod_i c_i \right)$ is the product of the weights associated to the boxes that must be removed from $R$ to
obtain $R'$. Some straightforward algebra gives
$${1\over d_{R'}} \sum_{\beta\in S_n} N^{C(\beta)} \Tr_{R'} ( \Gamma_R(\beta ))=f_R ,$$
where $f_R$ is the quantity defined in section 2.1. Note that the traces run over an irreducible representation $R'$. If the
trace were to run over an ``off the diagonal block" the result would be zero.

\section{Open String Correlators}

In this appendix, we will start by computing the two point function of the open string word $W^i_j=(Y^J)^i_j$. Recall that the
super Yang-Mills theory we study has a $U(N)$ symmetry. The Higgs field $Y$ transforms in the adjoint representation of this $U(N)$. 
Consistency with the conservation of this $U(N)$ charge implies that
$$\langle (Y^J)^i_j (Y^{\dagger J})^l_k\rangle = A\delta^i_j\delta^l_k +B\delta^i_k\delta^l_j .$$
Determining this two point function amounts to determining $A$ and $B$. This two point function may be contracted
in two different ways to obtain
$$ N^2 A+N   B = \langle \Tr (Y^J)\Tr (Y^{\dagger J})\rangle,$$  
$$ N   A+N^2 B = \langle \Tr (Y^J Y^{\dagger J})\rangle .$$
The free field theory Schwinger-Dyson equation
$$0=\int dY dY^\dagger {\partial\over\partial Y^i_j}\left[
\Tr (Y^{J+1})(Y^{J\dagger})^i_j e^{-S}\right],\qquad S=\Tr (YY^\dagger),$$
implies that
$$\langle \Tr (Y^{J+1})\Tr ((Y^{\dagger})^{ J+1})\rangle =(J+1)\langle \Tr (Y^J Y^{\dagger J})\rangle .$$
Thus, we need to solve
$$ N^2 A+N   B = \langle \Tr (Y^J)\Tr (Y^{\dagger J})\rangle,$$  
$$ N   A+N^2 B = {1\over J+1}\langle \Tr (Y^{J+1})\Tr((Y^{\dagger})^{J+1})\rangle .$$
It is now simple to obtain
$$A={1\over N^3-N}\left[ N\langle \Tr (Y^J)\Tr (Y^{\dagger J})\rangle-
{1\over J+1}\langle \Tr (Y^{J+1})\Tr(Y^{\dagger})^{J+1})\rangle\right],$$
$$B={1\over N^3-N}\left[ {N\over J+1}\langle \Tr (Y^{J+1})\Tr(Y^{\dagger})^{J+1})\rangle
-\langle \Tr (Y^J)\Tr (Y^{\dagger J})\rangle \right].$$
$A$ and $B$ are now easily obtained upon using the exact result\cite{Corley:2002mj},\cite{plefka}
$$\langle \Tr (Y^J)\Tr (Y^{\dagger J})\rangle ={N\over J+1}\left(
{(N+J)!\over N!}-{(N-1)!\over (N-J-1)!}\right).$$
Their large $N$ expansion is
$$A= (J-1)N^{J-2}+{(J-1)(J-2)(J-3)(J^2+3J+4)\over 24}N^{J-4}+O(J^9 N^{J-6}),$$
$$B=      N^{J-1}+{(J-1)(J-2)(J^2+5J+12)\over 24}N^{J-3}+O(J^8 N^{J-5}).$$
For the amplitude for closed string emission from an excited D-brane state, we need to compute the leading contribution to
the correlator $\langle \Tr(Y^J) (Y^{\dagger J})^l_k\rangle$. This is easily computed from the results we have so far,
as 
$$ \langle \Tr(Y^J) (Y^{\dagger J})^l_k\rangle =
\langle \delta_i^j (Y^J)^i_j (Y^{\dagger J})^l_k\rangle = (AN+B)\delta^l_k .$$

To study the splitting and joining of open strings attached to giant gravitons, we will need to evaluate three point functions
of open string words. Again, consistency with the global $U(N)$ symmetry allows us to write
\begin{eqnarray}
\langle (Y^{J_1})_i^j (Y^{J_2})_k^l ((Y^\dagger)^{J_1+J_2} )_p^q\rangle &=& 
  a\delta_i^j\delta_k^l\delta_p^q 
+ b\delta_i^l\delta_k^j\delta_p^q 
+ c\delta_i^j\delta_k^q\delta_p^l\cr
& & + d\delta_i^q\delta_k^l\delta_p^j
+ e\delta_i^l\delta_k^q\delta_p^j 
+ f\delta_i^q\delta_k^j\delta_p^l .
\nonumber
\end{eqnarray}
Not all of the coefficients we have introduced are independent. Indeed, under $J_1\leftrightarrow J_2$ we have
$c\leftrightarrow d$ and $e\leftrightarrow f$. This three point function may be contracted in six different ways
to obtain
$$ \langle \Tr (Y^{J_1}) \Tr (Y^{J_2}) \Tr ((Y^\dagger)^{J_1+J_2} )\rangle =N^3 a +N^2 (b+c+d)+N(e+f), $$
$$ \langle \Tr (Y^{J_1+ J_2}) \Tr ((Y^\dagger)^{J_1+J_2} )\rangle = N^3 b +N^2 (a+e+f)+N(c+d),$$
$$ \langle \Tr (Y^{J_1}) \Tr (Y^{J_2}(Y^\dagger)^{J_1+J_2} )\rangle =N^3 c +N^2 (a+e+f)+N(b+d),$$
$$ \langle \Tr (Y^{J_2}) \Tr (Y^{J_1}(Y^\dagger)^{J_1+J_2} )\rangle =N^3 d +N^2 (a+e+f)+N(b+c),$$
$$ \langle \Tr (Y^{J_1+J_2}(Y^\dagger)^{J_1+J_2} )\rangle =N^3 e +N^2 (b+c+d)+N(a+f),$$
$$ \langle \Tr (Y^{J_1+J_2}(Y^\dagger)^{J_1+J_2} )\rangle =N^3 f +N^2 (b+c+d)+N(e+a).$$
Solving these equations simultaneously, we find
\begin{eqnarray}
a&=&{1\over (N^2-1)(N^2-4)}\left[ 
{N^2-2\over N}\langle \Tr (Y^{J_1}) \Tr (Y^{J_2}) \Tr ((Y^\dagger)^{J_1+J_2} )\rangle\right. \cr
&-&\langle \Tr (Y^{J_1+ J_2}) \Tr ((Y^\dagger)^{J_1+J_2} )\rangle 
-\langle \Tr (Y^{J_1}) \Tr (Y^{J_2}(Y^\dagger)^{J_1+J_2} )\rangle\cr
&-&\left. \langle \Tr (Y^{J_2}) \Tr (Y^{J_1}(Y^\dagger)^{J_1+J_2} )\rangle 
 +{4\over N}\langle\Tr (Y^{J_1+J_2}(Y^\dagger)^{J_1+J_2} )\rangle \right],
\nonumber
\end{eqnarray}
\begin{eqnarray}
b&=&{1\over (N^2-1)(N^2-4)}\left[ 
-\langle \Tr (Y^{J_1}) \Tr (Y^{J_2}) \Tr ((Y^\dagger)^{J_1+J_2} )\rangle\right. \cr
&+&{N^2-2\over N}\langle \Tr (Y^{J_1+ J_2}) \Tr ((Y^\dagger)^{J_1+J_2} )\rangle 
+{2\over N}\langle \Tr (Y^{J_1}) \Tr (Y^{J_2}(Y^\dagger)^{J_1+J_2} )\rangle\cr
&+&\left. {2\over N}\langle \Tr (Y^{J_2}) \Tr (Y^{J_1}(Y^\dagger)^{J_1+J_2} )\rangle 
 -2\langle\Tr (Y^{J_1+J_2}(Y^\dagger)^{J_1+J_2} )\rangle \right],
\nonumber
\end{eqnarray}
\begin{eqnarray}
c&=&{1\over (N^2-1)(N^2-4)}\left[ 
-\langle \Tr (Y^{J_1}) \Tr (Y^{J_2}) \Tr ((Y^\dagger)^{J_1+J_2} )\rangle\right. \cr
&+&{2\over N}\langle \Tr (Y^{J_1+ J_2}) \Tr ((Y^\dagger)^{J_1+J_2} )\rangle 
+{N^2-2\over N}\langle \Tr (Y^{J_1}) \Tr (Y^{J_2}(Y^\dagger)^{J_1+J_2} )\rangle\cr
&+&\left. {2\over N}\langle \Tr (Y^{J_2}) \Tr (Y^{J_1}(Y^\dagger)^{J_1+J_2} )\rangle 
 -2\langle\Tr (Y^{J_1+J_2}(Y^\dagger)^{J_1+J_2} )\rangle \right],
\nonumber
\end{eqnarray}
\begin{eqnarray}
d&=&{1\over (N^2-1)(N^2-4)}\left[ 
-\langle \Tr (Y^{J_1}) \Tr (Y^{J_2}) \Tr ((Y^\dagger)^{J_1+J_2} )\rangle\right. \cr
&+&{2\over N}\langle \Tr (Y^{J_1+ J_2}) \Tr ((Y^\dagger)^{J_1+J_2} )\rangle 
+{2\over N}\langle \Tr (Y^{J_1}) \Tr (Y^{J_2}(Y^\dagger)^{J_1+J_2} )\rangle\cr
&+&\left. {N^2-2\over N}\langle \Tr (Y^{J_2}) \Tr (Y^{J_1}(Y^\dagger)^{J_1+J_2} )\rangle 
 -2\langle\Tr (Y^{J_1+J_2}(Y^\dagger)^{J_1+J_2} )\rangle \right],
\nonumber
\end{eqnarray}
\begin{eqnarray}
e=f&=&{1\over (N^2-1)(N^2-4)}\left[ 
{2\over N}\langle \Tr (Y^{J_1}) \Tr (Y^{J_2}) \Tr ((Y^\dagger)^{J_1+J_2} )\rangle\right. \cr
&-&\langle \Tr (Y^{J_1+ J_2}) \Tr ((Y^\dagger)^{J_1+J_2} )\rangle 
-\langle \Tr (Y^{J_1}) \Tr (Y^{J_2}(Y^\dagger)^{J_1+J_2} )\rangle\cr
&-&\left. \langle \Tr (Y^{J_2}) \Tr (Y^{J_1}(Y^\dagger)^{J_1+J_2} )\rangle 
 +N\langle\Tr (Y^{J_1+J_2}(Y^\dagger)^{J_1+J_2} )\rangle \right],
\nonumber
\end{eqnarray}
%
%
The only new correlators appearing are of the type 
$\langle \Tr (Y^{J_2}) \Tr (Y^{J_1}(Y^\dagger)^{J_1+J_2} )\rangle$ and
$\langle \Tr (Y^{J_1}) \Tr (Y^{J_2}) \Tr ((Y^\dagger)^{J_1+J_2} )\rangle$. 
The free field theory Schwinger-Dyson equation
$$0=\int dY dY^\dagger {\partial\over\partial Y^i_j}\left[
\Tr (Y^{J_1})\Tr (Y^{J_2})((Y^{J_1+J_2-1})^\dagger )^i_j e^{-S}\right],\qquad S=\Tr (YY^\dagger),$$
implies that
$$\langle \Tr (Y^{J_1})\Tr (Y^{J_2})\Tr ((Y^{\dagger})^{ J_1+J_2})\rangle 
=J_1\langle \Tr (Y^{J_1-1} Y^{\dagger J_1+J_2-1})\Tr (Y^{J_2})\rangle $$
$$ +J_2\langle \Tr (Y^{J_2-1} Y^{\dagger J_1+J_2-1})\Tr (Y^{J_1})\rangle . $$
The exact three point function is known
\begin{eqnarray}
\langle \Tr(Y^{J_1})\Tr(Y^{J_2})\Tr((Y^{\dagger})^{J_1+J_2})\rangle &=& {1\over J_1+J_2+1}\left(
{(N+J_1+J_2)! \over (N-1)!}-{(N+J_2)! \over (N-J_1-1)!}\right.\cr
& &\left. -{(N+J_1)! \over (N-J_2-1)!}+{N!\over (N-J_1-J_2-1)!}\right).
\nonumber
\end{eqnarray}
We will not need these exact expressions. Rather,
assuming that ${J_i\over\sqrt{N}}\ll 1$ for $i=1,2$ we can easily estimate the leading order behavior
$$ \langle \Tr (Y^{J}) \Tr ((Y^\dagger)^{J} )\rangle =JN^J, $$
$$ \langle \Tr (Y^{J} (Y^\dagger)^{J} )\rangle =N^{J+1}, $$
$$ \langle \Tr (Y^{J_1}) \Tr (Y^{J_2}) \Tr ((Y^\dagger)^{J_1+J_2} )\rangle = N^{J_1+J_2-1}J_1 J_2 (J_1+J_2), $$
$$ \langle \Tr (Y^{J_1}) \Tr (Y^{J_2}(Y^\dagger)^{J_1+J_2} )\rangle = J_1 (J_2 +1)N^{J_1 +J_2},$$
and thus,
$$a=\left( J_1 J_2 (J_1+J_2) +2J_1J_2 +4\right)N^{J_1+J_2-4},$$
$$b=(J_1 +J_2 -2) N^{J_1+J_2-3},$$
$$c=(J_1 J_2 +J_1 -2) N^{J_1+J_2-3},$$
$$d=(J_1 J_2 +J_2 -2) N^{J_1+J_2-3},$$
$$e=f=N^{J_1+J_2-2}.$$
The leading contribution comes from the terms $e$ and $f$, as expected from their index structure.

Finally, for the amplitude for closed string emission from an excited D-brane state to leave another excited D-brane state,
we need to compute the leading contribution to the correlator $\langle \Tr(Y^{J_1})(Y^{J_2})^i_j (Y^{\dagger (J_1+J_2)})^l_k\rangle$. 
This is easily computed from the results we have so far, as 
$$ \langle \Tr(Y^{J_1})(Y^{J_2})^i_j (Y^{\dagger (J_1+J_2)})^l_k = J_1 (J_2+1)N^{J_1+J_2-2}\delta^i_k \delta^l_j .$$

\section{String Splitting Correlator}

To evaluate the amplitude for string splitting, we needed to evaluate the correlator
$$I={1\over \left( (n-2)!\right)^2}\sum_{\sigma\in S_n}\sum_{\tau\in S_{n-1}}\Tr_{R''}\left(\Gamma_R(\sigma)\right)
\Tr_{S'}\left(\Gamma_S(\tau)\right)^* \langle 
\prod_{k=1}^{n-2}Z^{i_k}_{i_{\sigma (k)}}(Z^\dagger)^{j_\tau(k)}_{j_{k}}\rangle
\delta^{i_{n}}_{i_{\sigma (n-1)}}\delta^{i_{n-1}}_{j_{n-1}}\delta^{j_{\tau(n-1)}}_{i_{\sigma (n)}}.$$
$R''$ is obtained by removing two blocks from $R$; $S'$ by removing one block from $S$. Denote the chain of subgroups
used for the two block restriction by $R\to R'\to R''$. The details of the computation of the above correlator are summarized 
in this appendix. In section 3.3, two terms contribute to the correlator. Although we only explicitly deal with one of the 
terms, for the piece of the correlator that we consider in this appendix (i.e. the piece which is independent of the open string 
words), the two contributions are equal. 

Set $\sigma=\psi P$ where $P=(n,n-1)$. Clearly,
$$\matrix{&\sigma (k)=\psi P(k) &=\psi (k), &k\le n-2\cr
          &                     &=\psi (n), &k=n-1   \cr
          &                     &=\psi (n-1), &k=n}$$
Thus,
$$I={1\over \left( (n-2)!\right)^2}\sum_{\sigma\in S_n}\sum_{\tau\in S_{n-1}}\Tr_{R''}\left(\Gamma_R(\sigma P)\right)
\Tr_{S'}\left(\Gamma_S(\tau)\right)^* \langle 
\prod_{k=1}^{n-2}Z^{i_k}_{i_{\sigma (k)}}(Z^\dagger)^{j_\tau(k)}_{j_{k}}\rangle
\delta^{i_{n}}_{i_{\sigma (n)}}\delta^{i_{n-1}}_{j_{n-1}}\delta^{j_{\tau(n-1)}}_{i_{\sigma (n-1)}}.$$
Introducing
$$\tilde{\chi}_{R,R''}^{(1)}(Z,W)=
{1\over (n-1)!}\sum_{\sigma\in S_n}\Tr_{R''}\left(\Gamma_R(\sigma P)\right)
Z^{i_1}_{i_{\sigma (1)}}\cdots Z^{i_{n-1}}_{i_{\sigma (n-1)}}W^{i_n}_{i_{\sigma (n)}},$$
$$\chi_{S,S'}(Z)=
{1\over (n-1)!}\sum_{\sigma\in S_{n-1}}\Tr_{S'}\left(\Gamma_S(\sigma )\right)
Z^{i_1}_{i_{\sigma (1)}}\cdots Z^{i_{n-1}}_{i_{\sigma (n-1)}},$$
the above correlator can be interpreted as $I=(n-1)^2 \langle D_W \tilde{\chi}_{R,R''}^{(1)}\chi_{S,S'}^\dagger \rangle\Big|_{n-1}$.
The reduction formula for $\tilde{\chi}_{R,R''}^{(1)}$ can be derived in exactly the same way as for restricted Schur polynomials;
the result is the same:
$$(n-1)^2 \langle D_W \tilde{\chi}_{R,R''}^{(1)}\chi_{S,S'}^\dagger \rangle\Big|_{n-1}=c_{RR'}(n-1)^2
\langle \tilde{\chi}_{R,R''}\chi_{S,S'}^\dagger \rangle\Big|_{n-1},$$
where $c_{RR'}$ is the weight of the box that must be removed from $R$ to obtain $R'$ and
$$\tilde{\chi}_{R,R''}(Z)=
{1\over (n-1)!}\sum_{\sigma\in S_n|_{\alpha(n)=n}}\Tr_{R''}\left(\Gamma_R(\sigma P)\right)
Z^{i_1}_{i_{\sigma (1)}}\cdots Z^{i_{n-1}}_{i_{\sigma (n-1)}}=\tilde{\chi}_{R',R''}(Z).$$
Using the techniques of appendix E, we find
\begin{eqnarray}
\langle \tilde{\chi}_{R',R''}\chi_{S,S'}^\dagger \rangle\Big|_{n-1}
&=& {1\over \left((n-1)!\right)^2}\sum_{\tau,\beta\in S_{n-1}}
\sum_{\alpha\in S_{n-1}|_{\alpha (n-1)=n-1}}\Tr_{R''}\left(\Gamma_{R}(\alpha^{-1}\tau\alpha\beta P)\right)\cr
&&\qquad \times \Tr_{S'} \left(\Gamma_S (\tau )\right)^* N^{C(\beta )}.
\nonumber
\end{eqnarray}
Using the fundamental orthogonality relation exactly as in appendix E, it is not difficult to compute the sum
$$ \sum_{\tau\in S_{n-1}}\Tr_{R''}\left(\Gamma_{R}(\alpha^{-1}\tau\alpha\beta P)\right)
   \Tr_{S'} \left(\Gamma_S (\tau )\right)^* =\delta_{R'S}\delta_{R''S'} {(n-1)!\over d_{R'}}
\Tr_{R''}\left(\Gamma_{R}(\beta P)\right), $$
and hence to obtain (we have summed over $\alpha$)
\begin{equation}
\langle \tilde{\chi}_{R',R''}\chi_{S,S'}^\dagger \rangle\Big|_{n-1}
= \delta_{R'S}\delta_{R''S'}
{1\over (n-1)d_{R'}}\sum_{\beta\in S_{n-1}}\Tr_{R''}\left(\Gamma_{R}(\beta P)\right)
N^{C(\beta )}.
\label{RCorr}
\end{equation}
To compute this last sum, note that $P$ commutes with every element of $S_{n-2}$ and thus by Schur's Lemma, 
when acting on the $R''$ subspace we can replace
$$ \Gamma_R(P)|_{R''}=\Gamma_R\left( (n,n-1)\right)|_{R''} \to{\Tr_{R''}\left((n,n-1)\right)\over d_{R''}}{\bf 1}_{R''},$$
with ${\bf 1}_{R''}$ the identity on the $R''$ subspace. To use this replacement in (\ref{RCorr}), use the techniques of
appendix F to write
$$\sum_{\beta\in S_{n-1}}\Tr_{R''}\left(\Gamma_{R}(\beta P)\right)
N^{C(\beta )}=c_{R',R''}\sum_{\beta\in S_{n-2}}\Tr_{R''}\left(\Gamma_{R}(\beta P)\right)
N^{C(\beta )}.$$
Since $\beta\in S_{n-2}$, $\Gamma_{R}(\beta )$ is now diagonal in the $R''$ indices, that is, $P$ is acting only on the $R''$
subspace. It now follows that
\begin{eqnarray}
\langle \tilde{\chi}_{R',R''}\chi_{S,S'}^\dagger \rangle\Big|_{n-1}
&=& \delta_{R'S}\delta_{R''S'}{\Tr_{R''}\left((n,n-1)\right)\over d_{R''}}
{c_{R'R''}\over (n-1)d_{R'}}\sum_{\beta\in S_{n-2}}\Tr_{R''}\left(\Gamma_{R}(\beta )\right)
N^{C(\beta )}\cr
&=&
\delta_{R'S}\delta_{R''S'}\Tr_{R''}\left((n,n-1)\right)
{f_{R'}\over (n-1)d_{R'}},
\nonumber
\end{eqnarray}
Using the results of appendix D, we now obtain
\begin{eqnarray}
I&=&(n-1)\delta_{R'S}\delta_{R''S'}\Tr_{R''}\left((n,n-1)\right)
{f_{R}\over d_{R'}},\cr
&=&(n-1){f_{R}d_{R''}\over d_{R'}(c_{R,R'}-c_{R',R''})}\delta_{R'S}\delta_{R''S'}.
\nonumber
\end{eqnarray}

\section{Product Rule for Restricted Schur Polynomials} 

In this appendix we start by reviewing the known product rule for Schur polynomials. Our main interest in this
product rule is due to the fact that it allows a very simple computation of a class of three point functions.
We then explain how a similar rule can be obtained for the restricted Schur polynomials, allowing the
computation of a class of restricted Schur polynomial three point correlators. 

The product rule for Schur polynomials follows as a direct consequence of the duality between the symmetric groups and
the unitary groups. To develop this duality, consider the space ${\rm Sym}(V^{\otimes n})$ where $V$ denotes the fundamental 
representation of $U(N)$. This space is a representation of $U(N)$ and also admits an action of $S_n$. The duality between
the symmetric and unitary groups is a direct consequence of the fact that the actions of $U(N)$ and $S_n$ on
${\rm Sym}(V^{\otimes n})$ commute, allowing them to be simultaneously diagonalized. This also explains why Young diagrams 
label both representations of $U(N)$ and $S_n$. As a consequence of this duality, the Schur polynomials are the characters
of the unitary group in their irreducible representations. The decomposition of a product of characters into a sum over
characters is exactly the same as the decomposition of a product of two irreducible representations of $U(N)$ (which is
in general reducible) into its irreducible components. For irreducible representations $R,S$ of the unitary group, it is well
known that
$$ R\otimes S  =\oplus_T f(R,S;T) T,$$
where $f(R,S;T)$ are the Littlewood-Richardson coefficients. From the interpretation of the Schur polynomials as characters, we 
immediately obtain
$$\chi_R (Z)\chi_S (Z)=\sum_T f (R,S;T)\chi_T (Z).$$
Using the two point correlator
$$ \langle\chi_T(Z)\chi_U(Z^\dagger )\rangle =\delta_{TU}f_T,$$
we find
$$ \langle\chi_R (Z)\chi_S (Z)\chi_T(Z^\dagger )\rangle =f(R,S;T) f_T ,$$
in agreement with \cite{Corley:2001zk}.

Clearly, a product rule for the restricted Schur polynomials would allow an efficient evaluation of (a class of) three point functions
of restricted Schur polynomials. It is not obvious that such a rule even exists; indeed, the fact that the product of two Schur polynomials
is again a Schur polynomial is related to the fact that the Schur polynomials furnish a basis for the symmetric functions. 
To explore this issue,
consider the simplest case, when an ordinary Schur polynomial is multiplied by a restricted Schur polynomial with a single string
attached, $\chi_R(Z)\chi^{(1)}_{S,S'}(Z,W)$. Denote the number of boxes in $R$ by $n_R$, the number of boxes in $S$ by $n_S$ and the number of
boxes in $S'$ by $n_{S'}$. It is obvious that $n_S=n_{S'}+1$. The product $\chi_R(Z)\chi^{(1)}_{S,S'}(Z,W)$ will be a sum of terms of the form
\begin{equation}
\Tr (Z^{n_1-1}W)\prod_{i=2}^k \Tr (Z^{n_i}),\qquad {\rm where}\qquad \sum_{i=1}^k n_i =n_R+n_{S},
\label{mon}
\end{equation}
and $W$ is the open string word. It is clear that the number of such monomials is equal to the number of partitions of $n_R+n_S$
times the number of choices of which trace will contain $W$. The number of restricted Schur polynomials is given by the number
of representation $T$ of $S_{n_R+n_S}$ - which is equal to the number of partitions of $n_R+n_S$ times the number of ways we can
restrict $T$ to $T'$ by removing a box. It is now obvious that the number of restricted Schur polynomials 
with one string attached is equal to the number 
of monomials of the form (\ref{mon}). This suggests that it should indeed be possible to express $\chi_R(Z)\chi^{(1)}_{S,S'}(Z,W)$ 
as a sum over restricted Schur polynomials. To perform the required inversion, introduce the matrix
\begin{equation}
(M^{-1})_{\sigma\tau}=\sum_{(T,T')}\chi_{T,T'}(\sigma )\chi_{T,T'}(\tau),
\label{defM}
\end{equation}
where the sum on the right hand side runs over all possible labels $(T,T')$ allowed for a restricted Schur polynomial.
Using this matrix, define the ``dual character" (which we denote by a superscript)
$$ \chi^{T,T'}(\sigma )=\sum_{\tau}M_{\sigma\tau}\chi_{T,T'}(\tau ).$$
Let $\sigma$ denote the permutation for which
$$ \Tr (\sigma WZ^{\otimes n_R +n_{S'}})= \Tr (Z^{n_1}W)\prod_{i=2}^k \Tr (Z^{n_i}),\qquad\sum_{i=1}^k n_i =n_R+n_{S'} .$$
If two permutations $\sigma,\tau$ satisfy
$$\Tr (\sigma WZ^{\otimes n_R +n_{S'}}) = \Tr (\tau WZ^{\otimes n_R +n_{S'}}),$$
we say they are {\it restricted conjugate}. Restricted conjugate is an equivalence relation. Denote the number of
restricted conjugate classes by $N_{T,T'}$ and let $n^\sigma_{T,T'}$ denote the number of elements in restricted
conjugate class with representative $\sigma$. $N_{T,T'}$ is also equal to the number of restricted Schur polynomials.
The ``indices'' $\sigma$ and $\tau$ that appear in (\ref{defM}) run over the restricted conjugacy classes. 
Then, using the definition of $M$, we easily find
$$\sum_{(T,T')}\chi^{T,T'}(\sigma )\chi^{(1)}_{T,T'}(Z,W) = {n^\sigma_{T,T'}\over (n_T-1)!}\Tr (\sigma WZ^{\otimes n_R +n_{S'}}).$$
This proves that, as long as the matrix $(M^{-1})_{\sigma\tau}$ is invertible, the product $\chi_R\chi_{S,S'}$
can be expressed as a sum of restricted Schur polynomials. 

We have checked that $\det (M)$ is non-zero for $n_T=2,3,4$ and $5$. It is not difficult to check, for example, that
$$\chi_{\yng(1)}\chi_{\young({\,}1)}={2\over 3}\chi_{\young({\,}{\,}1)}+\chi_{\young({\,}1,{\,})}+{1\over 3}\chi_{\young({\,}{\,},1)},$$
$$\chi_{\yng(1)}\chi_{\young({\,},1)}={1\over 3}\chi_{\young({\,}1,{\,})}+\chi_{\young({\,}{\,},1)}+{2\over 3}\chi_{\young({\,},{\,},1)}.$$
It is now a simple matter to compute the three point correlators
$$\langle \chi_{\yng(1)}\chi_{\young({\,}1)}\chi^\dagger_{\young({\,}1,{\,})}\rangle ={3\over 2}N(N^2-1)F_0+(N+1)^2 N(N-1)F_1 ,$$
$$\langle\chi_{\yng(1)}\chi_{\young({\,},1)}\chi^\dagger_{\young({\,}1,{\,})}\rangle ={1\over 2}N(N^2-1)F_0+(N-1)^2 N(N+1)F_1 .$$

In this appendix we have provided strong evidence for the existence of a product rule which would resolve the product of a restricted
Schur polynomial (with a single string attached) with a Schur polynomial, in terms of restricted Schur polynomials with a single string 
attached. We have given two simple examples of these products and have used the products to compute two three point functions.
We hope to report on an explicit rule and to extend this analysis in a future article\cite{milena}.

\section{Numerical Results}

To provide a test of the technology developed in this article, we have developed a code to numerically evaluate the two
point functions of restricted Schur polynomials. In this appendix we will quote some of the results obtained using this
code. Our goal in quoting these results was both to give the reader confidence in our results and to provide some answers 
for specific computations that can be used by the reader who wishes to test her understanding. We have numerically 
computed all restricted Schur polynomials (and their correlators)
that can be obtained from representations with 5 boxes or less, by attaching one 
or two strings. We will not give an exhaustive list of these results.

Our notation for the representations of $S_4$, $S_3$ and $S_2$ is
$$R_1=\yng(4),\quad R_2=\yng(3,1),\quad R_3=\yng(2,2),\quad R_4=\yng(2,1,1),\quad R_5=\yng(1,1,1,1),$$
$$S_1=\yng(3),\quad S_2=\yng(2,1),\quad S_3=\yng(1,1,1),$$
$$T_1=\yng(2),\quad T_2=\yng(1,1).$$

\subsection{One String Attached}

$$\langle \chi^{(1)}_{A,B}(\chi^{(1)}_{C,D})^\dagger \rangle=\alpha_{AB,CD}F_0+\beta_{AB,CD}F_1$$
\begin{center}
\begin{tabular}{|c|c|c|c|}
\hline
$A=C$ & $B=D$ & $\alpha_{AB,CD}$ & $\beta_{AB,CD}$ \\
\hline
$R_1$ & $S_1$ & $4N^4+24N^3+44N^2+24N$ & $N^5+9N^4+29N^3+39N^2+18N$\\
\hline
$R_2$ & $S_1$ & ${4\over 3}N^4+{8\over 3}N^3-{4\over 3}N^2-{8\over 3}N$ & $N^5+N^4-3N^3-N^2+2N$\\
\hline
$R_2$ & $S_2$ & ${8\over 3}N^4+{16\over 3}N^3-{8\over 3}N^2-{16\over 3}N$ & $N^5+4N^4+3N^3-4N^2-4N$\\
\hline
$R_3$ & $S_2$ & $4N^4-4N^2$ & $N^5-N^3$\\
\hline
$R_4$ & $S_2$ & ${8\over 3}N^4-{16\over 3}N^3-{8\over 3}N^2+{16\over 3}N$ & $N^5-4N^4+3N^3+4N^2-4N$\\
\hline
$R_4$ & $S_3$ & ${4\over 3}N^4-{8\over 3}N^3-{4\over 3}N^2+{8\over 3}N$ & $N^5-N^4-3N^3+N^2+2N$\\
\hline
$R_5$ & $S_3$ & $4N^4-24N^3+44N^2-24N$ & $N^5-9N^4+29N^3-39N^2+18N$\\
\hline
\end{tabular}
\end{center}
\begin{center}
\begin{tabular}{|c|c|c|c|c|c|}
\hline
$A$ & $B$ & $C$ & $D$ & $\alpha_{AB,CD}$ & $\beta_{AB,CD}$ \\
\hline
$R_2$ & $S_1$ & $R_1$ & $S_1$ & $0$ & $N^5+5N^4+5N^3-5N^2-6N$\\
\hline
$R_2$ & $S_2$ & $R_3$ & $S_2$ & $0$ & $N^5+2N^4-N^3-2N^2$\\
\hline
$R_3$ & $S_2$ & $R_4$ & $S_2$ & $0$ & $N^5-2N^4-N^3+2N^2$\\
\hline
$R_2$ & $S_2$ & $R_2$ & $S_1$ & $0$ & $0$\\
\hline
\end{tabular}
\end{center}

\subsection{Two Strings Attached}

$$\langle \chi^{(1)}_{A,B}(\chi^{(1)}_{C,D})^\dagger \rangle=\alpha_{AB,CD}F_0^1 F_0^2
+\beta_{AB,CD}F_0^1 F_1^2 +\gamma_{AB,CD}F_1^1 F_0^2 +\delta_{AB,CD}F_1^1 F_1^2 .$$
For the two string examples, we describe the chain of subgroups participating in the restriction. 
We always assume that
$$\langle (W^{(i)})^a_b(W^{(i')\dagger})^{a'}_{b'}\rangle\propto\delta^{ii'}.$$
\begin{center}
\begin{tabular}{|c|c|c|}
\hline
Restriction of $\chi^{(1)}_{A,B}$ & Restriction of $(\chi^{(1)}_{C,D})^\dagger$ & $\alpha_{AB,CD}$ \\
\hline
$R_2\to S_2\to T_2$ & $R_2\to S_2\to T_2$ & $4N^4+8N^3-4N^2-8N$\\
\hline
$R_2\to S_2\to T_1$ & $R_2\to S_2\to T_1$ & $4N^4+8N^3-4N^2-8N$\\
\hline
$R_2\to S_2\to T_2$ & $R_2\to S_2\to T_1$ & $0$\\
\hline
\end{tabular}
\end{center}
\begin{center}
\begin{tabular}{|c|c|c|}
\hline
Restriction of $\chi^{(1)}_{A,B}$ & Restriction of $(\chi^{(1)}_{C,D})^\dagger$ & $\gamma_{AB,CD}$ \\
\hline
$R_2\to S_2\to T_2$ & $R_2\to S_2\to T_2$ & ${3\over 2}N^5+6N^4+{9\over 2}N^3-6N^2-6N$\\
\hline
$R_2\to S_2\to T_1$ & $R_2\to S_2\to T_1$ & ${3\over 2}N^5+6N^4+{9\over 2}N^3-6N^2-6N$\\
\hline
$R_2\to S_2\to T_2$ & $R_2\to S_2\to T_1$ & $0$\\
\hline
\end{tabular}
\end{center}
\begin{center}
\begin{tabular}{|c|c|c|}
\hline
Restriction of $\chi^{(1)}_{A,B}$ & Restriction of $(\chi^{(1)}_{C,D})^\dagger$ & $\beta_{AB,CD}$ \\
\hline
$R_2\to S_2\to T_2$ & $R_2\to S_2\to T_2$ & ${3\over 2}N^5+6N^4+{9\over 2}N^3-6N^2-6N$\\
\hline
$R_2\to S_2\to T_1$ & $R_2\to S_2\to T_1$ & ${43\over 18}N^5+{22\over 9}N^4-{127\over 18}N^3-{22\over 9}N^2+{14\over 3}N$\\
\hline
$R_2\to S_2\to T_2$ & $R_2\to S_2\to T_1$ & $0$\\
\hline
\end{tabular}
\end{center}
\begin{center}
\begin{tabular}{|c|c|c|}
\hline
Restriction of $\chi^{(1)}_{A,B}$ & Restriction of $(\chi^{(1)}_{C,D})^\dagger$ & $\delta_{AB,CD}$ \\
\hline
$R_2\to S_2\to T_2$ & $R_2\to S_2\to T_2$ & $N^6+5N^5+7N^4-N^3-8N^2-4N$\\
\hline
$R_2\to S_2\to T_1$ & $R_2\to S_2\to T_1$ & $N^6+3N^5-N^4-7N^3+4N$\\
\hline
$R_2\to S_2\to T_2$ & $R_2\to S_2\to T_1$ & $0$\\
\hline
\end{tabular}
\end{center}
In our last two string example, we will consider the correlator
$$\langle\chi_{\young({\,}{\,}{\onetwo},{\twoone})}\chi_{\young({\,}{\,}{\onetwo},{\twoone})}^\dagger\rangle=
F_0^1 F_0^2 (4N^4+8N^3-4N^2-8N)+F_0^1 F_1^2 ({4\over 9}N^5+{8\over 9}N^4-{4\over 9}N^3-{8\over 9}N^2).$$

\subsection{String Splitting Correlators}
We use the following notation for representations of $S_5$
$$Q_1=\yng(5),\quad Q_2=\yng(4,1),\quad Q_3=\yng(3,2),\quad Q_4=\yng(3,1,1),$$
$$ Q_5=\yng(2,2,1),\quad Q_6=\yng(2,1,1,1),\quad Q_7=\yng(1,1,1,1,1).$$
The correlator is
$$\langle \chi^{(2)}_{R,R''}(\chi^{(1)}_{S,S'})^\dagger \rangle=\alpha_{RR'',SS'}F .$$
We have assumed that the open string factor in each of the two terms which contribute to the leading result, coming from contracting 
the words describing the open strings, (the terms denoted $e$ and $f$ in appendix $G$) are equal.
In the tables below, we describe the chain of subgroups participating in the restriction. 
\begin{center}
\begin{tabular}{|c|c|c|}
\hline
Restriction of $\chi^{(1)}_{R,R''}$ & Restriction of $(\chi^{(1)}_{S,S'})^\dagger$ & $\alpha_{RR'',SS'}$ \\
\hline
$Q_1\to R_1\to S_1$ & $R_1\to S_1$ & $8N^5+80N^4+280N^3+400N^2+192N$\\
\hline
$Q_2\to R_1\to S_1$ & $R_1\to S_1$ & $-2N^5-10N^4-10N^3+10N^2+12N$\\
\hline
$Q_2\to R_2\to S_1$ & $R_1\to S_1$ & $0$\\
\hline
$Q_2\to R_2\to S_2$ & $R_2\to S_2$ & ${16\over 3}(N^5+5N^4+5N^3 -5N^2 -6N)$\\
\hline
$Q_3\to R_2\to S_2$ & $R_2\to S_2$ & $-{8\over 3}(N^5+2N^4-N^3-2N^2)$\\
\hline
$Q_4\to R_4\to S_2$ & $R_4\to S_2$ & ${4\over 3}\left( N^5 -5N^3 +4N\right) $\\
\hline
\end{tabular}
\end{center}

\subsection{Three Point Correlators}

$$\langle\chi_{\yng(1)}\chi^{(1)}_{T,{}_{\yng(1)}}\chi^{(1)\dagger}_{R,R'}\rangle =\alpha_{T,RR'}F_0 +\beta_{T,RR'}F_1 .$$

\begin{center}
\begin{tabular}{|c|c|c|c|}
\hline
$T$ & Restriction of $(\chi^{(1)}_{R,R'})^\dagger$ & $\alpha_{T,RR'}$ & $\beta_{T,RR'}$ \\
\hline
$T_1$ & $S_1\to T_1$ & $2N^3+6N^2+4N$ & $N^4+4N^3+5N^2+2N$\\
\hline
$T_1$ & $S_2\to T_1$ & ${1\over 2}N^3-{1\over 2}N$ & $N^4+N^3-N^2-N$\\
\hline
$T_1$ & $S_2\to T_2$ & ${3\over 2}N^3-{3\over 2}N$ & $N^4+N^3-N^2-N$\\
\hline
$T_1$ & $S_3\to T_2$ & $0$ & $N^4-2N^3-N^2+2N$\\
\hline
$T_2$ & $S_1\to T_1$ & $0$ & $N^4+2N^3-N^2-2N$\\
\hline
$T_2$ & $S_2\to T_1$ & ${3\over 2}N^3-{3\over 2}N$ & $N^4-N^3-N^2+N$\\
\hline
$T_2$ & $S_2\to T_2$ & ${1\over 2}N^3-{1\over 2}N$ & $N^4-N^3-N^2+N$\\
\hline
$T_2$ & $S_3\to T_2$ & $2N^3-6N^2+4N$ & $N^4-4N^3+5N^2-2N$\\
\hline
\end{tabular}
\end{center}

\subsection{Three String Correlators}

$$\langle\chi_{\young({\,}{\,}{1},{\,}{2},{3})}^\dagger \chi_{\young({\,}{\,}{1},{\,}{2},{3})}\rangle
=F_0^1F_0^2F_0^3 (15N^6-75N^4+60N^2)$$
$$+F_1^1F_0^2F_0^3 (8N^7 +16N^6-40N^5-80N^4+32N^3+64N^2)$$
$$+F_0^1F_1^2F_0^3({17\over 4}N^7+N^6-{85\over 4}N^5-5N^4+17N^3+4N^2)$$
$$+F_1^1F_1^2F_0^3 ({8\over 3}N^8+{16\over 3}N^7-{40\over 3}N^6-{80\over 3}N^5+{32\over 3}N^4+{64\over 3}N^3)$$
$$+F_0^1F_0^2F_1^3 ({1097\over 256}N^7-{247\over 32}N^6-{5485\over 256}N^5+{1235\over 32}N^4+{1097\over 64}N^3-{247\over 8}N^2)$$
$$+F_1^1F_0^2F_1^3 ({29\over 12}N^8+{1\over 3}N^7-{253\over 12}N^6-{5\over 3}N^5+{164\over 3}N^4+{4\over 3}N^3-36N^2)$$
$$+F_0^1F_1^2F_1^3 ({1145\over 768}N^8-{247\over 96}N^7-{6157\over 768}N^6+{1235\over 96}N^5+{1685\over 192}N^4-{247\over 24}N^3-{9\over 4}N^2)$$
$$+F_1^1F_1^2F_1^2 (N^9-9N^7+24N^5-16N^3).$$

$$ $$

$$\langle\chi_{\young({\,}{\,}{2},{\,}{1},{3})}^\dagger \chi_{\young({\,}{\,}{2},{\,}{1},{3})}\rangle
=F_0^1F_0^2F_0^3 (15N^6-75N^4+60N^2)$$
$$+F_1^1F_0^2F_0^3 ({20\over 3}N^7 -{100\over 3}N^5+{80\over 3}N^3)$$
$$+F_0^1F_1^2F_0^3({59\over 12}N^7+9N^6-{295\over 12}N^5-45N^4+{59\over 3}N^3+36N^2)$$
$$+F_1^1F_1^2F_0^3 ({8\over 3}N^8+{16\over 3}N^7-{40\over 3}N^6-{80\over 3}N^5+{32\over 3}N^4+{64\over 3}N^3)$$
$$+F_0^1F_0^2F_1^3 ({3371\over 768}N^7-{63\over 8}N^6-{16855\over 768}N^5+{315\over 8}N^4+{3371\over 192}N^3-{63\over 2}N^2)$$
$$+F_1^1F_0^2F_1^3 ({113\over 48}N^8-{14\over 3}N^7-{565\over 48}N^6+{70\over 3}N^5+{113\over 12}N^4-{56\over 3}N^3)$$
$$+F_0^1F_1^2F_1^3 ({1169\over 768}N^8-{7\over 96}N^7-{9733\over 768}N^6+{35\over 96}N^5+{6029\over 192}N^4-{7\over 24}N^3-{81\over 4}N^2)$$
$$+F_1^1F_1^2F_1^2 (N^9-9N^7+24N^5-16N^3).$$

\end{document}